\documentclass[12pt]{article}

\usepackage{authblk}
\usepackage{scicite}
\usepackage{float}
\usepackage{placeins}
\usepackage{url}
\usepackage{lscape}
\usepackage{longtable}
\usepackage{threeparttable} 
\usepackage[normalem]{ulem}
\usepackage{amsmath}
\newcommand\blueout{\bgroup\markoverwith
{\textcolor{blue}{\rule[.5ex]{2pt}{2pt}}}\ULon}
\newcommand\redout{\bgroup\markoverwith
{\textcolor{red}{\rule[.5ex]{2pt}{2pt}}}\ULon}
\usepackage{xcolor}
\usepackage{ulem}

\definecolor{myblue}{RGB}{130,130,255} 
\newcommand\REV[1]{\textcolor{black}{#1}}
\newcommand{\REVb}[1]{\textcolor{black}{#1}}

\def\aap{Astron. \& Astrophys.} 
\newcommand{\aj}{Astron. J.}
\newcommand{\apjs}{Astrophys. J.}
\newcommand{\apj}{Astrophys. J.}
\newcommand{\apjl}{Astrophys. J.}
\newcommand{\mnras}{Mon. Not. R. Astro. Soc.}
\newcommand{\pasa}{Publ. Astron. Soc. Australia}
\newcommand{\nat}{Nature}

\newcommand{\araa}{Annu. Rev. Astron. Astrophys.}
\newcommand{\raa}{Rev. Astron. Astrophys.}
\newcommand{\aapr}{Astronomy and Astrophysics Reviews}
\newcommand{\aaps}{Astronomy and Astrophysics Supplement}
\newcommand{\solphys}{Solar Physics}
\newcommand{\ssr}{Space Science Reviews}

\def \arcsec {''}
\newcommand{\ion}[2]{[#1\,\textsc{#2}]}

\usepackage{graphicx}
\usepackage{amsmath,amssymb} 
\usepackage{times}

\newenvironment{sciabstract}{%
\begin{quote} \bf}
{\end{quote}}

\title{\REV{A sudden dramatic change and recovery of magneto-environment of a repeating fast radio burst}}

\author
{Y. Li,$^{1,2\dagger}$ S. B. Zhang,$^{1,3\dagger}$ Y. P. Yang,$^{4\dagger}$
C. W. Tsai,$^{5,6}$ X. Yang,$^{1,2}$ C. J. Law,$^{7,8}$ R.~Anna-Thomas,$^{9,10,11,12}$ X. L. Chen,$^{5}$ K. J. Lee,$^{13,5,14}$ Z. F. Tang,$^{1,2}$ D. Xiao,$^{1}$
H.~Xu,$^{5}$ X. L. Yang,$^{15}$ G. Chen,$^{1}$ Y. Feng,$^{16,17}$ D. Z. Li,$^{18,19}$ R. Mckinven,$^{20,21}$ J.~R.~Niu,$^{5}$ K. Shin,$^{22,23}$ B. J. Wang,$^{5}$ C. F. Zhang,$^{24,13,5}$ Y. K. Zhang,$^{5}$ D. J. Zhou,$^{5}$ Y. H. Zhu,$^{5}$ Z. G. Dai,$^{25}$ C. M. Chang,$^{1,2}$ J. J. Geng,$^{1}$ J. L. Han,$^{5}$ L. Hu,$^{26, 27}$ D.~Li,$^{28,5}$ R. Luo,$^{29}$ C. H. Niu,$^{5}$ D. D. Shi,$^{30}$ T. R. Sun,$^{1}$ 
X. F. Wu,$^{1,2\ast}$ W.~W.~Zhu,$^{5,6\ast}$ P. Jiang,$^{5,31\ast}$ B. Zhang$^{32,33,34,35\ast}$\\
\normalsize{$^{1}$Purple Mountain Observatory, Chinese Academy of Sciences, Nanjing, P.~R.~China}\\
\normalsize{$^{2}$School of Astronomy and Space Sciences, University of Science and Technology of China, Hefei, P.~R.~China}\\
\normalsize{$^{3}${Australia Telescope National Facility, Commonwealth Scientific and Industrial Research Organisation–Space and Astronomy, Epping, Australia}}\\ 
\normalsize{$^{4}${South-Western Institute for Astronomy Research, Yunnan Key Laboratory of Survey Science, Yunnan University, Kunming, P.~R.~China}}\\
\normalsize{$^{5}$National Astronomical Observatories, Chinese Academy of Sciences, Beijing, P.~R.~China}\\
\normalsize{$^{6}$Institute for Frontiers in Astronomy and Astrophysics, Beijing Normal University, Beijing, P.~R.~China }\\
\normalsize{$^{7}$Cahill Center for Astronomy and Astrophysics, California Institute of Technology, Pasadena, USA }\\
\normalsize{$^{8}$Owens Valley Radio Observatory, California Institute of Technology, Big Pine, USA }\\
\normalsize{$^{9}$Department of Physics and Astronomy, West Virginia University, Morgantown, USA }\\
\normalsize{$^{10}$Center for Gravitational Waves and Cosmology, West Virginia University, Morgantown, USA }\\
{\normalsize{$^{11}$ Netherlands Institute for Radio Astronomy, Dwingeloo, The Netherlands}}\\
{\normalsize{$^{12}$Anton Pannekoek Institute for Astronomy, University of Amsterdam, Amsterdam, The Netherlands}}\\
\normalsize{$^{13}$Department of Astronomy, Peking University, Beijing, P.~R.~China}\\
\normalsize{$^{14}$Kavli Institute for Astronomy and Astrophysics, Peking University, Beijing, P.~R.~China }\\
\normalsize{$^{15}$Shanghai Astronomical Observatory, Chinese Academy of Sciences, Shanghai, P.~R.~China}\\
{\normalsize{$^{16}$Research Center for Astronomical Computing, Zhejiang Laboratory, Hangzhou, P.~R.~China }}\\
{\normalsize{$^{17}$Institute for Astronomy, School of Physics, Zhejiang University, Hangzhou, P.~R.~China }}\\
\normalsize{$^{18}$Department of Astrophysical Sciences, Princeton University, Princeton, USA }\\
\normalsize{$^{19}$Department of Astronomy, Tsinghua University, Beijing, P.~R.~China}\\
\normalsize{$^{20}$Department of Physics, McGill University, Montr\'eal, Canada}\\
\normalsize{$^{21}$Trottier Space Institute, McGill University, Montr\'eal, Canada}\\
\normalsize{$^{22}$Kavli Institute for Astrophysics and Space Research, Massachusetts Institute of Technology, Cambridge, USA }\\
\normalsize{$^{23}$Department of Physics, Massachusetts Institute of Technology, Cambridge, USA }\\
\normalsize{$^{24}$National Time Service Center, Chinese Academy of Sciences, Xi’an, China}\\
\normalsize{$^{25}$Department of Astronomy, University of Science and Technology of China, Hefei, P.~R.~China}\\
\normalsize{$^{26}$Department of Physics and Astronomy, University of Pennsylvania, Philadelphia, USA}\\
\normalsize{$^{27}$McWilliams Center for Cosmology, Department of Physics, Carnegie Mellon University, Pittsburgh, USA }\\
\normalsize{$^{28}$New Cornerstone Science Laboratory, Department of Astronomy, Tsinghua University, Beijing, P.~R.~China}\\
\normalsize{$^{29}$Department of Astronomy, School of Physics and Materials Science, Guangzhou University, Guangzhou, P.~R.~China }\\
\normalsize{$^{30}$Center for Fundamental Physics, School of Mechanics and Optoelectric Physics, Anhui University of Science and Technology, Huainan, P.~R.~China}\\
\normalsize{$^{31}$Guizhou Radio Astronomical Observatory, Guizhou University, Guiyang, P.~R.~China }\\
{\normalsize{$^{32}$The Hong Kong Institute for Astronomy and Astrophysics, The University of Hong Kong, Pokfulam, Hong Kong, P.~R.~China}}\\
{\normalsize{$^{33}$Department of Physics, The University of Hong Kong, Pokfulam, Hong Kong, P.~R.~China}}\\
\normalsize{$^{34}$Nevada Center for Astrophysics, University of Nevada, Las Vegas, USA}\\
\normalsize{$^{35}$Department of Physics and Astronomy, University of Nevada, Las Vegas, USA}\\
\normalsize{$^\ast${Corresponding author. Email:}  xfwu@pmo.ac.cn (X.F.W.); zhuww@nao.cas.cn(W.W.Z.); pjiang@nao.cas.cn (P.J.); 
{bzhang1@hku.hk(B.Z.)}} \\
\normalsize{$^\dagger$These authors contributed equally to this work.} 
}
\date{}

\begin{document} 


\baselineskip24pt


\maketitle 


\begin{sciabstract} 
Fast radio bursts (FRBs) are millisecond-duration radio bursts with unidentified extra-galactic origin. 
Some FRBs exhibit mild magneto-ionic environmental variations, possibly attributed to plasma turbulence or 
binary configuration.
We report an abrupt magneto-ionic 
variation of FRB 20220529, a repeating FRB from a disk galaxy at redshift \REVb{$0.1839 \pm 0.0001$}. 
Initially, the Faraday rotation measure (RM) had a median of $17~{\rm rad~m^{-2}}$ and a scatter of $101~{\rm rad~m^{-2}}$ over 17 months. 
In December 2023, it jumped to \REVb{$1977 \pm 84~{\rm rad~m^{-2}}$}, 
and returned to typical values within two weeks. 
\REV{This drastic RM variation suggests that a dense magnetized clump enters and exits the line of sight in week timescales. 
\REVb{One} plausible scenario invokes a coronal mass ejection from a companion star, 
\REVb{while} other scenarios invoking extreme turbulence or binary orbital motion are \REVb{also possible}.}
\end{sciabstract}


One sentence summary: 
\REV{ \REVb{a sudden} substantial variance in the Faraday rotation measure over several weeks indicates an eruptive magneto-ionic environment surrounding a fast radio burst.} 


Fast radio bursts (FRBs) are cosmological radio bursts with millisecond durations \cite{Lorimer07Sci,Petroff19AAR,Cordes19ARAA}.
Some repeat, while others apparently do not \cite{Spitler16Nat,chime2021}. 
\REV{Although their origin remains unknown \cite{ZhangB2023}, the association of FRB~20200428D with Milky Way magnetar SGR 1935+2154 (a highly magnetized neutron star) \cite{CHIME20,Bocheneck20,Li21} suggests that at least some FRBs are produced by magnetars. A fraction of repeating FRBs \cite{Spitler16Nat,CHIMErepeaters} display very high repetition rates \cite{LiDi2021,xuheng2022}, suggesting an active central engine. At least some active repeaters reside in star-forming host galaxies and local environments consistent with the birth site of a young magnetar \cite{CHIME20Nat,Michilli18Nat,Marcote20Nat}. A 16-day periodicity was observed in FRB~20180916B \cite{CHIME20Nat}, \REVb{hinting that the source is possibly in a binary system.}}

The local magneto-ionic environment \REV{can} be investigated through the polarization properties of FRBs, especially the rotation measure (RM), which is the convolution of the electron density $n_e$ and the magnetic field parallel to the line of sight $B_{||}$, ${\rm RM} \propto \int n_e(l)B_{||}(l)dl$. 
\REV{The observed RM amplitude of FRBs spans $\sim (1-10^5) \ \rm rad \ m^{-2}$ and often exhibits both long-term and short-term variations. 
FRB~20121102A, the first known repeater, showed a long-term decay of RM, from $1.03 \times 10^5~{\rm rad~m^{-2}}$ \cite{Michilli18Nat} initially to $6.7 \times 10^4{\rm~rad~m^{-2}}$ over two and a half years \cite{Hilmarsson2021}. This behavior can be explained by the expansion of a young supernova remnant (SNR \cite{Piro18}) or a magnetar nebula \cite{Margalit2018}. Another active FRB with large RMs, FRB~20190520B, exhibited a sign reversal of RM \cite{annathomas2023} which \REVb{can be} explained by invoking a binary system. FRB~20201124A displayed irregular RM variations, including a sudden disappearance of variation \cite{xuheng2022}. This RM variation may be accounted for via plasma turbulence, but a binary system progenitor may also \REVb{be} capable of producing it\cite{wangfy22}. }
In summary, the RM variations are diverse and carry important clues about 
plasma turbulence or the existence of a companion star \cite{Michilli18Nat,Luo20Nat,Hilmarsson2021,annathomas2023, Mckinven2023}. \REVb{However, there is still no smoking-gun signature for the existence of a companion star of the FRB engine.}

\section*{FAST and Parkes observations}

FRB 20220529 was discovered by the Canadian Hydrogen Intensity Mapping Experiment (CHIME) on 29 May 2022 
and was reported via the Virtual Observatory Event (VOEvent) service. It had a dispersion measure (DM) of $246.3 \pm 0.4~{\rm pc~cm^{-3}}$ and a Milky Way contribution of $\rm DM_{\rm MW}=39.93~pc~cm^{-3}$\cite{ne2001,YMW16}. 
\REVb{Triggered by the report of its repetition on 15 June 2022,} we \REV{observed} the source using the Five-hundred-meter Aperture Spherical radio Telescope (FAST), with the 19-beam receiver covering a frequency range of $1000-1500~{\rm MHz}$\cite{Jiang20RAA}.  
Two bursts were detected during our initial two-hour observation, 
which began at 23:09:10.131 UTC on 22 June 2022.
Since then, we have conducted a continuous monitoring campaign of FRB 20220529 using FAST, as well as the ultra-wide-bandwidth low-frequency (UWL) receiver of the Parkes telescope, which covers frequencies from $704~{\rm MHz}$ to $4032~{\rm MHz}$\cite{Hobbs20}. 
Up to \REV{5 September 2024, 112 observations totaling 53.9 hours were conducted with FAST, including 47.9 hours on-source tracking, while Parkes completed 59 observations totaling 132.0 hours.}
In total, we detected \REV{1156} bursts in FAST observations, with an average burst rate of about 23 bursts per hour \cite{mm}, and 56 bursts in Parkes observations.
Notably, FRB 20220529 continues to \REVb{be detected} in nearly every one-hour FAST observation \REV{during our 2.2 years of monitoring. Such an active episode is extremely long compared with those of most other active repeaters that have been \REVb{monitored on months–to-years timescales}, such as FRB 20201124A and FRB 20121102A, whose active episodes last for $2-3$ months \cite{Rajwade2020, xuheng2022}. 
This makes FRB 20220529 stand out as one of the longest actively repeating FRBs among all repeaters.} At the end of August 2022 and March 2023, it experienced two \REVb{of its} most active episodes, each lasting $1-2$ months. The peak rate was 204 bursts per hour. \REV{After excluding these two active episodes, the average burst rate in FAST observations is 7.5 bursts per hour.}
The temporal evolution of the properties of FRB 20220529 is presented in Fig. \ref{fig:rmflare}.

\section*{Polarization and Rotation Measure}

Before December 2023, the polarization analysis of FRB 20220529 revealed that the brightest bursts had a high level of linear polarization, with polarization fractions typically $> 90 \%$. 
The RM varied between $-300~\rm rad~m^{-2}$ and $+300~\rm rad~m^{-2}$ in general, with a median $\REVb{\rm RM} = 17~\rm rad~m^{-2}$ and a root-mean-square (rms) value $\sigma_{\rm RM}=101~{\rm rad~m^{-2}}$. 
FRB 20220529 showed RM reversals many times, regardless of whether the expected Milky Way contribution of $-35 \pm 9~\rm rad~m^{-2}$ was subtracted \cite{mwrm15}. 
The RM seemed to show two similar fluctuation patterns before MJD 60200 (13 September 2023), but \REV{no significant periodicity was detected in a period search}. 
In general, the RM evolution during this episode is consistent with fluctuations from a turbulent environment near the FRB source. 
By calculating the structure function during this episode and fitting it with the form of $D_{\rm RM}(\tau)\propto \tau^\alpha$, we obtained \REVb{$D_{\rm RM}(\tau)\propto \tau^{0.09 \pm 0.02}$}, implying that the index of the turbulence power spectrum is about $-(\alpha+2)\sim-2.09$.
Thus, the turbulent medium has a shallow spectrum in the inertial range, \REVb{i.e., the range where neither energy injection nor energy dissipation is important, and the fluctuation spectrum is characterized by a single power law and is self-similar}. This result suggests that the variation is possibly dominated by small-scale RM density fluctuations \cite{Yang23}, which could naturally arise from supersonic turbulence \cite{Hennebelle12,XuZhang17}. 
Albeit having a lower median value of RM, FRB 20220529 exhibits an RM variation amplitude similar to that of FRB 20201124A (\REV{$77.2~\rm rad~m^{-2}$})\cite{xuheng2022} before December 2023, suggesting that these two FRBs may reside in a similar magneto-ionic environment. 
The RM reversal behavior of FRB 20220529 is also similar to that of FRB \REV{20190520B}, although FRB \REV{20190520B} has a much larger amplitude in RM, with \REV{RM $\sim (10^4-10^5)~\mathrm{rad~m^{-2}}$}\cite{annathomas2023}.

An abrupt and prominent boost of the RM appeared at the end of 2023. Prior to this event, the last RM detection was recorded on 18 October 2023, with a value of \REVb{$-139 \pm 19~{\rm rad~m^{-2}}$}. Two 20-minute FAST observations were conducted on 7 and 22 November 2023, but no \REVb{FRBs were} detected. 
On 14 December 2023, another 20-minute observation with FAST detected four bursts, two of which provided successful RM measurements of \REVb{$1976 \pm 109~{\rm rad~m^{-2}}$} 
and \REVb{$1977 \pm 84~{\rm rad~m^{-2}}$} (Table~\ref{table:RM}).
Remarkably, within just 57 days, the RM value jumped from \REVb{$-139 \pm 19~{\rm rad~m^{-2}}$ to $1977 \pm 84~{\rm rad~m^{-2}}$}. 
We conducted a series of monitoring sessions with FAST and Parkes at intervals of $1-5~{\rm days}$ during the first 20 days following the discovery \cite{mm}. The burst rate in FAST observations during this episode ranged from 0 to 15 bursts per hour, \REV{with an average rate of 7.3 bursts per hour}, which is similar to the normal low state of this source. 
%
During this time,
we obtained 13 bursts with RM measurements in 6 observations 
(Table~\ref{table:RM}, Fig.~\ref{fig:rmflare}, fig.~\ref{fig:RM_spectrum} and fig.~\ref{fig:burst_pol}). 
It was found that the RM decreased nearly monotonically and dropped to the baseline of $\rm (-300, +300)~rad~m^{-2}$ on 28 December 2023, i.e., in 14 days. 
\REV{After this epoch, the RM values stayed in the baseline, and fluctuated within the normal range as before, at least up to September 2024. }
The RM evolution is presented in Fig. \ref{fig:rmflare}F.
\REVb{We refer to it as an ``RM flare'' in the following, indicating the sudden and dramatic increase of RM, followed by a rapid recovery to the baseline level. Please note that the term ``flare'' applies specifically to RM variation, not to brightness variation throughout this paper.}
The significance of such an ``RM flare'' could be quantified by ${\rm SNR_{\rm RM}}=\rm (RM_{\rm peak}-\overline{RM})/\sigma_{\rm RM}=20.4$, where $\sigma_{\rm RM}$ is the rms of RM before December 2023. Even if we define the $\overline{\mathrm{RM}}$ and $\sigma_{\rm RM}$ with all the data, the significance is still as high as 7.6 (fig. \ref{fig:rmflare_snr}). 
This is much more significant than RM variations of all other active repeaters \cite{mm}.
Such an abrupt variation of RM is unprecedented and \REV{has never been reported before.}

During the rapid return of RM, a decrease of the linear polarization fraction was observed\REVb{\cite{mm}}. Specifically, on 28 December 2023, the linear polarization fraction dropped to $27 \pm 3~\%$, and recovered to $81 \pm 6~\%$ on 17 January 2024, consistent with the nearly full linear polarization observed in the normal state of FRB 20220529. 
\REVb{The DMs and RMs during the ``RM flare'' episode have a Spearman's correlation probability of ${\rm log}~P=-4.9^{+1.8}_{-2.2}$, indicating a plausible positive correlation (fig. \ref{fig:flks_dmks})\cite{mm}.}

\section*{Localization and Redshift Identification}

During the active episode of FRB 20220529 in 2023, we carried out observations with the Karl G. Jansky Very Large Array (VLA) using the {\it realfast} fast transient detection system\cite{law2018}. We observed FRB 20220529 three times at frequencies ranging from $1$ to $2~{\rm GHz}$, 3 hours each. 
There are two bursts detected and FRB~20220529 is localized at right ascension $\alpha=01^{\rm h}16^{\rm m}25.014^{\rm s}$, declination $\delta=+20^{\circ}37^{\prime}56.6^{\prime\prime}$ in the International Celestial Reference Frame (ICRF, J2000), with a positional uncertainty (1 $\sigma$) of $0.3^{\prime\prime}$ \cite{mm}. 
We searched for the host galaxy in the Dark Energy Spectroscopic Instrument (DESI) Legacy Imaging Surveys (Fig. \ref{fig:host}A), and identified a disk galaxy PSO J019.1046+20.6327 as the most likely host galaxy, 
\REVb{with an association probability of 0.999, as estimated with the Probabilistic Association of Transients to their Hosts (PATH) \cite{aggarwal2021,mm}.} 
An optical spectral observation of the host galaxy was performed with the 10.4 m Gran Telescopio Canarias (GTC) telescope with its OSIRIS+ and R500R grism. The  obtained spectrum is presented in Fig. \ref{fig:host}B. H$\alpha$, H$\beta$, and \ion{{\rm O}}{iii}\,$\lambda\lambda4959,5007$ doublets are identified. The redshift of the host galaxy is identified as $z=0.1839 \pm 0.0001$. 
\REVb{The estimated stellar mass, star formation rate (SFR), half-light radius and the offset of the host galaxy from the FRB are $(2.7\pm0.7)\times10^9~M_{\odot}$, $0.13~M_{\odot}~\rm yr^{-1}$, $3.60\pm0.08~\rm kpc$, and $4.4\pm0.6~\rm kpc$,  respectively\cite{mm}. }
The basic properties of FRB~20220529 are summarized in Table \ref{tab:obs_prob}.

\section*{Physical origin of the ``RM flare''}

The prominent ``RM flare'' in a month timescale suggests the emergence and disappearance of magnetized plasma along the line of sight. 
The length scale $l$ of the ``object'' that produced the observed abrupt RM variation could be estimated with an assumed velocity $v$ and the observed recovery timescale $t$, i.e., $l \sim vt\simeq0.6~{\rm AU}(v/100~{\rm km~s^{-1}})(t/10~{\rm day})$. 
The velocity may range from $\sim 10~{\rm km~s^{-1}}$ (relative velocity of stars) to the speed of light $c$ (relativistic ejecta from catastrophic events). Consequently, the length scale could be estimated as $l\sim (0.06-1700)~{\rm AU}$, corresponding to the size of \REVb{a stellar system including planets or the distance between stars in} a binary system. 
\REV{
Several scenarios that might be proposed to explain the ``RM flare'' are summarized as follows, with details in the supplementary text.
1) Magnetar outflows: The observed ``RM flare'' is unlikely to be contributed by an outflow associated with typical X-ray bursts, because no RM changes have been observed in different epochs for the Galactic FRB-emitting magnetar SGR J1935+2154 \cite{CHIME200428, Bocheneck20, Mereghetti20, Li21}. \REVb{However, most FRBs are more luminous than 
bursts from SGR 1935+2154, which might 
be associated with giant flares emitted from magnetars \cite{Mazets1979,Palmer2005}.} To produce the observed ``RM flare'', the giant flare is required to be more powerful than the most powerful giant flares known for Galactic magnetars. \REVb{Even so, a more severe issue is that the RM evolution caused by the magnetar outflow is expected to show power-law decay with a positive second derivative, $d^2{\rm RM}/dt^2>0$, which is inconsistent with the observed $d^2{\rm RM}/dt^2<0$ at the beginning of the ``RM flare'' and results in a poor fit to the model (Fig. \ref{fig:rmflare_caseIaII}c, table \ref{table: model_nobkg})}. 
Thus, the magnetar outflow does not naturally explain the observed ``RM flare'' \cite{mm}.
2) Turbulence in an SNR or a PWN: The RM evolution from a young SNR with an unchanged magnetic geometry is supposed to be monotonic over a timescale longer than a few years \cite{Piro18,yang2023}. No rapid variations are expected. Thus, if the ``RM flare'' is attributed to a young SNR, it must involve an extreme turbulent clump along the line of sight, characterized by an extremely small scale size and/or a short variation timescale. 
Due to the self-similarity of turbulence, the structure functions that include and exclude the “RM flare” should be consistent with each other. 
However, observationally, the structure function with the ``RM flare'' significantly deviates from that without the ``RM flare'' (fig. \ref{fig:rm_sf}), which is quite unusual for turbulence. The small scale size and short lifetime of the turbulent clump also raise concerns about the low probability of detecting such clumps. Similarly, the PWN scenario faces the same issues as well. 
3) Orbital motion of a binary system: The observations of a Galactic binary system PSR B1259-63/LS 2883 show prominent RM variations when the pulsar PSR B1259-63 approaches the periastron \cite{Johnston1996,Johnston05}. Such a scenario may produce an ``RM flare'' similar to that observed in FRB 20220529. However, this scenario requires an extremely elliptical orbit for the binary system. Our model fit requires the binary orbital period and eccentricity to be $P_{\rm orb}>1.5~{\rm yr}$ and $e>0.95$, respectively. The fraction of neutron star binaries that have such a configuration 
is only $0.6\%$ \cite{mm}. \REVb{If the ``RM flare'' is indeed produced by the binary configuration, it would periodically occur. This could be tested through long-term monitoring of FRB~20220529, as well as other repeating FRBs, in the future.}
}

\REVb{One plausible} scenario is that
the ``RM flare'' is caused by a clump of magnetized plasma ejected by a nearby object. The most likely scenario is a coronal mass ejection (CME), often associated with a stellar flare, from a companion star \cite{mm}. 
\REV{
CME candidates are detected in different types of stars, including pre-main sequence stars, binary stars, G-M type stars
\cite{Leitzinger2022}.} 
The RM variation from a CME can be modeled by considering the expansion and passing through the line of sight of the CME.
The sudden rise of RM can be attributed to the CME entering the line of sight, while the decay is due to plasma expansion, which reduces both the number density and magnetic field, and its exit from the line of sight.
Two geometric models are studied in detail (fig. \ref{fig:model}): Case I involves both the entrance and exit of the CME from the line of sight, while in Case II, the CME never exits the line of sight, with the decrease of RM attributed solely to expansion.
The best-fit models are presented in Fig. 3 and table S3, indicating that the observed RM evolution prefers the former scenario.
\REV{
The following two sets of parameters from the observed stellar CMEs can explain the observed ``RM flare'': 1) a CME with a mass of $\sim10^{17}~{\rm g}$, a surface magnetic field of $\sim10^4~{\rm G}$, \REVb{attainable for M dwarfs \cite{Reiners2022},} and a scale of $\sim10^{-1}R_\odot$ from a magnetized M dwarf; 2) a CME with a mass of $\sim10^{21}~{\rm g}$, a surface magnetic field of $\sim1~{\rm G}$ and a scale of $\sim10^{-1}R_\odot$ generated in a giant star or a mass-transfer binary system, similar to the Algol binary system 
\cite{moschou2017}. 
The predicted DM variation is consistent with the observed DM variation during the ``RM flare'' epoch \cite{mm}.}
The probability of a CME moving across the line of sight is extremely low for an isolated FRB source.
\REV{Therefore, if \REVb{the} CME hypothesis is right, this observation implies that FRB~20220529 is from in a binary system\cite{iokazhang20,wangfy22}.}

\REV{Observationally, we detected one ``RM flare'' in 2.2 years. Assuming a \REVb{Poisson} distribution, one gets an event rate $0.45^{+1.04}_{-0.38}$ per year per FRB and a 3 $\sigma$ range of $6 \times 10^{-4}- 4.0$ per year per FRB.}
This suggests that similar abrupt RM variations may occur in FRB 20220529 and potentially other repeating FRB sources.
While CMEs are common in certain low-mass stars with strong convection (approximately once a month per source for CME energies $\gtrsim 10^{34}~{\rm erg}$, which is needed to explain the “RM flare” \cite{Lacy1976, Kowalski2010, Shibayama13}), the detection rate of “RM flares” is likely low, as only line-of-sight CMEs create significant observational signatures. 
\REV{Because the physical parameters  (e.g., explosion energy, CME mass, etc.) of stellar flares have wide power-law distributions \cite{Shibayama13, Gunther20}, the maximum RM and the duration of similar events should have a lognormal distribution.}  
Future observations can test such a prediction. 

\clearpage

\clearpage
\begin{figure*}
\centering
\includegraphics[width=0.99\textwidth]{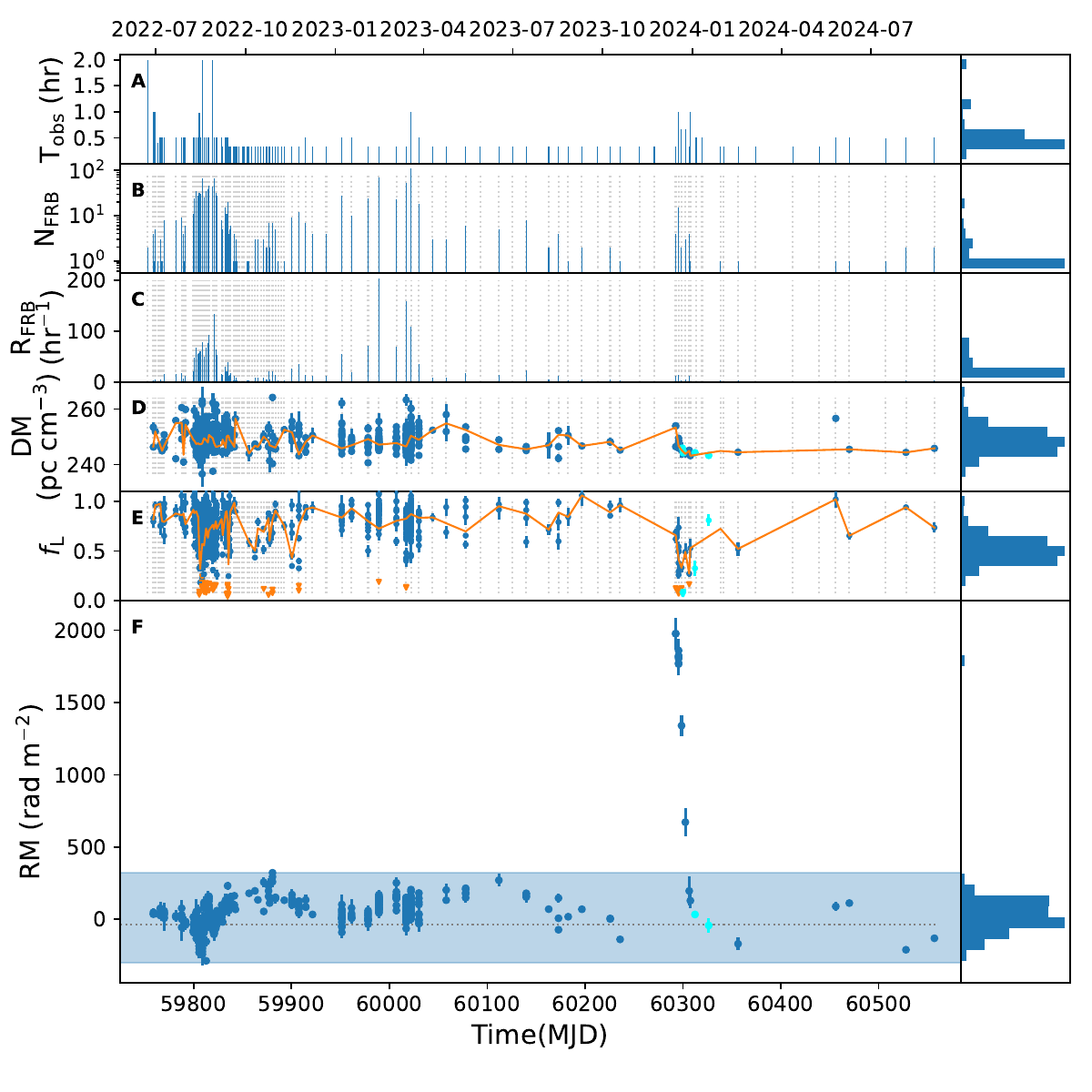}
\caption{\textbf{Temporal variations of various quantities of FRB~20220529.} {\bf (A)} Daily observing time. {\bf (B)} Daily number of bursts detected. {\bf (C)} Daily burst rate of FRBs. Vertical grey dotted lines indicate the days with observations, including those with non-detection. {\bf (D)} Dispersion measure of bursts. Blue dots present bursts detected with FAST and cyan dots are those detected with Parkes. \REV{The orange line shows the average DMs for each day.} {\bf (E)} Linear Polarization fraction of bursts. Orange triangles indicate the upper limits,  \REV{and the orange line shows the average $f_\mathrm{L}$ for each day.} {\bf (F)} Rotation measure of bursts. The horizontal dotted line indicates the \REV{estimated} contribution of the Milky Way. The filled blue region \REV{covers the minimum to maximum range of the low RM state}. A distinct RM flare is clearly visible. 
}
\label{fig:rmflare}
\end{figure*}

\clearpage
\begin{figure} 
\centering
\includegraphics[width=0.7\textwidth]{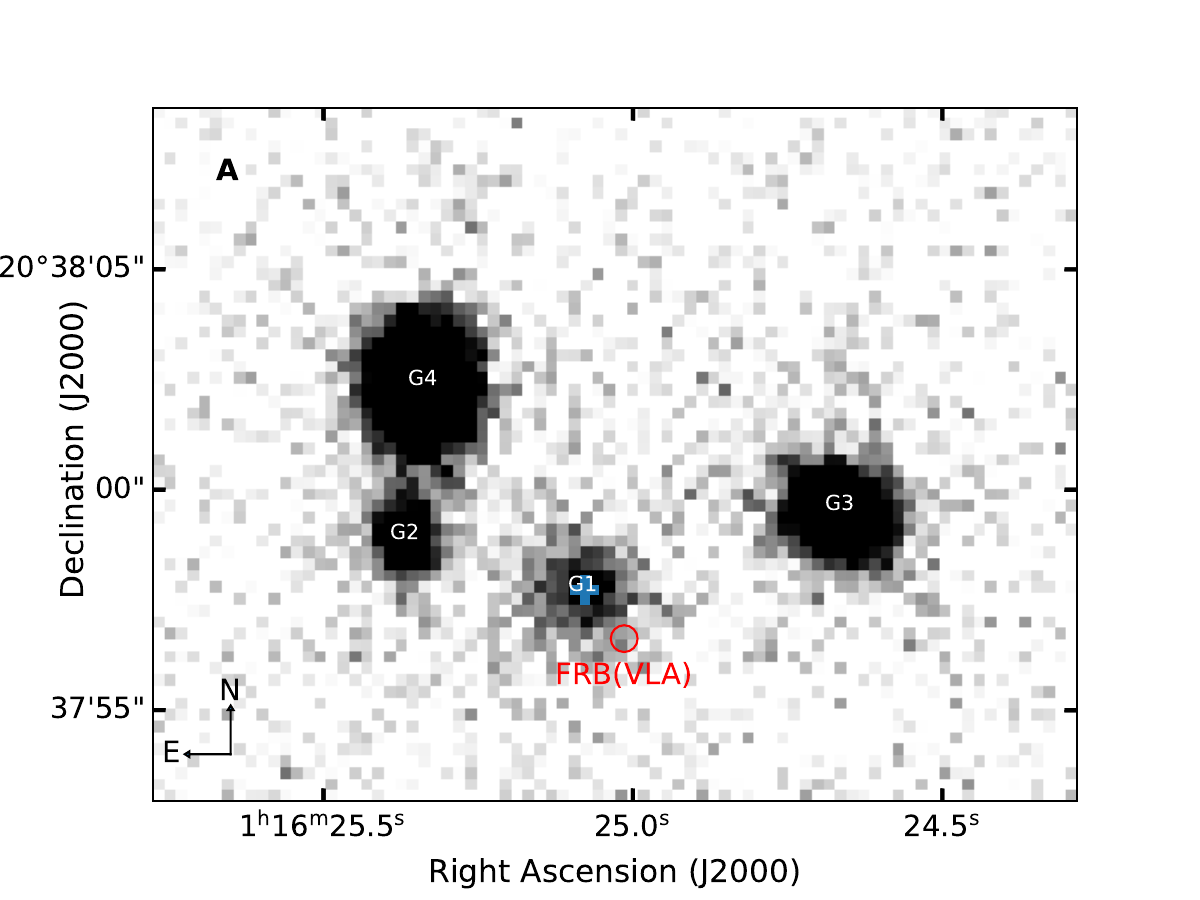}
\includegraphics[width=0.7\textwidth]{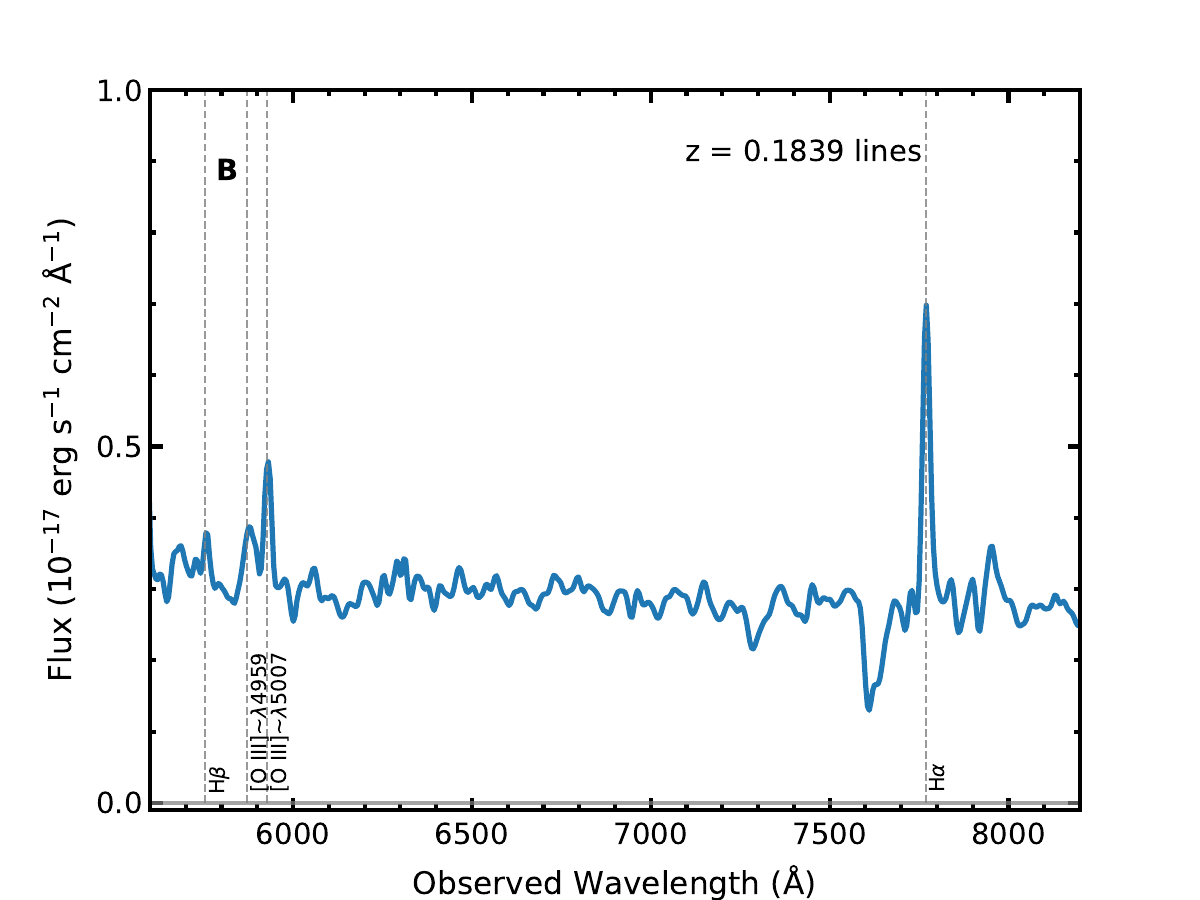}
\caption{
\textbf{Host-galaxy properties of FRB~20220529.} 
{\bf (A)} DESI $r$-band image of the host galaxy of FRB 20220529. The FRB position from the VLA observation is overplotted as a red circle, and the center of the host galaxy is marked as a blue cross. 
\label{fig:host}
{\bf (B)} GTC spectrum of the FRB~20220529 host galaxy (blue). Emission lines from the $z = 0.1839$ host galaxy are labeled with gray dashed lines.
}
\end{figure}

\clearpage
\begin{figure*}
\centering
\includegraphics[width = 0.7\linewidth, trim = 0 0 0 0, clip]{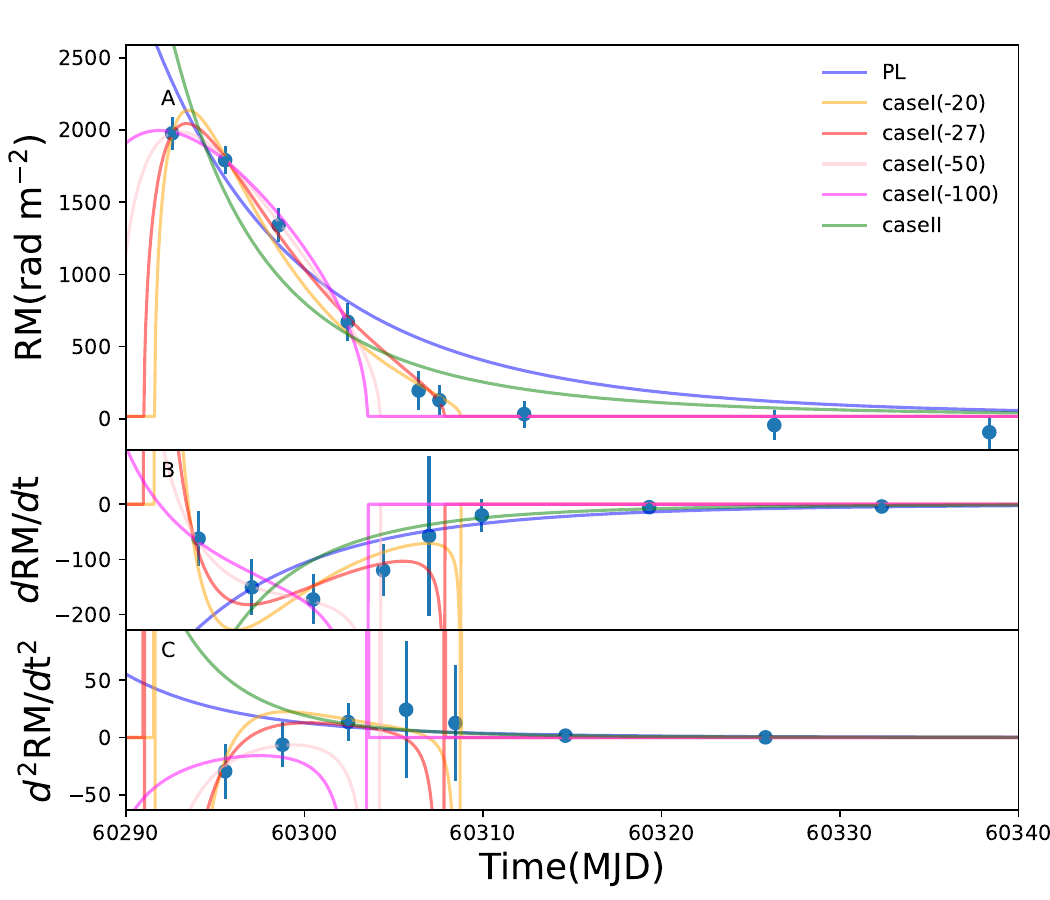}
\caption{\textbf{Best-fitting results of different models as well as the first and second derivatives}. {\bf (A)} The best fits for different models. CME Case I model with assumed $t_0=-20, -27, -50$ and $-100$ days before the first detection of the ``RM flare'' is presented as orange, red, pink and magenta lines. The daily averaged observed data are overplotted for comparison. The best-fitting parameters are provided in Table \ref{table: model_nobkg}. {\bf (B)} The first derivatives of different best-fitting models. {\bf (C)} The second derivatives of different best-fitting models. }
\label{fig:rmflare_caseIaII}
\end{figure*}

\clearpage
\begin{table*}
\centering
\renewcommand\arraystretch{0.8}
\begin{threeparttable}
\caption{\textbf{Basic Properties of FRB 20220529} All errors represent the 1$\sigma$ uncertainties.}
\label{tab:obs_prob}
\begin{tabular}{lc}
\hline
\textbf{Burst Parameters:} & \\
Right ascension (J2000) & $01^{\rm h}16^{\rm m} 25.014 ^{\rm s}$ \\
Declination (J2000) & $+20^{\circ}37^{\prime}56.6^{\prime\prime}$ \\
positional uncertainty ($\arcsec$) & $0.3$ \\
Galactic coordinates ($l,b$) & $(130.78767,-41.85802)$ \\
DM ($\rm pc~cm^{-3}$) & $249.4 \pm 3.5$ \\ 
DM$_{\rm MW,NE2001}$ \cite{ne2001}, DM$_{\rm MW,YMW16}$ \cite{YMW16} ($\rm pc~cm^{-3}$)& $39.95, 30.92$\\
DM$_{\rm MW,halo}$ ($\rm pc~cm^{-3}$) & $30$  \\
DM$_{\rm host}^*$ ($\rm pc~cm^{-3}$) & $0-24$ \\
RM$_{\rm MW}$ ($\rm rad~m^{-2}$) & $-35 \pm 9$ \\
\hline
\textbf{Host Galaxy:} & \\
Redshift ($z$) & $0.1839 \pm 0.0001$ \\
Half-light radius ($\rm kpc$) & $3.60 \pm 0.08$ \\
Offset ($\rm kpc$) & $4.4\pm 0.6$ \\
Normalized offset & $1.2 \pm 0.2$ \\
Host Association Probability\cite{aggarwal2021} & $0.999$ \\
Cumulative light fraction & $0.13^{+0.15}_{-0.09}$ \\
Stellar mass ($M_{\odot}$) & $(2.7\pm0.7) \times 10^9$ \\
Star-formation rate ($M_{\odot}~{\rm yr^{-1}}$) & $0.13$ \\
\hline
\end{tabular}
\begin{tablenotes}
\item [*] ${\rm DM_{host}=DM-DM_{MW}-DM_{MW,halo}}-fz$ and $850<f<1000$ is assumed.
\end{tablenotes}
\end{threeparttable}
\end{table*}

\clearpage
\begin{table*}
\centering
\renewcommand\arraystretch{0.8}
\begin{threeparttable}
\caption{{\color{black}\textbf{Properties of FRB 20220529 bursts with RM measurements during the ``RM flare'' phase and three bursts after it.} All errors represent the 1$\sigma$ uncertainties.}}
\begin{tabular}{lcccccccc}
\hline
$\rm MJD^a$ & S/N$^b$ & ${\rm DM}_{\rm det}^b$ & $f_{\rm low}^c$ &  $f_{\rm high}^c$  &  ${\rm DM}_{\rm stru}^d$ & $\rm RM$ & $f_{\rm L}$$^e$  & $f_{\rm V}$$^f$ \\
    &    & ($\rm pc~cm^{-3}$) &(MHz)   &  (MHz)         &   ($\rm pc~cm^{-3}$) & ($\rm rad~m^{-2}$) & ($\%$) & ($\%$) \\
\hline
60292.5811430 & 13 & 257.2 & 1200 & 1380 & 254.0 $\pm$ 0.1 & 1976 $\pm$ 109 & 62 $\pm$ 4 & -8 $\pm$ 5 \\
60292.5811432 & 13 & 257.2 & 1020 & 1180 & 254.0 $\pm$ 0.1 & 1977 $\pm$ 84 & 69 $\pm$ 4  & -3 $\pm$ 3\\
60295.5377214 & 13 & 248.5 & 1000 & 1130 & 245.9 $\pm$ 0.3 & 1816 $\pm$ 45 & 26 $\pm$ 3  &  9 $\pm$ 4 \\
60295.5396355 & 11 & 246.7 & 1000 & 1150 & 247.0 $\pm$ 0.6 & 1825 $\pm$ 84 & 55 $\pm$ 5  &-23 $\pm$ 6\\
60295.5440419 & 10 & 247.5 & 1000 & 1430 & 246.7 $\pm$ 0.2 & 1773 $\pm$ 31 & 74 $\pm$ 11 & 10 $\pm$ 12\\
60295.5535219 & 10 & 249.3 & 1020 & 1180 & 247.0 $\pm$ 1.0 & 1860 $\pm$ 83 & 29 $\pm$ 7  &  7 $\pm$ 10\\
60295.5587887 & 12 & 247.3 & 1040 & 1300 & 246.6 $\pm$ 0.4 & 1766 $\pm$ 44 & 50 $\pm$ 7  & 11 $\pm$ 9\\
60295.5606052 & 18 & 252.2 & 1000 & 1200 & 248.6 $\pm$ 0.5 & 1816 $\pm$ 78 & 49 $\pm$ 3  & -4 $\pm$ 4 \\
60295.5625614 & 8  & 248.9 & 1300 & 1450 & 249.5 $\pm$ 1.2 & 1804 $\pm$ 117 & 38 $\pm$ 9 & 18 $\pm$ 13\\
60298.5465724 & 10 & 247.9 & 1000 & 1170 & 245.2 $\pm$ 0.8 & 1340 $\pm$ 74 & 33 $\pm$ 5  &  8 $\pm$ 7 \\
60302.4196357 & 7  & 249.1 & 1300 & 1500 & 243.9 $\pm$ 0.3 & 672 $\pm$ 96 & 49 $\pm$ 8   &  4 $\pm$ 10\\
60306.3962105 & 11 & 245.3 & 1010 & 1120 & 245.2 $\pm$ 0.5 & 196 $\pm$ 99 & 27 $\pm$ 3   &  6 $\pm$ 4\\
60307.5594708 & 7  & 247.3 & 1050 & 1250 & 243.2 $\pm$ 0.3 & 129 $\pm$ 53 & 53 $\pm$ 10  & 19 $\pm$ 10 \\
\hline
60312.3030815$^g$ & 9  & 245.0 & 790 & 860 & 244.3 $\pm$ 0.1 & 33 $\pm$ 25 & 32 $\pm$ 7  & -2 $\pm$ 10 \\
60326.3185239$^g$ & 15 & 246.0 & 790 & 920 & 243.3 $\pm$ 0.2 & -43 $\pm$ 48 & 81 $\pm$ 6 &  6 $\pm$ 5\\
60338.3648098 & 7  & 258.8 & 1000 & 1200 & 245.0  $\pm$ 1.7  & -93 $\pm$ 69 & 72 $\pm$ 11& 7 $\pm$ 12 \\
\hline
\end{tabular}
\begin{tablenotes}
\item [a] MJDs are in barycentric dynamical time (TDB) and are referenced to infinite frequency.
\item [b] The detection DM at the maximum S/N. 
\item [c] $f_{\rm low}$ and $f_{\rm high}$ are the minimum and maximum frequency extent of the bursts, determined visually.
\item [d] The structure-maximizing DM obtained by \emph{\sc DM\_phase} with 1$\sigma$ uncertainties.
\item [e] Percentage of unbiased linear polarization along with 1$\sigma$ uncertainty.
\item [f] Percentage of circular polarization along with 1$\sigma$ uncertainty.
\item [g] Bursts observed by Parkes. 
\end{tablenotes}
\label{table:RM}
\end{threeparttable}
\end{table*}
\clearpage

\clearpage

\clearpage





\bibliographystyle{Science}

\section*{Acknowledgements}
\REVb{We thank the anonymous referees for their helpful suggestions.}
We are grateful to Qiang Yuan, \REV{Jie Hu, Lei Lu, Jun Yang, Xiao-Ting Fu, Min Fang, Jie Yu, Hong-Peng Lu, Bo-Yang Liu} and Jingjing Li for helpful discussions.
This work made use of data from the FAST. FAST is a Chinese national mega-science facility, built and operated by the National Astronomical Observatories, Chinese Academy of Sciences. The Parkes radio telescope (Murriyang) is part of the Australia Telescope National Facility, which is funded by the Australian Government for operation as a National Facility managed by CSIRO. This study is partially based on data collected at Karl G. Jansky Very Large Array, operated by the National Radio Astronomy Observatory (NRAO). The National Radio Astronomy Observatory is a facility of the National Science Foundation operated under cooperative agreement by Associated Universities, Inc. We acknowledge the use of public data from the DESI Legacy Survey. We gratefully acknowledge the support and assistance provided by the staff of the Gran Telescopio Canarias (GTC).

\subsection*{Funding}
This work is partially supported by the National Natural Science Foundation of China (Grant Nos. 12321003, 12041306, 12103089, 11988101, 12303042, 12203045, 1247304, 12041303, 12303015, 11988101), the National SKA Program of China (2022SKA0130100, 2022SKA0130101, 2022SKA0120103, 2020SKA0120200), the Natural Science Foundation of Jiangsu Province (Grant No. BK20211000, BK20231106), International Partnership Program of Chinese Academy of Sciences for Grand Challenges (114332KYSB20210018), the CAS Project for Young Scientists in Basic Research (Grant No. YSBR-063), the CAS Organizational Scientific Research Platform for National Major Scientific and Technological Infrastructure: Cosmic Transients with FAST, the Youth Innovation Promotion Association (2023331), 
the Science \& Technology Champion Project (202005AB160002) and from two Team Projects -- the Innovation Team (202105AE160021) and the Top Team (202305AT350002) all funded by the Yunnan Revitalization Talent Support Program. DL is a new Cornerstone investigator.

\subsection*{Author Contributions}
Y.L., S.-B.Z., X.-F.W., and B.Z. coordinated the observational campaign, co-supervised data analyses and interpretations, and led the paper writing. S.-B.Z. and X.Y. led the radio data analyses. Y.-P.Y. led the interpretations and the paper writing. W.-W.Z. and B.Z. coordinated the regular monitoring program through the FAST FRB Key Science Project. P.J. coordinated the prompt FAST observation during the "RM flare" state. C.L., Y.L., C.-W.T., R.A.-T., X.-L.Y., and R.L. conducted the VLA observation and data analysis. C.-W.T. led the GTC observation campaign, and X.-L.C., G.C., Y.L., T.-R.S., D.-D.S. and L.H. contributed to the GTC spectrum campaign and optical data analysis. K.-J.L. contributed to the localization of the source. 
C.-M.C., J.-R.N., Y.-K.Z., B.-J.W., D.-J.Z., Z.-F.T., Y.F., C.-F.Z., D.-Z.L., R.M., K.S., C.-H.N., W.-W.Z., Y.-H.Z., H.X., R.L. aided the radio data analysis. D.X., J.-L.H., J.-J.G., Z.-G.D. and D.L. helped with the data interpolation. 

\subsection*{Competing Interests} 
The authors declare no competing financial interests.

\subsection*{Data and materials availability:}
FAST observational data (ZD2021\_5, ZD2022\_5, DDT2022\_4, DDT2022\_6, ZD2023\_5, PT2023\_0148, PT2024\_0095) can be accessed through the FAST data center at \url{http://fast.bao.ac.cn} , while Parkes raw data (P1105 and P1183) are available via CSIRO's data archive at \url{https://data.csiro.au/}.
VLA raw data (project 23A-385) are available via VLA data archive site at \url{https://data.nrao.edu/portal/}.
The raw GTC data (GTCMULTIPLE1A-23ACNT) is available via GTC data archive \url{gtc.sdc.cab.inta-csic.es/gtc/jsp/searchform.jsp}, and the reduced GTC spectrum is available through \url{https://github.com/Astroyx/FRB20220529}\REVb{, and archived at Zenodo \cite{zenodo}}.
Due to the large data volume, interested users are encouraged to contact the corresponding author to arrange the data transfer. 

\REVb{Our derived properties of the bursts are listed in table S4 for the unpolarized bursts within the ``RM flare'' phase, and in table S5 and table S6 for bursts with and without RM measurements outside the “RM flare” phase. }



\section*{Supplementary materials}
Materials and Methods\\
Supplementary Text\\
Figs. S1 to S11\\
Tables S1 to S6\\
References \textit{(51-116)}


\clearpage


\renewcommand\thesection{S\arabic{section}} 
\setcounter{section}{0} 
\renewcommand\thetable{S\arabic{table}} 
\setcounter{table}{0} 
\renewcommand\thefigure{S\arabic{figure}}
\setcounter{figure}{0}
\renewcommand\theequation{S\arabic{equation}}
\setcounter{equation}{0}

\newcommand{\ypy}[1]{\textcolor{red}{YPY: #1}}
\newcommand{\note}[1]{\textcolor{magenta}{(#1)}}

\newpage
\pagenumbering{arabic}
    \setcounter{page}{1}

\begin{center}
{\Huge
Supplementary Materials for:\\}
{\Large
\REV{A sudden dramatic change and recovery of magneto-environment of
a repeating fast radio burst}}

\vspace{6mm}
\noindent 
\normalsize{Y. Li, S. B. Zhang, Y. P. Yang,
C. W. Tsai, X. Yang, C. J. Law, R. Anna-Thomas,
X. L. Chen, K. J. Lee, Z. F. Tang, D. Xiao, H. Xu, X. L. Yang, G. Chen,
Y. Feng, D. Z. Li, R. Mckinven, J. R. Niu, K. Shin, B. J. Wang, C. F. Zhang,
Y. K. Zhang, D. J. Zhou, Y. H. Zhu, Z. G. Dai, C. M. Chang, J. J. Geng, J. L. Han,
L. Hu, D. Li, R. Luo, C. H. Niu, D. D. Shi, T. R. Sun,
X. F. Wu, W. W. Zhu, P. Jiang, B. Zhang}
\\
To whom correspondence should be addressed; E-mail:  xfwu@pmo.ac.cn (X.F.W.); zhuww@nao.cas.cn(W.W.Z.); pjiang@nao.cas.cn (P.J.); zhang@physics.unlv.edu(B.Z.) \\
\vspace{3mm}

\end{center}

\noindent
{\bf This PDF file includes:}\\
Materials and Methods\\
Supplementary Text\\
Figs. S1 to S11\\
Tables S1 to S6\\
References \textit{(51-116)}

\clearpage
\section{Materials and Methods}

\subsection{FAST \& Parkes observations}

A brief summary of the radio observations during the sudden dramatic RM evolution is given in Table \ref{table:obs}. 

\subsubsection{FAST observations}
Our campaign began with two consecutive one-hour grid observations using all beams of the 19-beam receiver of the 500\,m-diameter FAST radio telescope, starting at 23:09:10.131 UTC on 22 June 2022. 
Two bursts were detected in different beams, yielding an initial localization of $\alpha = 01^{\rm h}16^{\rm m}23.35^{\rm s}$, $\delta = +20^\circ37^\prime34.7^{\prime\prime}$. 
Subsequent follow-up observations were conducted using only the central beam. We performed another two grid observations on 14 and 17 August 2022, along with an off-beam tracking observation on 28 August 2022. These observations, with bursts detected simultaneously in up to three beams, refined the source position to $\alpha = 01^{\rm h}16^{\rm m}24.24^{\rm s}$, $\delta = +20^\circ38^\prime27.6^{\prime\prime}$. 
From 23 August 2022 to 14 February 2023, observations were carried out using this refined position, followed by a new localization based on VLA detection (see details in the below section): $\alpha = 01^{\rm h}16^{\rm m}25.01^{\rm s}$, $\delta = +20^\circ37^\prime57^{\prime\prime}$.    

%
The 19-beam receiver covers a frequency range of 1000$-$1500\,MHz with 4096 channels. 
Dual linear polarization signals were 8-bit sampled and channelized\cite{Jiang20RAA} using the Reconfigurable Open Architecture Computing Hardware generation 2 (ROACH 2)\cite{Hickish16} and stored in PSRFITS search mode format\cite{Hotan04}. The sample time is $49.153~\mu$s. 
A 1K equivalent noise-switched calibration signal was recorded before each observation to calibrate the results.
The observation durations are presented in Fig. \ref{fig:rmflare}A. The observations on 22 June 2022 and 28 August 2022 lack FRB rate estimates as they were off-beam. For the observations on 14 and 17 August 2022, FRB rates were estimated based on bursts detected in the first half-hour when the beam was on the source.

\subsubsection{Parkes observations} 
FRB 20220529 was monitored by Parkes using the Ultra-Wideband Low (UWL) receiver from 27 June 2022 to 5 September 2024, following the pointing strategy of the FAST campaign. 
%
The UWL system covers frequencies from $704\,{\rm MHz}$ to $4032\,{\rm MHz}$\cite{Hobbs20}. Data were 2-bit sampled every 32 or 256 $\rm \mu s$, in frequency channels of 1 MHz or 0.125\,$\rm MHz$ wide, respectively.  
Coherent de-dispersion at a dispersion measure (DM) of $247\,\rm pc~cm^{-3}$ with only one polarization was conducted before 27 September 2022.
Full Stokes parameters have been recorded since then. 
A 2-minute noise diode signal was injected before each tracking observation for polarization calibration.

\subsection{Burst detection}
Data collected from the FAST and Parkes radio telescopes were processed using two independent search pipelines based on the pulsar/FRB single pulse searching packages \texttt{PRESTO}\cite{Ransom01} and \texttt{HEIMDALL}\cite{Petroff15a}. 
We processed the full-band data from FAST, but divided the Parkes UWL data into a series of sub-bands ranging from $128$ to $3328\,{\rm MHz}$ based on a tiered strategy\cite{Kumar21_11a}.  
In both pipelines, the data were dedispersed over a DM range of $200-300\,{\rm pc~cm^{-3}}$, with a step size of $0.1\,{\rm pc~cm^{-3}}$. Single pulse candidates with signal-to-noise ratio (S/N) greater than 7 were recorded and visually inspected. 

In total, \REV{1156} bursts were detected in the FAST observations, with \REV{1081} bursts detected when the source was on-beam. Thus, the average FRB rate is 22.6 bursts per hour. 
56 bursts were detected in the Parkes observations. 
\REV{The number of detected FRBs and the FRB rate are presented in Fig. \ref{fig:rmflare}B and \ref{fig:rmflare}C, respectively. They show two active episodes around August 2022 and March 2023, with peak rates of 134 and 204 bursts per hour, respectively. Excluding these two active episodes, the average rate is 7.5 bursts per hour. 
Fig.~\ref{fig:flux_wid_FAST} presents the distribution of peak flux density and effective width for bursts of FRB 20220529 detected by the FAST telescope.}
%

\subsection{Polarization properties} \label{sec:polarization}
\REV{Each burst from FRB 20220529 was dedispersed at the detection DM with the maximum S/N.}
The de-dispersed polarization data were calibrated using the {\sc psrchive} software package\cite{Hotan04}, with correction for differential gain and phase between the receivers achieved through the injection of a noise diode signal before each observation. 
Rotation measures (RMs) for all bursts were determined using the {\sc rmfit} program, searching for a peak in the linearly polarized flux $L = \sqrt{Q^2 + U^2}$, within the range of RM from $-4000$ to $4000\,{\rm rad\,m^{-2}}$, with a step of $1\,{\rm rad\,m^{-2}}$. {\sc rmfit} corrects for Faraday rotation at each trial RM, producing a total linear polarization profile and an RM spectrum. A Gaussian fit was then applied to determine the optimal RM and its 1$\sigma$ uncertainty.

To compute the polarization fraction $f_{\rm L}$ of each burst, the profile was de-rotated at the best-fit RM. The polarized pulse profile was generated by averaging over the frequency dimension. Due to the presence of noise, the linear polarization $L$ tends to be overestimated. To obtain an unbiased estimate, we used $L_{\rm unbias}$\cite{Everett01}:  
\begin{equation}
L_{\rm unbias} =
\begin{cases}
\sigma_{I}\,\sqrt{\frac{L}{\sigma_{I}}-1}  & \text{if } \frac{L}{\sigma_{I}} \geq 1.57 \\
0 & \text{otherwise}
\end{cases}
\end{equation}
where $\sigma_{I}$ is the off-pulse standard deviation in Stokes I.
\REV{ To reduce the effect of uncertain polarization fractions due to insufficient S/N, we compared $L_{\rm unbias}$ only for bursts with RM measurements that have a relatively high S/N (equal or above ten) in our analysis (Table \ref{table:RM} and table \ref{table:withRM_NOrf}). The upper limits of the linear polarization fraction for bursts without detectable RM are given in 3$\sigma$ confidence (table \ref {table:withoutRM_rf} and \ref{table:withoutRM_NOrf}). The temporal RM and linear polarization degree evolution are presented in Fig. \ref{fig:rmflare}F and \ref{fig:rmflare}E.}

\subsection{RM Variation} \label{sec:RM_vari}

The observed RM evolution of FRB 20220529 includes two stages: the low RM state
(before December 2023, lasting over 500 days), and a sudden dramatic increase and drop episode (at the end of December 2023). After the turn, the RM stays at the baseline as the low RM state.  
The RM in the low RM state displayed a random evolution, which seems to originate from the turbulent environment near the FRB source. We calculated the structure function, which \REV{ is defined as $D_{\rm RM}(\tau) = \sum\limits_i[{\rm RM}(t)-{\rm RM}(t + \tau)]^2_i/N$} and reveals the variability on different timescales, using the RM values in the low RM state (i.e., before 1 November 2023, or \REV{ $\rm MJD<60249$}). The result is presented in fig. \ref{fig:rm_sf}. The structure function was fitted with a power law function $D_{\rm RM}(\tau)\propto \tau^{\alpha}$. \REV{In order to have enough points in each time bin, the first bin was made wider than others. The orthogonal distance regression (ODR) was employed to account for the uncertainty of time.} The best-fit index is $\alpha=0.09\pm 0.02$, and the index of the turbulence power spectrum is about $-(\alpha+2)\sim-2.09$.
This suggests that the turbulent medium has a shallow spectrum in the inertial range, \REVb{where neither energy injection nor energy dissipation is important, and the fluctuation spectrum is characterized by a single power law and self-similar.} This indicates that the variation of the RM should be dominated by small-scale RM density fluctuations \cite{Yang23}. 
\REVb{Since the magnetic energy spectrum is usually steep as proposed in the literature \cite{Goldreich1995, Lazarian1999, Bowen2024}
the observed result implies that a shallow electron density spectrum is more likely to dominate the RM fluctuations for this particular FRB source, which could naturally arise in supersonic turbulence \cite{Yang23}.}
\REV{ In order to examine the possibility that the ``RM flare'' is also from density fluctuations, we calculate the structure function including the ``RM flare'' as well. It is presented as pink and red dots in fig. \ref{fig:rm_sf}. The contribution of ``RM flare'' boosts the points on day timescales, especially those in the timescale from tens to hundreds of days by more orders of magnitudes. It does not fit the structure function distribution of $\rm MJD < 60249$, and indicates a different origin of the ``RM flare''. }

The RMs observed in the ``RM flare'' state are significantly larger than those in the low RM state. The significance can be quantified as ${\rm SNR_{\rm RM}}=\rm (RM_{\rm peak}-\overline{RM})/\sigma_{\rm RM}=20.4$, where $\sigma_{\rm RM}$ is the standard deviation of RM in the background low RM state. 
In fig. \ref{fig:rmflare_snr}, we compare the significance of FRB 20220529 with that of other repeating FRBs with long-term RM measurements, including FRB 20121102A \cite{Hilmarsson2021}, 20180916B \cite{Mckinven2023}, 20190520B \cite{annathomas2023}, and 20201124A \cite{xuheng2022}. Since no ``RM flare'' has been reported in other FRBs, we estimate their mean RMs and RM standard deviations with all the RM values and denote them as $\overline{\mathrm{RM_{all}}}$ and $\sigma_{\rm RM,all}$, respectively. To be consistent, the mean RM and RM standard deviation of FRB 20220529 in fig. \ref{fig:rmflare_snr} are also calculated using all the RM values. It is obvious that the ``RM flare'' in FRB 20220529 is very significant and of short duration. For comparison, the RM variations of other FRBs are consistent with fluctuations within 3$\sigma$ values. Although FRB 20180916B also exhibited an RM increase after a prolonged stochastic period\cite{Mckinven2023}, the amplitude is much smaller and the timescale is \REV{more than half a year. Thus, its RM evolution stays within $3\sigma$ of the standard deviation. In order to examine the influence of the secular variations, such as those in FRB 20121102A and FRB 20180916B, we tried to fit the RM evolution of each FRB with a second-degree polynomial. After subtracting the best-fit polynomial, we recalculated the $\rm SNR_{RM}$. It turns out that the ``RM flare'' of FRB~20220529 is still as \REVb{significant} as $7\sigma$, while the RMs of other FRBs are still within $3\sigma$.}

\REV{In order to explore the origin of the ``RM flare'', we investigate the possible differences in the linear polarization degree and DMs between the ``RM flare'' and the normal low RM state in \ref{sec:fL_vari} and \ref{sec:DM_vari}. The ``RM flare'' episode is defined as the duration during which the RM exceeds 1$\sigma$ of the low RM state, that is, $60292<\mathrm{MJD}<60308$.}

\subsection{\REV{Linear Polarization Degree Variation}} \label{sec:fL_vari}

\REV{The average linear polarization degree $f_\mathrm{L}$ outside the ``RM flare'' is $0.78 \pm 0.17$, while during the ``RM flare'' it is $0.47 \pm 0.15$. It should be noted that these average values do not take upper limits into account. To incorporate the upper limits into our comparison, we employ the two-sample Kolmogorov-Smirnov (KS) test to analyze the differences between samples during and outside the ``RM flare''. The cumulative distributions are presented in fig. \ref{fig:flks_dmks}A. Orange and blue steps indicate the $f_\mathrm{L}$ distributions during the ``RM flare'' episodes and outside it, respectively. Gray region indicates the uncertainties. In order to use more information, we include all bursts with RM measurements during the ``RM flare'' here, regardless of their S/N values. Additionally, we simulate the samples based on the uncertainties associated with each data point to estimate the uncertainty of the null probability of the KS test. For upper limits, we uniformly sample values from 0 up to the specified upper limits. The resulting null probability distribution from the KS tests yields a value of $\mathrm{log~}P_\mathrm{KS}=-7.5^{+0.8}_{-1.0}$, indicating a lower linear polarization during the ``RM flare'' compared with normal time.}

\REV{However, there is another episode $59850<\mathrm{MJD}<59900$ where $f_\mathrm{L}$ seems to be smaller than the normal state, with a ${\rm log}~P_\mathrm{KS}=-1.3^{+0.2}_{-0.4}$, although not as significant as the ``RM flare'' episode. \REVb{It is unclear if} the observed linear polarization degree decrease is physically related to the ``RM flare''.}

\REV{We also examined the relation between the RM and $f_\mathrm{L}$ during the ``RM flare'' episode, as shown in fig. \ref{fig:flks_dmks}C. 
For the bursts without detectable RM, the $f_\mathrm{L}$ upper limits are presented as arrows and RMs are assumed to be the interpolated values of their temporal neighbors. 
It turns out that there is no correlation between them. We tested the relation with the Spearman's rank coefficient of correlation and obtained a null probability of \REVb{$\rm log\ P=-0.5^{+0.3}_{-0.8}$, with the uncertainties estimated by the bootstrap method.} }

\REVb{Since the detected RM values are significantly lower than the maximum $ \left| {\rm RM} \right|$ where the signal would be completely depolarized within a single frequency channel\cite{Hotan04} ($\sim 9\times 10^4 \rm~rad~m^{-2}$ for our FAST observation with 0.122\,MHz-wide frequency channels and Parkes observation with 0.125\,MHz-wide frequency channels, and $\sim 1\times 10^4 \rm~rad~m^{-2}$ for our Parkes observation with 1\,MHz-wide frequency channels), the reduced polarization observed in high RM bursts is unlikely to be caused by instrumental depolarization.}
\REV{ There are some possible reasons to explain the observed polarization drop: 1) The drop in polarization may be due to the evolution of the multi-path propagation. When a radio wave propagates in a turbulent magnetized plasma screen, it would be depolarized due to the RM scatter in different paths. We consider that there is a fluctuation in electron density $n_e$ and the strength of magnetic field $B_\parallel$ across the length scale $l$, then the RM scatter is $\sigma_{\rm RM}\propto\delta(n_eB_\parallel)_l$ and the fractional reduction in the linear polarization amplitude is $f=1-\exp(-2\lambda^4\sigma_{\rm RM}^2)$ \cite{Feng2022, Yang2022}. If the magnetic geometry remains unchanged during the plasma screen expansion, $\sigma_{\rm RM}$ would prominently decrease due to the decrease of $\delta(n_eB_\parallel)_l$, leading to an increase of the polarization degree. Such a scenario is inconsistent with the observed polarization drop. Thus, a cascading magnetic field might be expected to contribute to the drop in polarization. During the plasma screen expansion, the magnetic field might be cascaded due to reconnection, turbulence, or other processes, leading to the increase of $\delta(n_eB_\parallel)_l$ and the drop in polarization. 
Furthermore, the depolarization features of FRB 20121102A and FRB 20190520B are mainly at GHz as reported in \cite{Feng2022}. 
Since the RM of FRB 20220529B is much smaller than those of the two repeaters, according to the picture of the multi-path, the transfer frequency of the depolarization is expected to be smaller than GHz. Thus, it might be natural that there is no significant relation between the polarization degree and the wavelength/frequency in our measurement.
2) The drop in polarization is due to Faraday conversion. When the radio wave propagates in a region with a strong magnetic field or a perpendicular field, the polarized states (linear and circular) would be converted from each other but the total polarization degree remains unchanged. Thus, if the strength or the geometric configuration of the magnetic field evolves, the observed drop in the linear polarization is expected. However, such a scenario predicts that the \REVb{circular} polarization would evolve, which is not confirmed by the current measurement.
3) The drop in polarization is due to the intrinsic properties of the radiation mechanism of FRBs. The radiation mechanism of FRBs is still unknown, although it is required to be coherent. Since a coherent process usually requires a fine-tuned regulation of the phase and polarization angle of the electromagnetic waves, coherent emission is usually expected to be highly polarized. If the drop in polarization is due to the intrinsic properties of the radiation mechanism, it is expected that the coherence level evolves as the polarization degree drops, leading to decrease of the burst flux. However, such a scenario has not been confirmed by the current measurements.}

\subsection{\REV{DM Variation}} \label{sec:DM_vari}

\REV{
Complex morphologies, such as downward-drifting subbursts, are often observed in bursts from repeating FRBs~\cite{Hessels2019, Chawla2020}. Similar structures are also presented in FRB 20220529, as shown in fig.~\ref{fig:burst_pol}. In view of these structures, the DM determined by maximizing the S/N of the integrated pulse profile may misestimate the intrinsic DM. To address this, we use the \emph{\sc DM\_phase} package~\cite{Seymour2019} to estimate the structure-maximizing DM (${\rm DM}_{\rm stru}$), based on the coherent power over the emission bandwidth. 
The ${\rm DM}_{\rm stru}$ of FRB 20220529 bursts with RM measurements during the ``RM flare'' phase are listed in Table~\ref{table:RM} and Fig. \ref{fig:rmflare}D. Since this approach is more accurate for bursts with high S/N and narrow features, we only use bursts with an S/N equal to or above ten 
in the subsequent comparisons.  }

\REV{In general, the $\mathrm{DM_{stru}}$ outside and during the ``RM flare'' are $249.5 \pm 3.5~\mathrm{pc~cm^{-3}}$ and $247.1 \pm 2.8~\mathrm{pc~cm^{-3}}$, respectively. They are consistent with each other. 
\REVb{The DM estimation for FRBs is usually limited by the narrow frequency range of the bursts and yields different results with different methods, such as the highest S/N method or structure method. }
Moreover, DMs usually show substantial variations within one day, which is hardly explained by the variation of electron column density. Thus, it is believed that the detected DMs, regardless of whether it is estimated with the highest S/N method or structure method, are modified by some intrinsic frequency-delay time relations.
Thus, before the exploration of the DM variation, we average the DM daily. Weighted average $\overline{\mathrm{DM}}(\mathrm{MJD})=\Sigma~(w_\mathrm{i}\mathrm{DM_i})/\Sigma~w_\mathrm{i}$ is used to estimate the daily average DM, where $w_\mathrm{i}=1/\epsilon^2_\mathrm{i}$ is the weight and $\epsilon_\mathrm{i}$ is the uncertainty of each detected DM. In order to take into account both the observation uncertainties and the scattering between different bursts, we define the uncertainty of the average $\mathrm{\overline{DM}}$ in each day as 
$$\sigma^2_\mathrm{\overline{DM}}=
\sum \big(\frac{w_\mathrm{i}}{\Sigma~w_{\rm i}}\big)^2\epsilon^2_\mathrm{i}+\mathrm{RMS^2_{DM}}
=\frac{1}{\Sigma~w_\mathrm{i}}+\mathrm{RMS}^2_\mathrm{DM},
$$
where $\mathrm{RMS_{DM}^2}$ is the scattering of the detected DM. 
We explore the similarity of the daily average $\mathrm{DM_{stru}}$ during and outside the ``RM flare''. The distributions are presented in fig. \ref{fig:flks_dmks}B. It shows that they are statistically consistent with each other, although the $\mathrm{DM_{stru}}$ within the ``RM flare'' appears to be slightly smaller.}

\REV{Moreover, we also examined the relation between RM and $\rm DM_{stru}$ during the ``RM flare'' episode. The results are presented in fig. \ref{fig:flks_dmks}D. Similar to the examination of the relation between $f_\mathrm{L}$ and RM, we included all the bursts with $\rm DM_{stru}$, irrespective of its S/N, and assumed the RM values of the bursts without detectable RM by interpolation.  
It turns out that there seems to be a positive correlation between RM and $\rm DM_{stru}$ during the ``RM flare'' episode. The null probability of the Spearman correlation analysis is \REVb{$\rm log\ P=-4.9^{+1.8}_{-2.2}$, with the uncertainties are estimated with the bootstrap method}. If the bursts without detectable RM are excluded, the null probability is $\rm log\ P=-4.1^{+1.6}_{-2.1}$. 
\REVb{Consequently, although the correlation looks significant, the uncertainties are quite large, consistent with no correlation within 2.5 $\sigma$.}
}

\REV{In summary, the daily average $\rm DM_{stru}$ during the ``RM flare'' episode is not significantly different from the whole DM sample, while the DM \REVb{seems to be} positively correlated with the RM during the ``RM flare'' episode, 
We warn here \REVb{the larger uncertainty and} that DMs in the low RM episode also exhibit DM fluctuations. Thus, we are cautious to claim such a correlation being physical. 
}

\subsection{VLA observation and localization}

During the active interval between January 2023 and April 2023, we conducted VLA observations under the director’s discretionary time (DDT) project 23A-385 (PI: Ye Li). The FRB field was observed with three 3-hour observations, two on 2023-02-18 and one on 2023-02-24. The observations were in a frequency range of $1-2\,\rm GHz$, with 1024 channels. The VLA antennas were in the B array configuration, with a maximum baseline of 11.1 km. The nominal spatial resolution is 4.3 arcsec in $1.5~\rm GHz$ 
\cite{vlaangular}.

Besides the standard visibility data recorded with a $3\,\rm s$ sampling time, the data with a sampling time of 10 milliseconds were also streamed to the {\it realfast} search system to search for bursts from FRB 20220529 in our VLA observations \cite{law2018}. 
Two FRBs were detected with the {\it realfast} system during our nine-hour observation, \REV{with prompt signal-to-noise ratios 7.8 and 9.9, respectively. Usually, the bursts below $\rm S/N = 8$ are \REVb{difficult} to image and have a larger error on the position.} The first one, detected on 2023-02-18, was too faint, and we do not take its information into account here. The second burst was detected on 2023-02-24 at 22:12:41.97 UTC.
We make an image with the raw visibilities dedispersed at the real-time detected DM, $237.8~{\rm pc~cm^{-3}}$, using CASA 6.1.4.12. After converting the visibilities in the science data model (SDM) format to measurement set (MS) format using the CASA tasks \texttt{importasdm}, we calibrate it using the task \texttt{applycal} with the CASA calibration tables from the NRAO Archive for this observation. 
The quasar 3C48 was used as a flux and bandpass calibrator, and J0122+2502 was used as a phase and amplitude calibrator. 
The calibrated measurement set was imaged using the CASA task \texttt{tclean} and then the burst was fitted as an elliptical Gaussian with task \texttt{imfit}.
The resulting S/N is 12.6 in $1.39 - 1.43~\rm GHz$. We identify the position using \texttt{imfit} as $01^{\rm h}16^{\rm m}25.0124^{\rm s}$, $+20^{\circ}37^{\prime}56.8270^{\prime\prime}$ with uncertainties of $0.11^{\prime\prime}$ and $0.12^{\prime\prime}$ in RA and Dec, respectively. 

The systematic position error was estimated by comparing the objects in the radio continuum images with the PanSTARRS-DR2 stack catalog \cite{panstarrs16,panstarrsweb}. 
We reduced the standard visibility data with CASA 6.1.4.12 and made images for each observation individually with a resolution of 0.5 arcsec per pixel. The sources are extracted with PyBDSF \cite{PyBDSF}.
We selected true bright, compact radio sources using the following criteria: 1) Objects with other sources $10^{\prime\prime}$ nearby are excluded to avoid the sidelobes of very bright sources. 2) The signal-to-noise ratio (the ratio between the peak flux and the background root-mean-square) is required to be larger than 5. 3) The peak flux is required to be larger than 70\% of the total flux. 4) The source can be fitted with a single Gaussian, i.e., {\it S$\_$Code} is 'S'. There are 84 point radio sources without nearby objects detected in the observational session. The radio sources are then cross-matched with the PanSTARRS-DR2 catalog within a separation of $1^{\prime\prime}$. The trials for larger separations reveal that the cross-matched pairs with separations larger than $1^{\prime\prime}$ are dominated by chance coincidence. Due to the notably higher noise in the edge of the VLA field of view, only objects within 0.2 degrees of the center of the field of view are taken into account.
There are 23 pairs cross-matched.
We visually checked the radio images to ensure they were unresolved point sources and the optical counterparts were real. The median RA and Dec offsets between VLA and PanSTARRS catalogs are $-0.024^{\prime\prime}$ and $0.204^{\prime\prime}$, respectively, and the systematic position errors between the radio positions and PanSTARRS positions are $0.161^{\prime\prime}$ and $0.196^{\prime\prime}$. Thus, the final localization of FRB 20220529 is $\rm RA=01^{\rm h}16^{\rm m}25.014^{\rm s}$, $\rm Dec=+20^{\circ}37^{\prime}56.6^{\prime\prime}$ with a positional error of $0.3^{\prime\prime}$, dominated by the systematic uncertainty. 

\subsection{Optical image and host galaxy identification}

We explore host galaxy candidates in the DESI Legacy Survey. The DESI $r$-band image of the \REV{FRB 20220529 position} is presented as Fig. \ref{fig:host}A. 
\REVb{For objects within 5 arcminutes of FRB 20220529, we exclude stars with detected parallax in Gaia and then estimate their probabilities to be the host galaxy with the Probabilistic Association of Transients to their Hosts (PATH) formalism\cite{aggarwal2021}. The PATH method estimates the probability of an extragalactic transient source associated with host galaxy candidates using Bayes' rule. The probability is assumed to be inversely proportional to the angular surface density of the galaxies and proportional to the normalized offset distribution of the transients. The surface density is specified to the magnitudes of the host candidates. The brighter the galaxy is, the smaller the density is. The normalized offset is the offset normalized by the half-light radius of the galaxy, while the offset is the distance between the FRB and the galaxy center. The normalized offset distribution of FRB is assumed to be exponential, with a scaling factor 1, and have a maximum of 10. Note that different scaling factors such as 0.5 or 2.0 do not affect our results much. 
For galaxies within 10 arcseconds of FRB~20220529, we list the galaxy coordinates, $r$ band magnitudes $m_{\rm r}$, the half-light radii $R_{50}$, the offset between FRB 20220529 and the center of the galaxies $R_{\rm off}$, as well as their probabilities to be the host $\rm P(O|x)$ in table \ref{tb_host}. It turns out that the galaxy PSO J019.1046+20.6327 (G1) has the highest probability to be the host galaxy, nearly one. Consequently, PSO J019.1046+20.6327 is identified as the most plausible host galaxy. }


\subsection{GTC spectrum observation}
We observed the host galaxy of FRB 20220529 using OSIRIS+ Long Slit Spectroscopy (LSS) on the GTC telescope under project ID: GTCMULTIPLE1A-23ACNT (PI: C. W. Tsai) on 18 August 2023. The R500R grism with a slit width of $1^{\prime\prime}$, which covers a wavelength range from 4800 \AA\ to 10000 \AA~, was used in order to cover the possible redshift range of the host galaxy, up to redshift 0.5. The observations were conducted on a dark night with a seeing of $0.9^{\prime\prime}$. Three 800 s exposure observations were conducted. The data were bias-subtracted, flat-fielded, cleaned of cosmic rays, wavelength calibrated using comparison-lamp spectra, and combined with PyPeit \cite{pypeit1,pypeit2} under standard techniques. The standard star ROSS 640 is utilized for flux calibration. 

The combined spectrum is presented in Fig. \ref{fig:host}B. 
\REV{ H$\alpha$, H$\beta$, \ion{{\rm O}}{iii}\,$\lambda\lambda4959,\,5007$ doublet emission lines are identified.}
We fit the spectrum with a power law indicating the continuum and Gaussian functions indicating the emission lines. The central wavelengths of the emission lines are tied together and shifted for different redshifts. The median pixel size of our spectrum is 4.9 \AA. Therefore, we incorporate the wavelength uncertainties with the Orthogonal Distance Regression (ODR) method \cite{ODR}.
With these emission lines, the redshift of the host galaxy is identified as $z=0.1839 \pm 0.0001$.

\subsection{Host Galaxy Properties}

\paragraph*{Offset and $F_{\rm light}$:} 
We localize the center of the host galaxy with the $r$ band image of DESI Legacy Survey using \REVb{Source-Extractor} (SExtractor \cite{1996A&AS..117..393B,sextractorweb})
as $\rm RA = 01^{\rm h}16^{\rm m}25.078^{\rm s}$, $\rm Dec = +20^{\circ}37^{\prime}57.72^{\prime\prime}$. 
The offset between the FRB and the host galaxy center is calculated as $1.4 \pm 0.2~\rm arcsec$, corresponding to $R_{\rm off} = 4.4 \pm 0.6~\rm kpc$ at redshift $0.1839 \pm 0.0001$ using a Lambda cold dark matter ($\Lambda$CDM) cosmological model with standard parameters \cite{WMAP5}. 
The half-light radius of the host galaxy is provided in the
DESI Legacy Survey catalog, which is $1.17^{\prime\prime} \pm 0.02^{\prime\prime}$, corresponding to $3.60 \pm 0.08~\rm kpc$. The normalized offset is $r_{\rm off} = R_{\rm off}/R_{50} = 1.2 \pm 0.2$.
The offset information of FRB 20220529 is quite typical in FRBs \cite{2022AJ....163...69B}. 

In order to explore the environment of the FRB, we also estimate the cumulative light fraction $F_{\rm light}$, the fraction of the total brightness of the regions fainter than the FRB position to the total brightness of the host with the $r$-band DESI Legacy Survey image. Following Lyman et al. (2017), we get the region of the host galaxy from SExtractor and sort the brightness of the pixels. The brightness of the FRB region is estimated based on the position and uncertainty of the FRB with the ds9 region. The cumulative light fraction is then estimated to be $F_{\rm light} = 0.13^{+0.15}_{-0.09}$. Although the $F_{\rm light}$ is small, the uncertainty is quite large. It is consistent with core-collapse supernovae and binary mergers. 

\paragraph*{Stellar Mass:} 
We perform a broadband Spectral Energy Distribution (SED) fitting to explore the stellar masses of the host galaxy. The Code Investigating GALaxy Emission (CIGALE \cite{2009A&A...507.1793N, cigaleweb})
is utilized, with the Galactic extinction corrected \cite{2011ApJ...737..103S} $g$, $r$, and $z$ band model fluxes from the DESI Legacy Survey and the W1, W2 magnitudes from the Wide-Field Infrared Survey Explorer (WISE) telescope. 
The composite stellar populations are built from simple stellar populations \cite{2003MNRAS.344.1000B} combined with flexible star formation histories.
The Salpeter model is used as the initial mass function. An e-folding \texttt{sfhdelayed} model with an initial SFR of 0.1 is assumed as the star formation history. The \texttt{dustatt\_calzleit} dust attenuation model with the UV bump centroid at 217.5 nm, as well as the \texttt{casey2012} dust emission model\cite{2012MNRAS.425.3094C}, is used. No AGN is added. 
The resulting stellar mass is $M_* = (2.7 \pm 0.7) \times 10^9~\rm M_{\odot}$.

\paragraph*{Star Formation Rate (SFR):}
We use the luminosity of H$\alpha$ line $L_{\rm H\alpha}$ to estimate the star formation rate (SFR).
Firstly, we made the Galactic extinction corrections with $R_{\rm V}=3.1$ and $\rm E(B-V)_{\rm MW} = 0.0706~mag$ \cite{2011ApJ...737..103S, extinction}. 
Then, the GTC spectrum is fitted with a power law as the continuum and Gaussian functions as emission lines. The slit correction is estimated by the fraction between the light within the slit and the entire galaxy in the galaxy segmentation provided by SExtractor. After correction, the H$\alpha$ flux is $2.5 \times 10^{-16}~{\rm erg~s^{-1}~cm^{-2}}$, corresponding to $2.5 \times 10^{40}~{\rm erg~s^{-1}}$ at redshift $0.1839 \pm 0.0001$. The SFR can then be estimated with the H$\alpha$ \cite{2012ARA&A..50..531K} as ${\rm SFR} = L_{\rm H\alpha}/(10^{41.27}~{\rm erg~s^{-1}})= 0.13~ M_{\odot}~{\rm yr}^{-1}$. The specific SFR is subsequently calculated as $0.05~\rm Gyr^{-1}$, which is consistent with repeating FRBs\cite{2022AJ....163...69B} as well as supernovae and SGRBs.

\paragraph*{\REV{Comparison with other stellar transient host galaxies:}}

\REV{We compare the host galaxy properties of FRB 20220529 with those of other FRBs as well as well-known stellar transients in fig. \ref{fig:host_comparison}. The host galaxy property samples from \cite{liye2020} are used, including galaxy stellar mass ${\rm log}~M_*$, SFR, sSFR, half-light radius $R_\mathrm{50}$, offsets from the center of the host galaxies $R_\mathrm{off}$, and the cumulative light fraction $F_\mathrm{light}$ of long-duration gamma ray bursts (LGRBs), superluminous supernovae(SLSNe), Type Ib/Ic SNe (SN Ibc), Type II SNe (SN II), Type Ia SNe (SN Ia), as well as short-duration gamma ray bursts (SGRBs). The host galaxy properties of FRB 20220529 are presented as red vertical lines, consistent with those of other FRBs. The values are also consistent with core-collapse supernovae and SGRBs in all panels, but not with LGRBs and SLSNe in the $F_\mathrm{light}$ panel.} 

\REV{To be quantitative, we apply the Naive Bayes method to identify the similarity of FRB 20220529 host galaxy to other stellar transients \cite{liye2020}. The result is presented in the lower right panel of fig. \ref{fig:host_comparison}. The probabilities $P$ of FRB 20220529 to have the same origin of each type is estimated based on the host galaxy information. It is shown that FRB 20220529 has a similar environment with core-collapse supernovae and SGRBs. }
\REVb{If the source is in a globular cluster like FRB 20200120E, the host galaxy properties would not be representative
of the FRB environment.}

\section{Supplementary text}

\subsection{Possible astrophysical origins of the ``RM flare''}

The observed feature of the ``RM flare'' is significantly different from that in the low RM state. It displays a continuous decline from $2000~{\rm rad~m^{-2}}$ to $200~{\rm rad~m^{-2}}$ over two weeks.
\REV{ In the following subsections, we will discuss several possible astrophysical scenarios that might produce the observed ``RM flare'', including magnetar outflows, turbulence in a supernova remnant or a pulsar wind nebula, orbital motion in a binary system, and coronal mass ejection from a companion star.} 

\subsubsection{``RM flare'' from magnetar outflows} \label{sec:giantflare}

\REV{ It is natural to consider that the ``RM flare'' is generated by an outflow from the central engine of the FRB source.
For a magnetar as the FRB central engine, the outflow could be associated with a typical X-ray burst or a giant flare.}
We first note that the observed ``RM flare'' is unlikely contributed by a magnetar outflow associated with a typical X-ray burst due to the following observational evidence. 
Over 200 X-ray bursts were detected during the active phase of the Galactic FRB-emitting magnetar SGR 1935+2154. 
This implies that a large number of the outflows associated with typical X-ray bursts have been launched during a short period of time. However, no significant RM variation was observed from the source \cite{CHIME20,Zhu23}. For example, the RM of FRB 20200428D is almost the same as that of the radio pulses from SGR J1935+2154 in the non-active period. Also, typical X-ray bursts usually occur in the active phases of magnetars.
If the ``RM flare'' were related to a high FRB burst rate \cite{Metzger19MN, Lu20, YangZhang21}, one would expect to see a correlation between RM and burst rate. However, the burst rate in the ``RM flare'' state is normal and even slightly lower than that in the low RM state (as shown in Fig. \ref{fig:rmflare} and discussed in the main text). 
Based on these arguments, we conclude that the ``RM flare'' from FRB 20220529 is not related to magnetar outflows associated with typical X-ray bursts. 

\REV{Another possibility is that the observed ``RM flare'' is from a magnetar outflow associated with a giant flare that has a much larger explosion energy than that of typical X-ray bursts. 
We consider that in the outflow of a giant flare, the electron number density is $n_e$, the bulk Lorentz factor of the outflow is $\gamma$, and the particles' thermal Lorentz factor is $\gamma_{\rm th}$. Then the RM contributed by the relativistic outflow is \cite{yang2023}
\begin{align}
{\rm RM}
\sim\frac{e^3}{2\pi m_e^2c^4\gamma_{\rm th}^2}n_{e,{\rm net}}B \Delta R, 
\end{align}
where $n_{e,{\rm net}}$ is the net charge density in the outflow (only the net charges can contribute to RM), and $\Delta R$ is the thickness of the magnetar outflow.  
We notice that the RM contribution is suppressed by a factor of $\gamma_{\rm th}$ due to the relativistic mass $m_e\rightarrow\gamma_{\rm th} m_e$. 
If there is no baryon loading in the magnetar outflow, the net charge would be related to the Goldreich--Julian density and the corresponding RM would be extremely small \cite{yang2023}.
The radio afterglow associated with the giant flare of SGR 1806-20 requires a relatively high baryon ejection \cite{Gelfand2005, Granot2006}. 
Substantial baryon ejection may accompany at least the most powerful magnetar flares as a result of a shock wave driven by sudden magnetospheric energy release from the neutron star crust \cite{Cehula2024}. In the following discussion, we assume that the net charges in the outflow are dominated by baryon loading in the giant flare. The number density of the net charges in the outflow is then given by
\begin{align}
n_{e,{\rm net}}\simeq\frac{M}{4\pi  m_p(v_{\rm ej}t)^2\Delta R},
\end{align}
where $M$ is the total ejecta mass, $m_p$ is the proton mass, and $v_{\rm ej}$ is the ejecta velocity. We assume that the magnetar outflow is generated at the magnetosphere with a scale of the light cylinder, $R_{\rm LC}=cP/2\pi$, where $P$ is the magnetar spin period. Meanwhile,
the magnetic field in the outflow is dominated by that of the magnetar wind due to its relatively slow decay with radial distance (i.e., $B\propto r^{-1}$).
At distance $r=v_{\rm ej}t$, the magnetic field strength could be estimated by
\begin{align}
B(r)=\frac{B_p}{2}\left(\frac{R}{R_{\rm LC}}\right)^3\left(\frac{R_{\rm LC}}{r}\right).
\end{align}
Therefore, the RM contributed by the magnetar outflow associated with a magnetar giant flare is estimated by
\begin{align}
{\rm RM}\sim1.6~{\rm rad~m^{-2}}~M_{24}v_{\rm ej,0.1c}^{-3}t_{10{\rm day}}^{-3}B_{p,14}P_{0}^{-2}\gamma_{\rm th,0}^{-2}\label{GFRM}
\end{align}
where the convention $Q_x = Q/10^x$ is adopted in cgs units, $v_{\rm ej,0.1c}=v_{\rm ej}/0.1c$ and $t_{10{\rm day}}=t/10~{\rm day}$. 
A magnetar outflow with a mass of $M\sim10^{25}-10^{26}~{\rm g}$ and a velocity of $v_{\rm ej}\sim(0.3-0.6)c$ is compatible with the properties
inferred from the afterglow of the giant flare from SGR 1806-20 \cite{Cehula2024}.
Thus, in order to explain the observed ``RM flare'' of FRB 20220529 with a maximum value of ${\rm RM}\sim2000~{\rm rad~m^{-2}}$, the required giant flare should be at least more powerful than the most powerful giant flare known. Furthermore, considering that the electrons could be accelerated to a relativistic speed by magnetic dissipation processes, leading to $\gamma_{\rm th}\gg1$, one requires an even more powerful giant flare with an even larger baryon loading to account for the observation. }

\REV{ A severe problem of the giant flare scenario is that the predicted RM evolution curve does not match the observation. This is because the plasma contributing to the ``RM flare'' is always on the line of sight and becomes diluted as the outflow expands. Its RM evolution is predicted to have a power-law decay (as shown in Eq. (\ref{GFRM})), corresponding to an evolution with a positive second time derivative of RM, i.e., $d^2{\rm RM}/dt^2>0$. However, even though the observed ``RM flare'' of FRB 20220529 showed a decreasing evolution, the evolution of its second derivative appears to be negative, i.e., $d^2{\rm RM}/dt^2<0$, especially near the maximum value at MJD 60292. We explore the first and second derivatives, i.e., $d{\rm RM}/dt$ and $d^2{\rm RM}/dt^2$, in Fig. \ref{fig:rmflare_caseIaII}. The observed RM evolution, as well as the first and second derivatives are presented in panels A, B and C. The RM values are averaged on a daily basis and the uncertainties are estimated with the same method described in Section \ref{sec:DM_vari}. The predictions of different models are also presented. The best fit values from Section \ref{sec:model_fit} are used here. One can see that the power law model deviates from the data badly. We therefore, conclude that the ``RM flare'' of FRB 20220529 likely did not originate from a magnetar giant flare.}

\subsubsection{``RM flare'' from turbulence in a supernova remnant or a pulsar wind nebula}

If the FRB source is a young magnetar, as proposed by some previous works \cite{Metzger19MN, Lu20, YangZhang21}, it may be surrounded by a supernova remnant (SNR) and/or a young pulsar wind nebula (PWN).
We first discuss whether the ``RM flare'' of FRB 20220529 could be attributed to an SNR. 
\REV{ The RM evolution from a young SNR has been proposed to be monotonic over a timescale much longer than a few years \cite{Piro18}, if the magnetic geometry along the line of sight does not experience substantial changes. However, the observed week-timescale variation of the ``RM flare'' of FRB 20220529 is inconsistent with such a long-term monotonic evolution. Thus, if the ``RM flare'' is attributed to a young SNR, it is more likely related to intrinsic time fluctuations of some turbulent clumps or the relative motion between a clump and the FRB source \cite{yang2023}. The distribution of the polarization position angle from radio observations of some SNRs suggests that magnetic fields in SNRs are largely disordered \cite{Milne1987, Reynolds2012}. In this case, due to the intrinsic time fluctuations of turbulent clumps or the relative motion between the FRB source and the turbulent clump, the RM could show random variations in short terms. 
However, there are some issues with this explanation: 
1) The maximum value of the ``RM flare'' episode is significantly higher than that in the low RM state and the ``RM flare'' eventually falls back to the baseline level of the low RM state, as shown in Fig. \ref{fig:rmflare}. If both the ``RM flare'' and the RM fluctuations in the low RM state are attributed to the same turbulent medium, it is not easy to explain why the ``RM flare'' can fall back to the baseline, because fluctuations in the turbulent medium usually are random. 
2) The turbulence properties could be tested by some statistical methods. For example, although the RM evolution caused by a turbulent medium could be variable at different times or in different regions, for a certain medium, the statistical properties (e.g., the structure function) should remain more or less unchanged. We calculated the structure functions of RM evolution for both the low RM state alone and that including the ``RM flare''. If the ``RM flare'' and the RM fluctuations in the low RM state are attributed to turbulence in the same medium, the structure functions in the two scenarios should be almost the same, because the power spectrum of turbulence fluctuations usually satisfies a certain power-law distribution \cite{yang2023}.
However, as shown in fig. \ref{fig:rm_sf}, at the delay time of a few hundred days, the structure function involving the ``RM flare'' (pink dots) becomes one order of magnitude higher than that of the low RM state only (blue dots). This suggests that the ``RM flare'' should be from another component that is significantly different from the turbulence contribution in the low RM state. One possibility is that there are some special intermittent structures deviating from the power-law (e.g., Kolmogorov) scaling. However, the small scale size of the special turbulent clump raises the issue that the chance probability of detecting such clumps is very small as discussed below. 
3) The week-timescale of the ``RM flare'' leads to an extremely small scale of the turbulence clump. We consider that the turbulent medium might have a typical velocity of $v_{\rm tur}\sim(10-1000)~{\rm km~s^{-1}}$ in an SNR, then the scale of the turbulent clump is estimated to be $l\sim v_{\rm tur}t\sim0.6~{\rm AU}(v_{\rm tur}/100~{\rm km~s^{-1}})(t/10~{\rm day})$. Such a scale is much smaller than the observed minimum scale of the magnetic fluctuations in some SNRs \cite{Milne1987, Landecker1999, Reynolds2012}. Thus, SNRs older than a few hundred years would be less likely to account for the ``RM flare''. 
Furthermore, in order to contribute to the ``RM flare'', the electron density and the magnetic field in the clump are required to be much higher than the rest of the medium on the large scale. We assume that the SNR has a typical scale $L$ and contributes to the RM in the low state, while the small clump has a typical scale of $l_m$ and contributes to the ``RM flare''. Since the ``RM flare'' is about one order of magnitude more significant than that in the low RM state, the ratio of $n_eB_\parallel$ between the small clump and the average value of the SNR is $(n_eB_\parallel)_{\rm clump}/(n_eB_\parallel)_{ave}\sim10(L/l_m)\sim2\times10^6(L/1~{\rm pc})(l_m/1~{\rm AU})^{-1}$, leading to an extreme value of $n_eB_\parallel$ for the small clump.
A young SNR with an age of less than a few hundred years might have a smaller minimum fluctuation scale. In addition to the variation in short terms, it would also show an observable, secular evolution in both DM and RM \cite{yang2023}. However, these features have not been shown in the observation of FRB 20220529. In conclusion, based on the above three points, the SNR scenario needs to involve some special small intermittent structure with a scale of AU, which means that the chance probability of detecting such clumps is very small and the value of $n_eB_\parallel$ of the clump is extreme.}

\REV{ Next, we discuss whether a PWN could contribute to the observed ``RM flare''. A PWN usually includes a relativistic pair wind and a termination shock. The pair wind is mainly composed of relativistic electron/positron pairs. The RM contributions from electrons and positrons would be canceled out due to the symmetry of the charged particles. Although there are some net charges in the pair wind, the relativistic thermal motion of these particles would also significantly suppress the RM value \cite{yang2023}. Besides, in the pair wind, the magnetic fields are likely almost perpendicular to the wind velocity \cite{Becker2009}, leading to a much smaller RM contribution. Thus, the RM from the pair wind may be ignored. A more likely location is at the termination shock region, where a reverse shock (i.e., the pair wind is shocked) and a forward shock (i.e., the SNR is shocked) coexist.
In the reverse shock, the interaction between the pair wind and the SNR transfers the kinetic energy of the particles in the pair wind to thermal energy, with the number of net charges remaining unchanged. Thus, the RM contribution from the reverse shock may still be negligible.
In the forward shock, the RM is mainly contributed by the shocked medium in the SNR. Such a case is similar to the above discussion about the SNR scenario, but the RM region is mainly at the inner radius of the SNR. In order to generate the observed ``RM flare'', the intrinsic time fluctuations or the relative motion along the line of sight between the FRB source and the turbulent clump is still required. This raises the same issue pointed out for the SNR scenario that we will not repeat.}

\REV{In summary, even though turbulence from either a SNR or a PWN termination shock cannot be ruled out as the mechanism for the ``RM flare'', extreme physical conditions are required. 
}

\subsubsection{``RM flare'' from orbital motion of a binary system}

\REV{ There is some evidence suggesting that prominent RM variations could arise from orbital motion of a binary system.
Observationally, a Galactic binary system PSR B1259-63/LS 2883, containing the pulsar PSR B1259-63 and the Be star LS 2883, shows substantial RM variations when the pulsar approaches the periastron \cite{Johnston1996,Johnston05}. The apparent RM of the pulsar reaches a maximum of $-14800~{\rm rad~m^{-2}}$, which is some 700 times the value measured away from the periastron. The behavior is similar to the ``RM flare'' of FRB 20220529. 
In the binary scenario, the electron density and the magnetic field strength that contribute to RM are supposed to be from the stellar wind {(e.g., from an O/B star or a giant star\cite{yang2023})} or the disk (e.g., surrounding a Be star \cite{wangfy22}) of the companion. The electron density and the magnetic field strength near the periastron would be much higher than those away from the periastron, which could naturally explain the RM evolution of PSR B1259-63. 
Such observational evidence implies that the ``RM flare'' of FRB 20220529 might also be caused by orbital motion of a binary system. Within such a scenario, the following physical conditions should be satisfied: 
1) The orbital period of the source of FRB 20220529 should satisfy 
\begin{align}
P_{\rm orb}\gtrsim 1.5~{\rm yr},
\end{align}
because there is only one ``RM flare'' detected since the repeating FRB was discovered in May 2022. 
We consider that the central engine of FRB 20220529 has a mass of $m\sim1.4~M_\odot$ (assuming that it is a neutron star), and the companion has a mass of $M$.
According to Kepler's third law, the orbital separation is
\begin{align}
a=\left[\frac{G(M+m)P_{\rm orb}^2}{4\pi^2}\right]^{1/3}=2.8~{\rm AU}\left(\frac{M+m}{10M_\odot}\right)^{1/3}\left(\frac{P_{\rm orb}}{1.5~{\rm yr}}\right)^{2/3}.\label{separation} 
\end{align}
To satisfy $P_{\rm orb}\gtrsim1.5~{\rm yr}$, the orbital separation is required to be $a>2.8~{\rm AU}$ for a companion with a mass of $M\sim10~M_\odot$.
2) The orbit should have a large eccentricity $e$, so that the RM near the periastron is prominently larger than elsewhere in the orbit, and the duty cycle of the high-RM state is very small. For the Galactic binary system PSR B1259-63/LS 2883, the orbital period is measured to be $P_{\rm orb}\simeq1237~{\rm day}$ and the eccentricity is measured to be $e\simeq0.87$. Within one orbital period, the high-RM state of PSR B1259-63/LS 2883 lasts for about $T_{\rm high}\sim100~{\rm day}$ \cite{Johnston1996,Johnston05}. Thus, the duty cycle of the high-RM state is $T_{\rm high}/P_{\rm orb}\sim0.1$. For FRB 20220529, its ``RM flare'' lasts for over two weeks, $T_{\rm high}\sim14~{\rm day}$, and the orbital period should be larger than 1.5 years at least, $P_{\rm orb}\gtrsim1.5~{\rm yr}$, thus, the duty cycle of the high-RM state is
\begin{align}
\frac{T_{\rm high}}{P_{\rm orb}}\lesssim0.026,
\end{align}
which is much smaller than that of the Galactic binary system PSR B1259-63/LS 2883. }

Consider that the RM is mainly contributed by the medium from the companion. Both the electron density and the magnetic field from the companion should have a power-law drop with distance $r$, giving
\begin{align}
{\rm RM} (r) \propto r^{-\alpha},
\end{align}
where $\alpha=2$ and $3$ correspond to a toroidal field and a radial field, respectively \cite{yang2023}. 
Thus, the relative RM variation is
\begin{align}
\frac{\delta {\rm RM}}{{\rm RM}}=-\alpha\frac{\delta r}{r}.
\end{align} 
Since the RM reaches its maximum value ${\rm RM_{max}}$ at the periastron distance of $r_0=a(1-e)$, for an RM variation from ``a fraction of ${\rm RM_{max}}$'' to ${\rm RM_{max}}$, the region contributing to the ``RM flare'' should be from a distance scaled by the periastron distance $r_0$, say at $\eta r_0$ with $1<\eta\leqslant(1+e)/(1-e)$, 
as shown in fig. \ref{fig:model_binary}. The Kepler motion of the FRB source gives 
a relation between radial distance and time,
\begin{align}
t=\sqrt{\frac{\mu}{2GMm}}\int_{r_0}^{ r}\frac{r'dr'}{\sqrt{r'-r'^2/(2a)-a(1-e^2)/2}},
\end{align}
where $t$ is the time of the Kepler motion from $r_0$ to $r$, and $\mu=Mm/(M+m)$ is the reduced mass. The orbital period of Kepler motion is
\begin{align}
P_{\rm orb}=2\pi a^{3/2}\sqrt{\frac{\mu}{GMm}}.
\end{align}
Therefore, the duty cycle of the high-RM state in one orbital period is given by
\begin{align}
\frac{T_{\rm high}}{P_{\rm orb}}=\frac{2t(r=\eta r_0)}{P_{\rm orb}}
=\frac{1}{\sqrt{2}\pi}(1-e)^{3/2}\int_1^\eta \frac{xdx}{\sqrt{x-x^2(1-e)/2-(1+e)/2}},
\end{align}
where $x\equiv r'/r_0$ is defined, and $r_0=a(1-e)$ is used for an elliptical orbit. For a high-eccentricity case with $e\sim1$, the integral term in the above equation is roughly independent of $e$ and can be approximated as $\sim(2/3)\sqrt{\eta-1}(\eta+2)$, leading to
\begin{align}
\frac{T_{\rm high}}{P_{\rm orb}}\sim\frac{\sqrt{2}}{3\pi}\sqrt{\eta-1}(\eta+2)(1-e)^{3/2}.
\end{align}
We assume that the companion properties of the binary system including FRB 20220529 are similar to those of the Galactic binary system PSR B1259-63/LS 2883. This ensures that both  companions have a similar $\eta$, so that
\begin{align}
\frac{(T_{\rm high}/P_{\rm orb})_{\rm FRB}}{(T_{\rm high}/P_{\rm orb})_{\rm PSR}}\sim\frac{(1-e_{\rm FRB})^{3/2}}{(1-e_{\rm \REVb{PSR}})^{3/2}}\lesssim0.26 .   
\end{align}
Using $e_{\rm PSR}\simeq0.87$ of the PSR B1259-63/LS 2883 system, one can derive the eccentricity of the binary system including FRB 20220529 as
\begin{align}
e_{\rm FRB}\gtrsim0.95.
\end{align}
Therefore, if the ``RM flare'' of FRB 20220529 is attributed to orbital motion of a binary system, the orbital eccentricity is required to be very high. 

It is necessary to check whether the FRB source collides with the companion under such a high eccentricity. Assuming that the companion has a radius of $R$, then $R<r_0=a(1-e)$ is required. For typical values of $a\sim3.4~{\rm AU}$ (Eq.~(\ref{separation})) and $e\sim0.96$, the companion star radius is required to be smaller than $R\sim 2\times10^{12}~{\rm cm}$, which is 
substantially larger than the radii of most main sequence stars.

However, it should be noted that the binary systems with $e\gtrsim0.95$ and $P_{\rm orb}\gtrsim 1.5~{\rm yr}$ are rare among the binary sample containing a neutron star. 
We conducted a statistical analysis of the eccentricities and orbital periods of binary systems containing neutron stars and plotted their two-dimensional distribution in fig. \ref{fig:ecc}.
The sample, derived from the Australia Telescope National Facility \cite{Manchester2006} catalogs and relevant literature \cite{Fortin2023}, includes 352 sources with both eccentricity and orbital period information (after excluding 2 overlapping sources). Among them, two sources have both $e\gtrsim0.95$ and $P_{\rm orb}\gtrsim 1.5~{\rm yr}$. 
The first one is PSR J2032+4127, which was identified as part of a binary system with the Be star MT91 213 in 2015 \cite{Lyne2015}. 
It has the longest known orbital period of neutron star binaries, approximately 17,000 days, and an eccentricity of up to 0.96. Extensive multi-wavelength observations were conducted when it reached periastron in 2017 \cite{Camilo2009}. It is also one of the two sources showing both radio and X-ray emission characteristics \cite{Camilo2009}.
The second source is PSR J1638-4275, a radio pulsar binary source with a long orbital period of 1940.9 days and an eccentricity of 0.955.
In conclusion, the fraction of binary systems with $e\gtrsim0.95$ and $P_{\rm orb}\gtrsim1.5~\rm yr$ is $2/352=0.57\%$. 
This result suggests that the probability of an ``RM flare'' originating from binary orbital motion is quite low. We do not rule out this possibility, but regard it as less likely compared with the CME scenario we propose later.

\REV{ It is worth noting that the DM variations from the orbital motion of a binary system might be small based on the observations of radio pulsars in binary systems. Physically, the DM variation mainly depends on the orbital geometry and the properties of the companion. For the Galactic binary system PSR B1259-63/LS 2883, its observed DM variation is about $\delta {\rm DM}\sim{\rm a~few~pc~cm^{-3}}$\cite{Johnston05}. The observed DM variation of PSR B1744-24A (Ter5A) is about $\delta {\rm DM}\sim0.1{\rm~pc~cm^{-3}}$, although it shows a secular evolution and a maximum value at the superior conjunction \cite{Li2023Nature}. For FRB 20220529, there appears to be a positive correlation between the DM and RM in the ``RM flare'' episode (see fig. \ref{fig:flks_dmks}), which seems to support this scenario. However, the uncertainties are quite large and, as pointed out above, the orbital parameters are required to be extreme in this case.}

\REV{ Finally, we should note that since orbital motion is periodic, this scenario suggests that the ``RM flare'' would occur periodically, although the amplitude, sign, or duration might change somewhat. Such a picture could be tested by continued monitoring of the source in the future. } 

\subsubsection{``RM flare'' from coronal mass ejection} 
\label{sec:cme}

Coronal mass ejections (CMEs), often accompanied by stellar flares, are catastrophic releases of magnetic energy and plasma.
CME-induced RM changes are occasionally observed when solar CMEs are between Earth and a background radio source \cite{Kooi2017}.
\REV{ CME candidates are detected in different types of stars, including pre-main sequence stars, binary stars, G-M type 
\cite{Leitzinger2022}.
Observationally, there are several methods to identify stellar CMEs, such as asymmetric emission and absorption lines in stellar spectra, coronal dimming, and variabilities of continuous X-ray absorption 
\cite{Vida2024}.
Many stellar CME candidates have been observed in dwarf M stars, with ejection mass of $10^{14}-10^{18}$ g and bulk velocity ranging from several hundred to $\sim 3000$ $\rm km~s^{-1}$ 
\cite{Houdebine1990,Vida2019}. \REVb{Due to strong convection near their surfaces, the surface magnetic field of M dwarfs could be $10^3-10^4$ G \cite{Reiners2022}.}
There are also a few CME candidates detected in dwarf K and G stars. A CME with a mass of $10^{15}~\rm g$ was detected in V471 Tauri, an eclipsing system with a hot DA white dwarf (WD) and a dK2 star, with an estimated rate of $100-500$ CMEs per day 
\cite{Bond2001}. 
The young solar analog EK Dra showed a CME with a maximum bulk velocity of $\sim 510~\rm km~s^{-1}$ and a mass of $\sim 10^{18}~\rm g$ \cite{Namekat2022}. 
Moreover, stellar CMEs have also been detected in giant stars \cite{argiroffi2019}, and binary stars \cite{moschou2017, Pandey2012}. For example, Algol, a binary system with a B8~V primary and a K2~IV secondary which underwent a period of mass transfer, showed a CME with a mass of $10^{21}-10^{22}~\rm g$, much greater than those of dwarf M stars \cite{moschou2017}. Besides, a CME with a mass of $1.2 \times 10^{21} \rm~g$ from the G giant star HR 9024 was detected within a 98-ks-long Chandra observation\cite{argiroffi2019}.}

\REV{Due to the limited sample of stellar CMEs, stellar flares are usually used as a probe of the CME event rate. The empirical relationship between flare energy in the X-ray band, $E_X$, and the CME mass $M$ is given by $\log M=0.63\log E_X-2.57$ \cite{Aarnio2012}. The energy of stellar flares ranges from $10^{23}~{\rm erg}$ for nanoflares to $10^{31}-10^{38}~{\rm erg}$ for superflares \cite{ParnellJupp00,Shibayama13,Gunther20}. The ratio between the stellar flare X-ray energy $E_X$ and the CME kinetic energy $E_{\rm CME}$ is approximately $\epsilon_X \equiv E_X/E_{\rm CME} \sim0.01$ \cite{OstenWolk15, Gunther20}. Thus, the relation between the CME mass and the CME kinetic energy is \cite{yang2023}
\begin{align}
M\simeq 2.1\times10^{16}~{\rm g}\left(\frac{E_{\rm CME}}{10^{32}~{\rm erg}}\right)^{0.63},\label{cmemass}
\end{align}
and the CME velocity is $v\simeq(2E_{\rm CME}/M)^{1/2}\simeq1000~{\rm km~s^{-1}}(E_{\rm CME}/10^{32}~{\rm erg})^{0.185}$. 
The duration $\Delta t$ of a stellar flare depends on the U-band flare energy $E_U$, i.e., $\log\Delta t=0.3\log E_U-7.5$ \cite{Pettersen1989}, and the ratio between the stellar flare U-band energy $E_U$ and the CME kinetic energy $E_{\rm CME}$ is approximately $\epsilon_U\equiv E_U/E_{\rm CME} \sim0.1$. Thus, the relation between the flare duration and the CME kinetic energy is $\Delta t\simeq63~{\rm s}(E_{\rm CME}/10^{32}~{\rm erg})^{0.3}$, which means that the time scale of more energetic flares is longer than that of less energetic flares.
On the other hand, frequent flaring occurs on stars with an outer convection zone, which includes both young and evolved stars, as well as single stars and members of multiple systems. 
For low-mass stars, strong convection near their surfaces could allow their surface magnetic field to reach $10^3-10^4~{\rm G}$ \cite{Reiners2022}. Furthermore, multi-wavelength observations of stellar flares \cite{Kowalski2024} suggest that flares and CMEs are usually frequent for stars with large magnetic loops that have sizes comparable to the stellar radius.}

We consider a two-dimensional toy model for simplicity, in which the CME could be treated as a blob that expands during its propagation. 
We assume that the blob has a mass of $M$, an expanding speed of $c_s$, a bulk-motion velocity of $v$, an initial size of $l_0$, and an initial magnetic field of $B_0$. The blob's expanding velocity depends on its sound speed, 
\begin{align}
c_s\sim\left(\frac{kT}{m_p}\right)^{1/2}\simeq10~{\rm km~s^{-1}}~T_4^{1/2}.
\end{align}
We note that the expanding speed $c_s$ may be much smaller than the bulk-motion velocity 
$v\sim(10^2-10^3)~{\rm km~s^{-1}}$ for a typical CME.
\REV{ Physically, the CME ejection speed 
depends on gravitational binding and magnetic field binding on the surface of stars, leading to
\begin{align}
v\sim\max\left(v_{\rm esc},v_B\right), 
\end{align}
where $v_{\rm esc}$ is the escape velocity with
\begin{align}
v_{\rm esc}=\sqrt{\frac{2GM_c}{R_c}}\simeq620~{\rm km~s^{-1}}M_{c,\odot}^{1/2}R_{c,\odot}^{-1/2}  
\end{align}
for a star with a mass of $M_c=M_{c,\odot} M_\odot$ and a radius of $R_c=R_{c,\odot}R_\odot$. The term
$v_B$ is the velocity depending on the balance between the magnetic pressure and the ram pressure, $\rho v_B^2\sim B^2/8\pi$, i.e.,
\begin{align}
v_B\simeq\frac{B}{\sqrt{8\pi\rho}}\simeq630~{\rm km~s^{-1}}B_3\rho_{-11}^{-1/2},
\end{align}
where $B$ and $\rho$ are the magnetic field strength and the gas density at the star's surface, respectively.}

At the distance $r=vt$ from the companion, the blob size is estimated as
\begin{align}
l\simeq l_0+c_s t\simeq\frac{c_s}{v}r.
\end{align}
Since $c_s t\sim8\times10^{10}~{\rm cm}~t_{\rm day}$ is of the order of the stellar radius $R_\ast$ for the observing timescale $t\sim\text{a few days}$, one always has $c_st\sim R_\ast> l_0$.
Assuming that the electron number density $n_e$ and the magnetic field strength $B$ are uniform within the CME plasma, one has
\begin{align}
&n_e(r)\simeq\frac{M}{\mu_m m_p(4\pi/3)l^3}\simeq\frac{3M}{4\pi\mu_m m_p r^3}\left(\frac{v}{c_s}\right)^3,\\
&B(r)\simeq B_0\left(\frac{l}{l_0}\right)^{-2}\simeq B_0\left(\frac{v}{c_s}\right)^{2}\left(\frac{r}{l_0}\right)^{-2},
\end{align}
where $\mu_m=1.2$ is the mean molecular weight for a solar composition.

We define the angle between the blob velocity and the line of sight (LOS) as $\theta$ and consider three scenarios. The schematic configurations are presented in fig. \ref{fig:model}. Case I: The blob can enter and eventually exit the LOS. Case II: The moving direction is close to the LOS, with $\sin\theta<c_s/v$, and the blob can enter but never exit the LOS eventually. Case III: The blob never enters the LOS, and the moving direction is far away from the LOS, with $\sin\theta>c_s/v$. In Case III, the RM of FRBs remains unchanged. We will discuss the first two scenarios in detail in the following discussion.

Case I requires $v\sin\theta>c_s$. Meanwhile, at the distance $r_c=v t_c$ from the star, the blob center reaches the LOS, corresponding to the peak time of the observed ``RM flare''. 
We define the transverse distance from the blob center to the LOS as $x$, as shown in fig. \ref{fig:model}. One then has
\begin{align}
x(t)=v|t-t_c|\sin\theta.
\end{align}
We define the RM rising time as $t_i$. When the blob enters the LOS, one has $x(t_i)=l(t_i)=c_st_i$ with $t_i<t_c$, leading to
\begin{align}
t_i=\frac{vt_c\sin\theta}{v\sin\theta+c_s}.
\end{align}
We define the time when RM decreases to the pre-flare value as $t_f$. Then one has $x(t_f)=l(t_f)=c_st_f$ with $t_f>t_c$, leading to
\begin{align}
t_f=\frac{vt_c\sin\theta}{v\sin\theta-c_s}.
\end{align}
Therefore, the total duration of the RM flare is
\begin{align}
\Delta t=t_f-t_i=\frac{2(v/c_s)\sin\theta}{(v/c_s)^2\sin^2\theta-1}t_c.
\end{align}
The observed variation of RM depends on the properties of the CME along the line of sight,
\begin{align}
{\rm RM}(t)\sim\frac{e^3}{2\pi m_e^2c^4}n_eBd,
\end{align}
where $d$ is the LOS scale of the blob, which can be estimated as
\begin{align}
d(t)=2\sqrt{l(t)^2-x(t)^2},~~~\text{with}~x(t)<l(t).
\end{align}
Therefore, the RM evolution during the flaring state satisfies
\begin{align}
{\rm RM}(t)
&\sim \frac{e^3}{2\pi m_e^2c^4}\frac{3MB_0l_0^2}{4\pi\mu_m m_p(c_st)^5}d(t)\nonumber\\
&\sim5.5\times10^3~{\rm rad~m^{-2}}M_{17}B_{0,4}l_{0,-1,\odot}^2c_{s,6}^{-4}t_{10{\rm day}}^{-4}\nonumber\\
&\times\left[1-(v^2\sin^2\theta/c_s^2)(1-t_c/t)^2\right]^{1/2}\label{rmcasei}
\end{align}
for $x(t)<l(t)$, where the convention $Q_x = Q/10^x$ is adopted in cgs units, $t_{10{\rm day}}=t/10~{\rm day}$ and $l_{0,-1,\odot}=l_0/0.1R_\odot$. Otherwise, ${\rm RM}(t)\sim0$. The RM evolution is shown in fig. \ref{result}.

Next, we consider that the blob moving direction satisfies $v\sin\theta<c_s$ and the transverse distance from the blob center to the LOS as $x_\ast$ at $t_\ast$, corresponding to Case II in fig. \ref{fig:model}. At time $t$, the transverse distance from the blob center to the LOS satisfies
\begin{align}
x(t)=x_\ast+v(t-t_\ast)\sin\theta.
\end{align}
For the late-time evolution shown in the observation of FRB 20220529, one approximately has $x(t)\simeq vt\sin\theta$.
The line-of-sight scale of the blob is estimated as 
\begin{align}
d(t)=2\sqrt{l(t)^2-x(t)^2}\simeq2t\sqrt{c_s^2-v^2\sin^2\theta}.
\end{align}
Therefore, the RM evolution approximately satisfies
\begin{align}
{\rm RM}(t)\sim t^{-4}. \label{rmcaseii}
\end{align}
Such a scenario predicts that the RM variation in the high state follows a power-law decay.

Based on the above discussion, one can estimate the typical RM value from the CME from the companion star.
At the observed timescale, for both Case I and Case II, one has
\begin{align}
{\rm RM}(t)
&\sim \frac{e^3}{2\pi m_e^2c^4}\frac{3f_{\rm geo}MB_0l_0^2}{2\pi\mu_m m_p}(c_st)^{-4}\nonumber\\
&\sim5.5\times10^3~{\rm rad~m^{-2}}f_{\rm geo}M_{17}B_{0,4}l_{0,-1,\odot}^2c_{s,6}^{-4}t_{10{\rm day}}^{-4},\label{cmerm}
\end{align}
where $f_{\rm geo}$ is a geometric factor depending on the geometry of Case I and Case II, and a factor of 2 is involved from $d\sim 2 c_s t$. For Case I, one has $f_{\rm geo}\sim [1-(v^2\sin^2\theta/c_s^2)(1-t_c/t)^2]^{1/2}$. For Case II, one has $f_{\rm geo}\sim1$. 
\REV{ Based on Eq. (\ref{cmerm}), there are two sets of parameters consistent with the CME data that can generate the observed ``RM flare'': 1) A CME with $M\sim10^{17}~{\rm g}$, $B_0\sim10^4~{\rm G}$ and $l_0\sim10^{-1}R_\odot$ from a magnetized M dwarf.
2) A CME with $M\sim10^{21}~{\rm g}$, $B_0\sim1~{\rm G}$ and $l_0\sim10^{-1}R_\odot$ generated in a giant star \cite{argiroffi2019}, or a binary system with mass transfer, which is similar to the case of the Algol binary system where a CME mass could reach $10^{21}-10^{22}~{\rm erg}$ 
\cite{moschou2017}. }

\REV{ The DM contributed by the CME could be estimated by
\begin{align}
{\rm DM}(t)\sim n_ed(t)\simeq0.01~{\rm pc~cm^{-3}}M_{17}c_{s,6}^{-2}t_{10{\rm day}}^{-2}. 
\end{align}
If the ``RM flare'' is caused by a CME with $M\sim10^{17}~{\rm g}$ from an M dwarf, the DM contribution is estimated to be ${\rm DM}\sim 0.01~{\rm pc~cm^{-3}}$, which is much smaller than the observed DM scatter of $\delta {\rm DM}\sim{\rm a ~few~pc~cm^{-3}}$, as shown in Table \ref{table:RM}. If the ``RM flare'' is caused by a CME with $M\sim10^{21}~{\rm g}$ from an Algol-like binary system \cite{moschou2017} or a gaint star, the DM contribution is estimated to be ${\rm DM}\sim 10~{\rm pc~cm^{-3}}$, which is of the order of the observed DM scatter, $\rm 3.5~pc~cm^{-3}$. Besides, as shown in fig. \ref{fig:flks_dmks}, the DM seems to have a positive correlation with the RM during the ``RM flare'', which is also consistent with the CME picture.}

\REV{We then estimate the occurrence rate of the ``RM flare'', $\mathcal{R}_{\rm RM}$, which is related to the occurrence rate of CMEs, $\mathcal{R}_{\rm CME}$, and the geometric configuration of CMEs.
As pointed out above, the observed ``RM flare'' suggests that the stellar flare is relatively large, with a CME mass of $M\gtrsim 10^{17}~{\rm g}$, leading to $E_{\rm CME}\gtrsim 10^{34}~{\rm erg}$ based on Eq. (\ref{cmemass}).
It is noteworthy that the frequency of superflares with $\sim10^{34}~{\rm erg}$ in the dM4.5e star YZ CMi is about $\mathcal{R}_{\rm CME}\sim1$ per month 
\cite{Lacy1976,Kowalski2010}
, which is comparable to those of the most active G-dwarfs and Sun-like stars, once in 10–100 days \cite{Shibayama13}. Since the ``RM flare'' can only be observed when the CME blob passes through the LOS, as shown in fig. \ref{fig:model}, the occurrence rate of the ``RM flare'' (which is comparable to or even larger than that reported by this work) satisfies $\mathcal{R}_{\rm RM}\ll\mathcal{R}_{\rm CME}\sim 1$ per month. Besides, due to the randomness of the magnetic field in the CME, we predict that the ``RM flare'' of FRB 20220529 could be either positive or negative.
Furthermore, the solid angle of the CME blob related to the flaring star is about $\Omega_{\rm CME}\sim\pi(c_s/v)^2$. In order to generate an ``RM flare'' comparable to or even larger than the ``RM flare'' reported in this work, the CME must pass through the LOS in a region close to the FRB source, e.g., $r_c\lesssim 10~{\rm AU}$. Thus, the occurrence rate of the ``RM flare'' in M dwarfs is roughly of the order of
$\mathcal{R}_{\rm RM}\sim(c_s/4\pi v)\mathcal{R}_{\rm CME}\sim(0.01-0.1)~{\rm yr^{-1}}$  
for $c_s\sim10~{\rm km~s^{-1}}$ and $v\sim(100-1000)~{\rm km~s^{-1}}$. 
Above discussion assumes the companion star of the FRB is a M dwarf. 
In other cases, such as giant stars or binaries, the event rate of CMEs with masses larger than $10^{21}~\rm g$ is poorly known. A rough estimation based on the detection of a CME with a mass of $1.2 \times 10^{21}~\rm g$ in a 98-ks Chandra observation of the G giant star HR 9024 suggests a rate of about 1 per day \cite{argiroffi2019} for giant stars. This might increase the rate to $\mathcal{R}_{\rm RM}\sim(0.3-3)~{\rm yr^{-1}}$. }

\REV{Observationally, we detected one ``RM flare'' in 2.2 years. Assuming a \REVb{Poisson} distribution, we obtain an event rate of $0.45^{+1.04}_{-0.38}$ per year per FRB source and a 3 $\sigma$ range of $6 \times 10^{-4}- 4.0$ per year per FRB source, which is roughly consistent with the theoretical discussion above. To investigate the chance of having a ``RM flare'' occurring but unnoticed during our monitoring, we conducted a simulation. We randomly selected a time between 2022-06-23 and 2024-09-04 as the starting time $\tau$ of the ``RM flare'', and assumed that the ``RM flare'' had a similar amplitude and duration as we observed. Here, we use the best-fit CME Case I model to present the increasing part of the ``RM flare''. Our actual observation cadence is assumed to ``observe'' the simulated RM evolution, and the maximum RM is recorded. We simulated this process 10,000 times. The chance of having a maximum RM smaller than 3$\sigma$ of the low RM episode is 15\%. It indicates the possibility of having missed an ``RM flare'' similar to what we observed is less than 15\%. 
}


\REV{ Since the CME is close to the FRB source in the above picture, one may be concerned about whether the nonlinear effect of strong waves may render the usual RM and DM expressions no longer applicable, because the strong electric field in the waves may accelerate electrons to relativistic speed. 
In order to describe the nonlinear effect of the strong wave, one generally defines the strength parameter as \cite{yangzhang20}
\begin{align}
a=\frac{eE}{m_ec\omega}=\frac{eS_\nu^{1/2}d}{\pi^{1/2}m_ec^{3/2}\nu^{1/2}r}=1.2\left(\frac{S_\nu}{{\rm Jy}}\right)^{1/2}\left(\frac{\nu}{{\rm GHz}}\right)^{-1/2}\left(\frac{d}{{\rm Gpc}}\right)\left(\frac{r}{{\rm AU}}\right)^{-1},
\end{align}
where $S_\nu$ is the peak flux of an FRB at frequency $\nu$, $E=(4\pi\nu S_\nu/c)^{1/2}(d/r)$ is the electric intensity at radius $r$, and $d$ is the FRB source distance. FRB 20220529 has a distance of 880 Mpc, and the bursts’ typical flux is $\sim13$ mJy, as presented in fig. \ref{fig:flux_wid_FAST}.  
The CME has a typical velocity of $v\sim1000~{\rm km~s^{-1}}$ and passes through the LOS at $t_c\sim10~{\rm day}$ after a stellar flare (see the following fitting result). Thus, the CME blob distance from the companion is about 
$r_c=vt_c\sim5.8~{\rm AU}$. The distance between the CME blob and the FRB source is of the order of $r_c$, unless the CME blob velocity points directly to the FRB source. Therefore, one always has $a<1$ in most cases, implying that the nonlinear effect of strong waves is not significant in the CME scenario.}

\REV{ In order to keep the radio waves transparent, the plasma frequency of the CME blob should be smaller than $\sim$GHz, which requires that the electron density is smaller than $10^{10}~{\rm cm^{-3}}$. We estimate the electron number density as $n_e\sim2\times10^7~{\rm cm^{-3}}\ll 10^{10}~{\rm cm^{-3}}$ when the blob crosses the LOS. Thus, The FRBs can propagate through the blob without being absorbed. 
We further check whether free-free absorption and induced Compton scattering lead to substantial attenuation of the FRB waves in light of the electron density requirements on the CME blob. When the CME passes through the LOS, it has a typical scale of $l\sim c_s t_c\sim8.6\times10^{11}~{\rm cm}$, a temperature of $T\sim10^4~{\rm K}$, and an electron density of $n_e\sim 3M/(4\pi \mu_m m_pl^3)\sim1.8\times10^4~{\rm cm^{-3}}$ for $M\sim10^{17}~{\rm g}$. Thus, the optical depth of free-free absorption is
\begin{align}    
\tau_{\rm ff}&=\alpha_{\rm ff}L\simeq0.018T^{-3/2}Z^2n_en_i\nu^{-2}\bar g_{\rm ff}l\nonumber\\
&=5.3\times10^{-6}\left(\frac{T}{10^4~{\rm K}}\right)^{-4}\left(\frac{M}{10^{17}~{\rm g}}\right)^{2}\left(\frac{t_c}{10~{\rm day}}\right)^{-5}\left(\frac{\nu}{1~{\rm GHz}}\right)^{-2},
\end{align}
where $\bar{g}_{\rm ff}\sim1$ is the Gaunt factor, and we assume that $n_e=n_i$ and $Z=1$ for a fully ionized hydrogen-dominated composition. Since $\tau_{\rm ff}\ll1$, free-free absorption from the CME could be ignored. The optical depth for induced scattering on short, bright radio pulses passing through a plasma screen at large distances can be estimated as 
\cite{Lyubarsky2008,Lyubarsky2016}
\begin{align}
\tau_{\rm ind}&\simeq\frac{3\sigma_{\rm T}}{8\pi}\frac{n_eS_\nu}{m_e\nu^2}\left(\frac{D}{r}\right)^{2}c\Delta t\nonumber\\
&\simeq2.0\times10^{-3}\left(\frac{S_\nu}{10~{\rm mJy}}\right)\left(\frac{t_c}{10~{\rm day}}\right)^{-3}\left(\frac{r}{10~{\rm AU}}\right)^{-2}\left(\frac{D}{1~{\rm Gpc}}\right)^{2}\left(\frac{\Delta t}{1~{\rm ms}}\right)\left(\frac{\nu}{1~{\rm GHz}}\right)^{-2},
\end{align}
where $D$ is the distance of the FRB source and $\Delta t$ is the typical FRB duration.
Again, since $\tau_{\rm ind}\ll1$, the effect of induced scattering from the CME could be ignored.}

\subsubsection{Model fitting of RM evolution} \label{sec:model_fit}

\REV{We performed a Bayesian inference of the RM evolution using standard MCMC techniques for both phenomenological and physical models to explore its origin. The emcee \cite{emcee}
software package was applied. Only bursts in the ``RM flare'' episode (here defined as the interval with RMs larger than 1 $\sigma$ of the low RM episode, $\rm 60290<MJD<60310$) are taken into account. The background estimated from the low RM state data, $17\pm 101~\rm rad~m^{-2}$, is subtracted from the data. The scattering of the low RM state, $101 \rm\ rad~m^{-2}$, is taken into account by the root-sum-square method. }

The power law, CME Case I (Eq. (\ref{rmcasei})), and CME Case II model (Eq. (\ref{rmcaseii})) are explored. To fit the observed data, we correct the observational time with a zero point $t_0$ for each model. 
For power law, 10 independent chains of 500,000 samples are run, and the first 250,000 samples are discarded. 
The best fit is $t_0=-56.1^{+15.6}_{-1.4}$ days, and the index $\Gamma=6.67^{+0.03}_{-1.79}$. 
For CME Case I, the initial time $t_0$ is not well constrained. We thus fit the model with different assumed $t_0$ values. For each run, 30 independent chains of 500,000 samples are operated, and the first 250,000 samples are discarded. For initial times $t_0$ of MJDs = 60272, 60265, 60242, 60192 (-20, -27, -50, -100 days before the first detection of the abrupt RM variance), the times when the blob center reaches the LOS are $t_{\rm c}= 25.6^{+1.1}_{-0.3}, 32.4^{+0.3}_{-0.6}, 53.9^{+0.5}_{-0.6}, 102.3^{+0.7}_{-2.7}$ days, and the best-fitting $v^2 {\rm sin}^2 \theta/c_{\rm s}^2 = 10.8^{+0.9}_{-3.4}, 16.8^{+4.0}_{-2.0}, 55.5^{+5.4}_{-18.3}, 146.6^{+21.6}_{-70.9}$, respectively. 
For the CME case II model, the best-fit zero time point is $-20.6^{+0.6}_{-0.6}$ days before MJD 60292, and the logarithmic normalization is $8.72^{+0.04}_{-0.05}$. 
 
\REV{Fig. \ref{fig:rmflare_caseIaII}A shows the best-fit RM evolution curves with the best fitting parameters, and Table \ref{table: model_nobkg} gives the best-fitting parameters of each model, as well as the 
statistical criteria to distinguish the models. 
 $-2~{\rm ln}~L$ is equivalent to $\chi^2$, which is expected to be similar to the degree of freedom (dof) for a proper fit. It is obvious that power law and CME Case II models have $-2~{\rm ln}~L/\mathrm{dof}=44.5/11, 106.8/12$, respectively, indicating the models are inconsistent with the data. Akaike information criterion (AIC, defined as $2k-2{\rm ln}L$) and the Bayesian information criterion (BIC, defined as ${\rm BIC}= -2 {\rm ln} L + k{\rm ln}N$) are also calculated for model comparison, in which $L$ is the maximum likelihood for each model, $k$ is the number of parameters of each model and $N$ is the number of data points. It turns out that although CME Case I model needs more parameters, it gives much smaller AIC and BIC than power law as well as CME Case II. It describes the RM increase we observed well and is better than the other two models. }

\clearpage

\begin{figure*}
\centering
\includegraphics[width=0.65\textwidth]{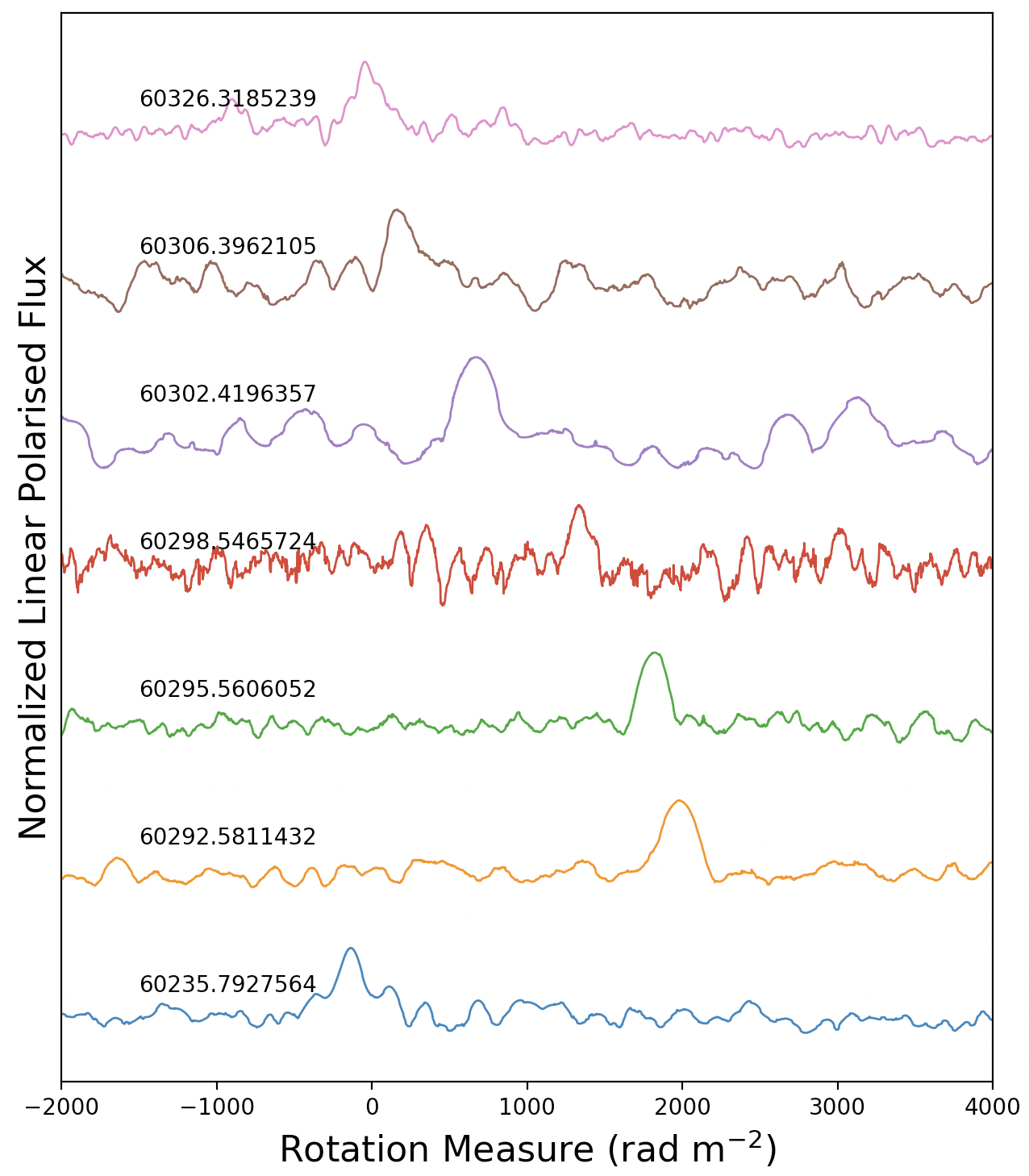}
\caption{\textbf{RM Variability for bursts of FRB 20220529.} The normalized linear polarized flux is shown as a function of RM for seven bursts on different days (sorted by MJD). The RM of each burst is determined by its maximum linear polarization value. 
}
\label{fig:RM_spectrum}
\end{figure*}

\begin{figure*}
\centering
\begin{tabular}{cccc}
\includegraphics[width=3.75cm]{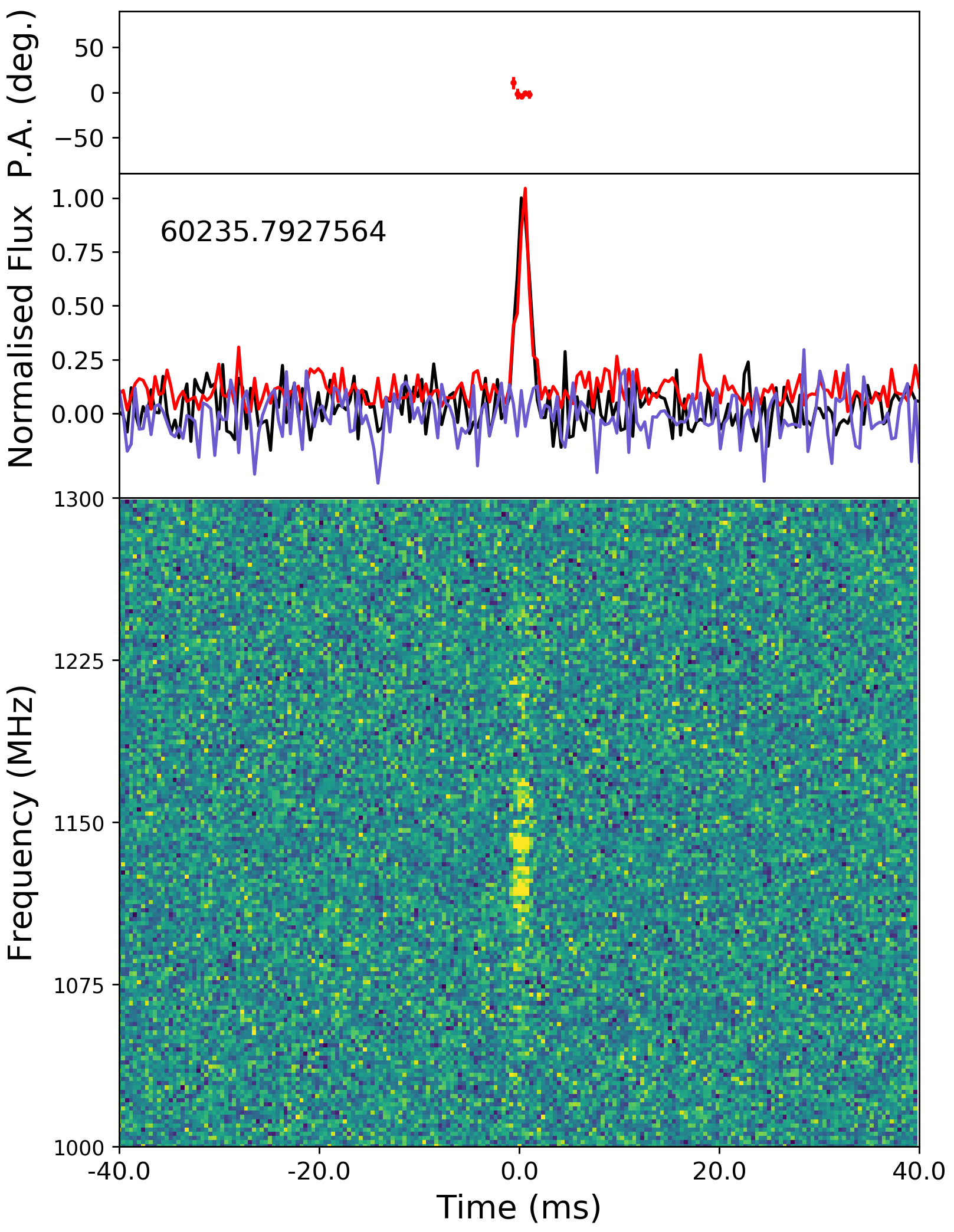} &
\includegraphics[width=3.75cm]{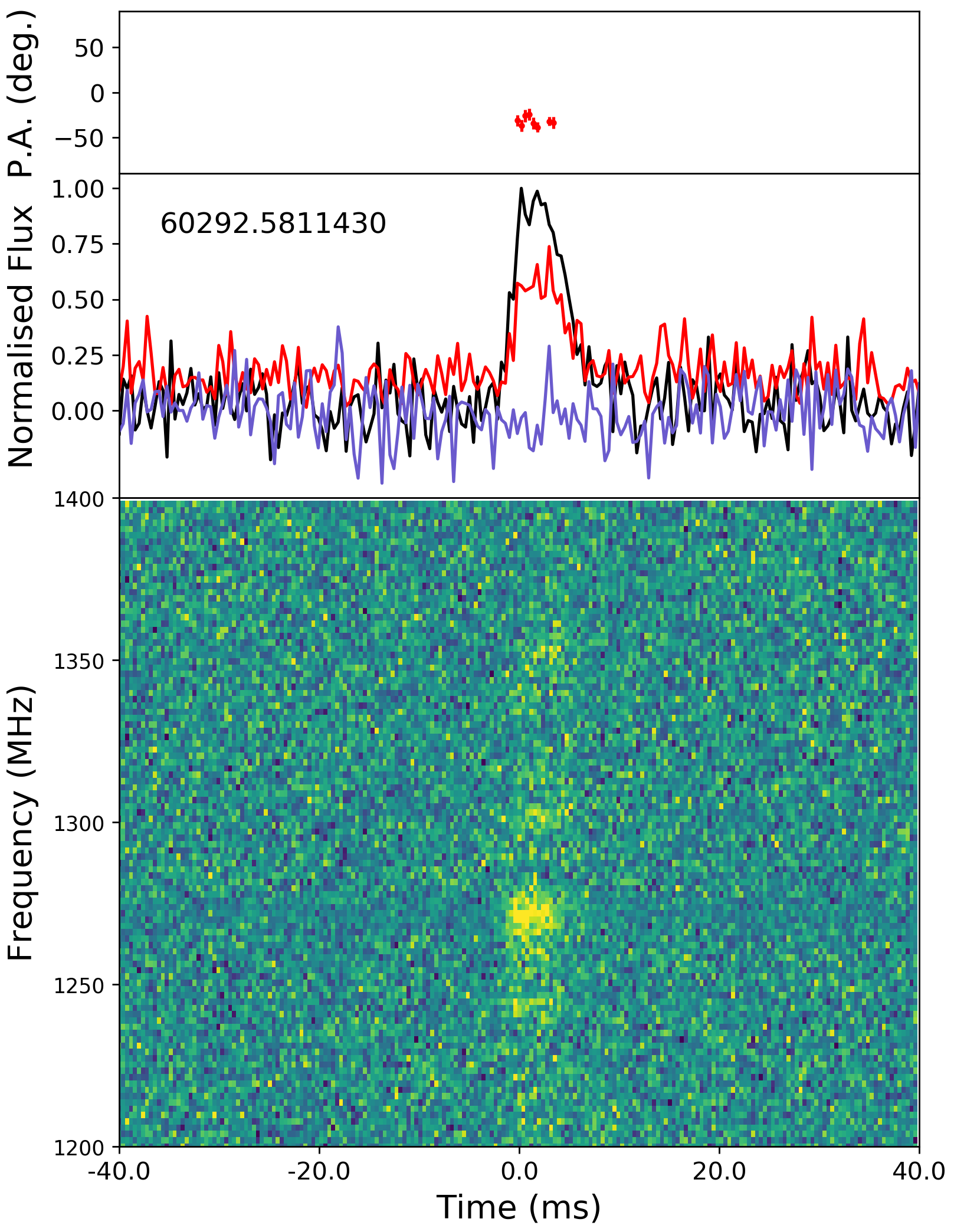}  &
\includegraphics[width=3.75cm]{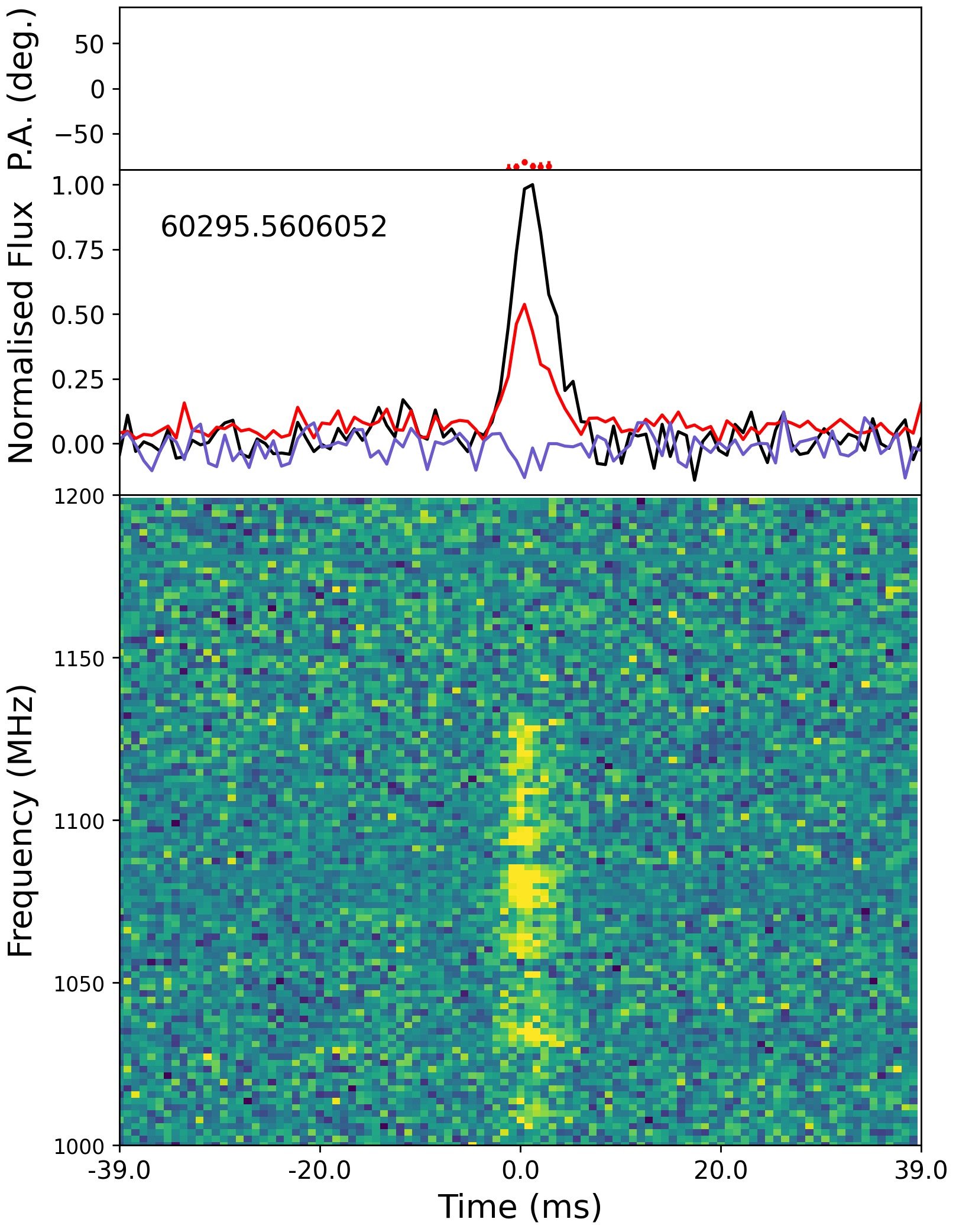} &
\includegraphics[width=3.75cm]{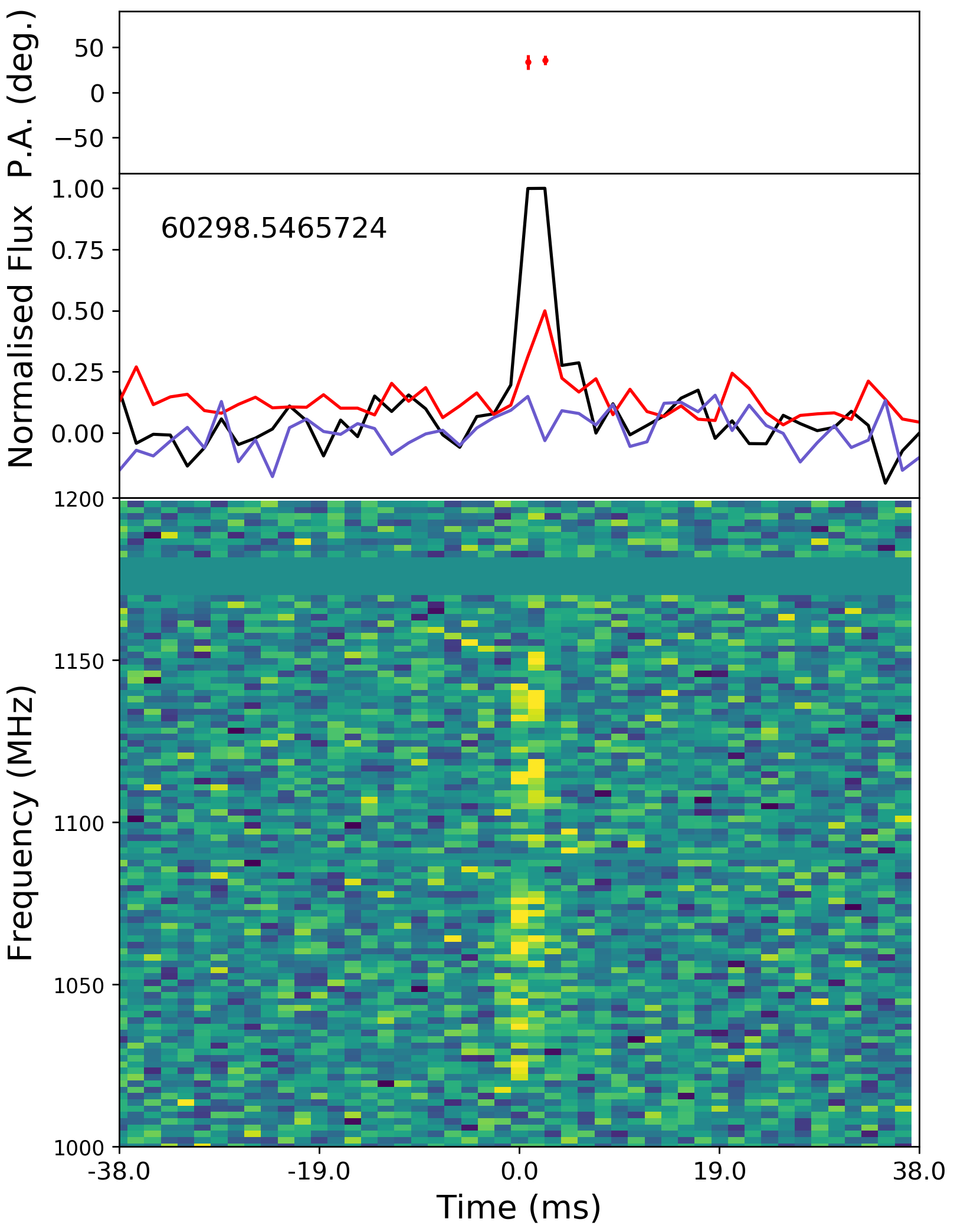}  \\
\includegraphics[width=3.75cm]{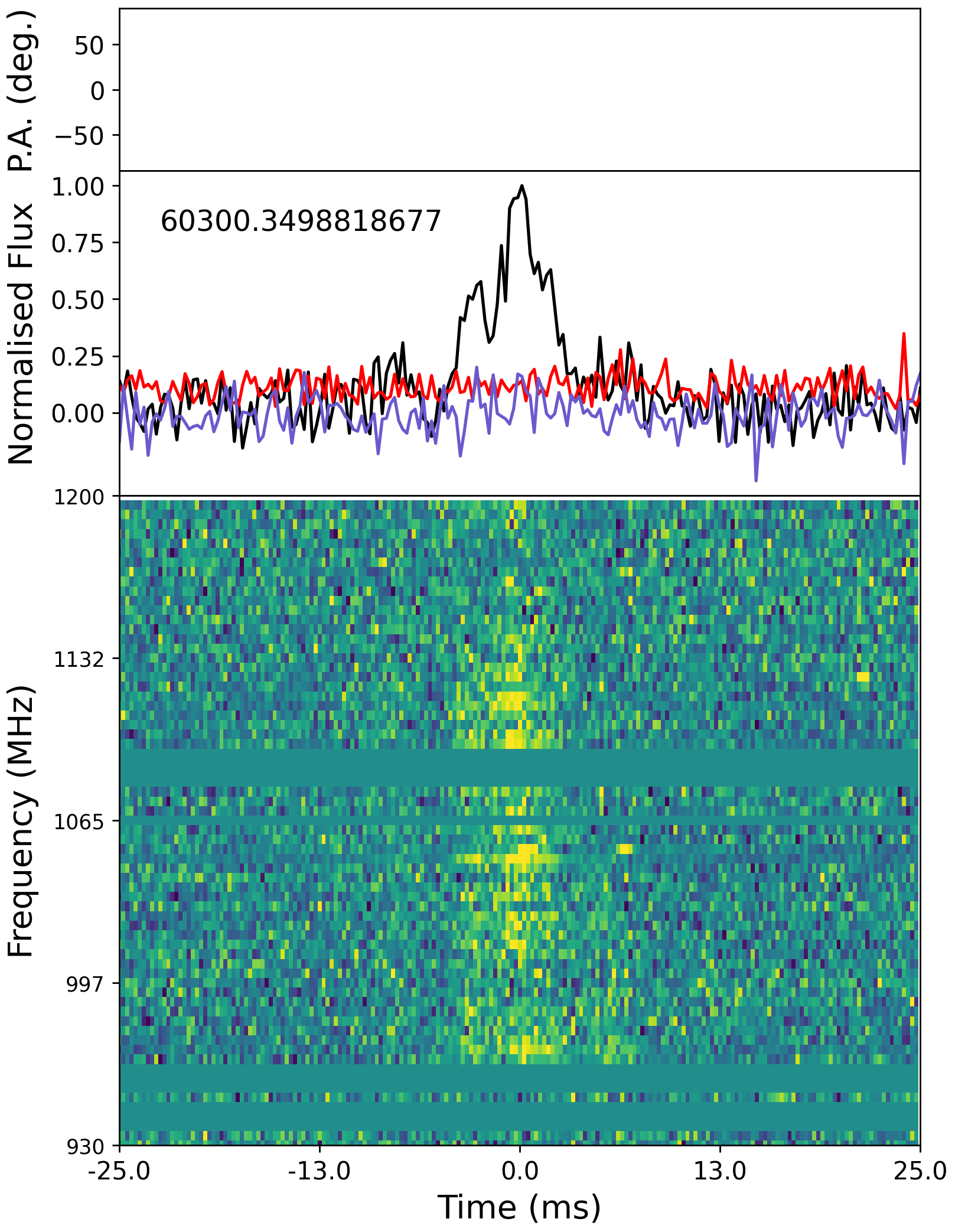}  &
\includegraphics[width=3.75cm]{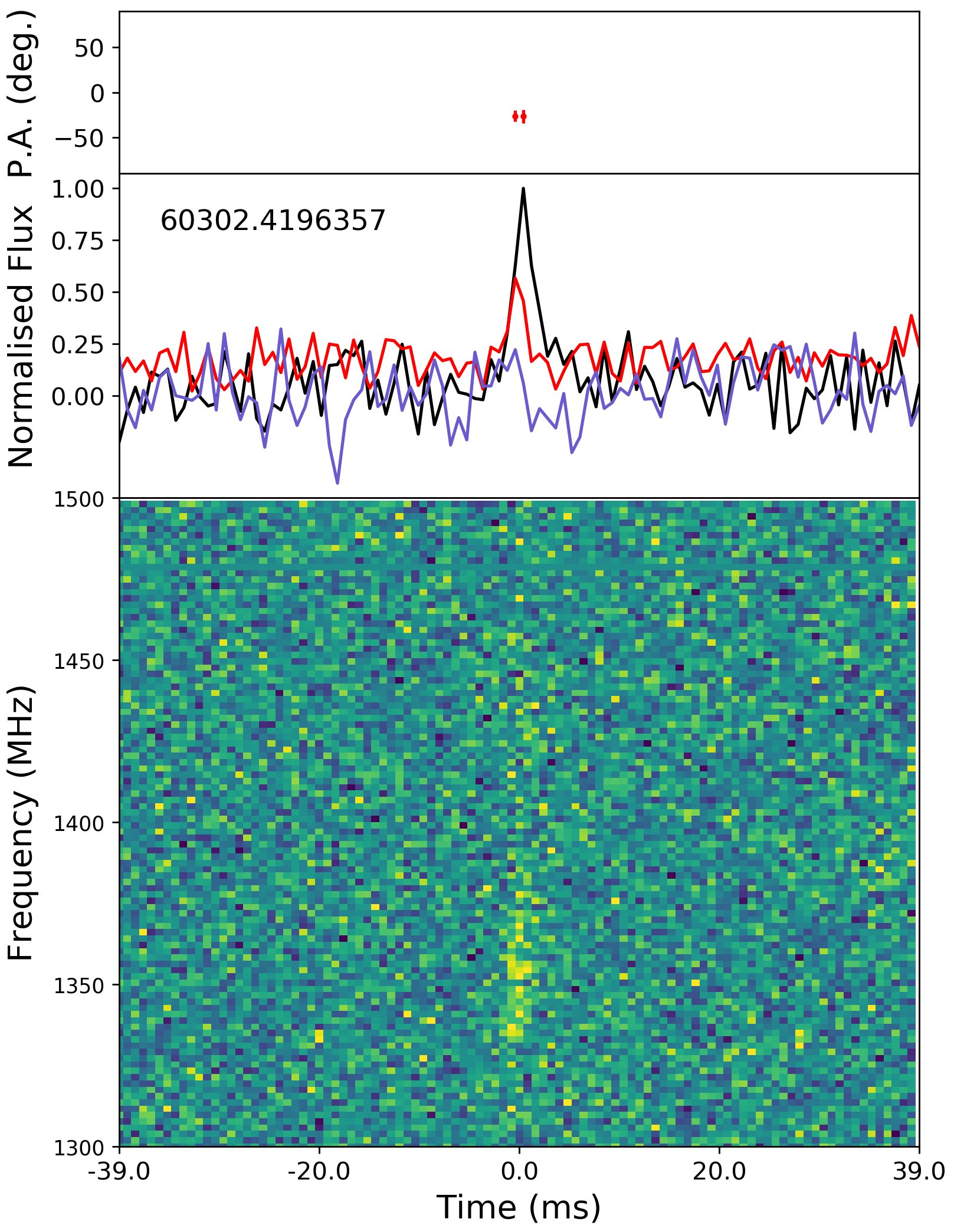}  &
\includegraphics[width=3.75cm]{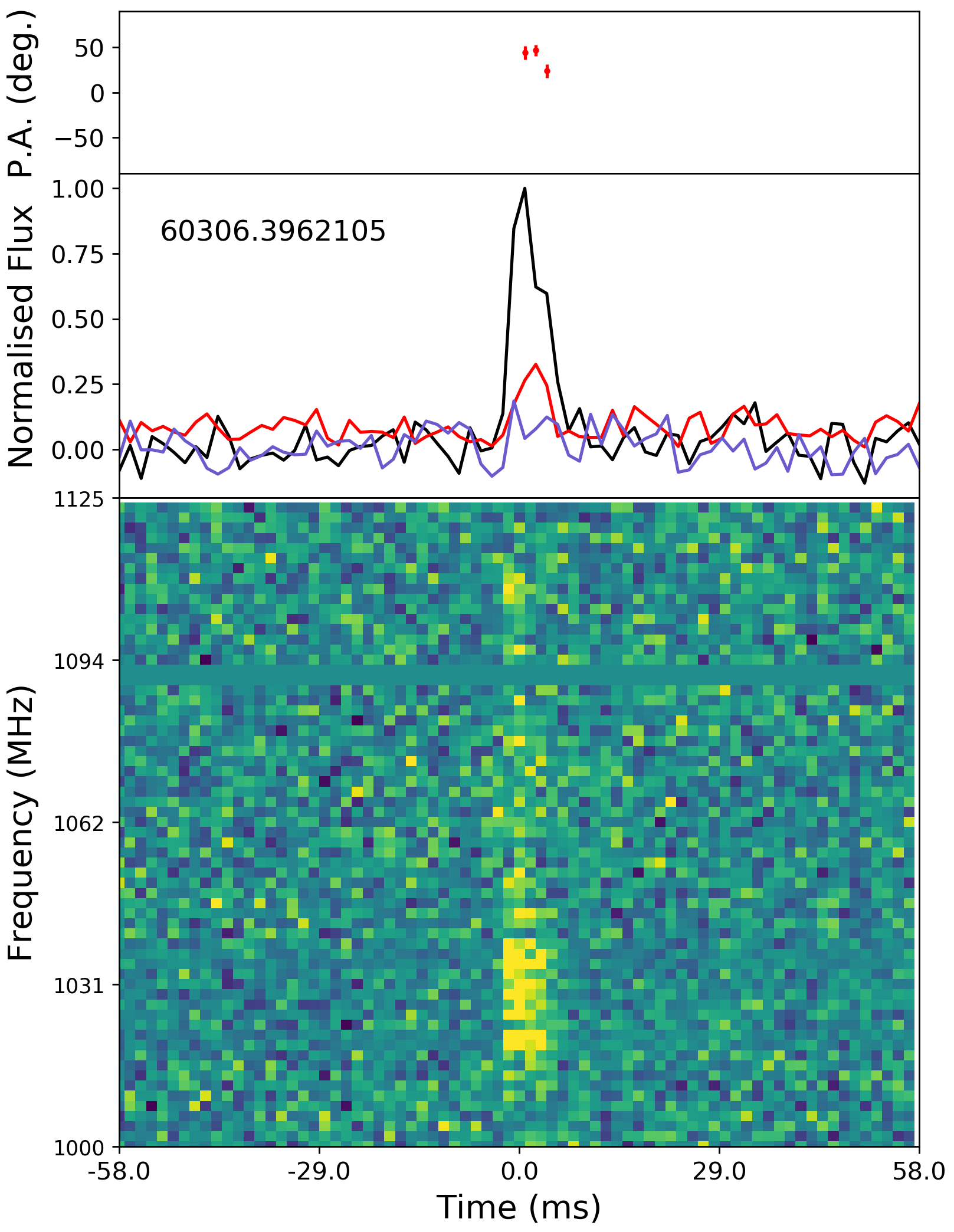}  &
\includegraphics[width=3.75cm]{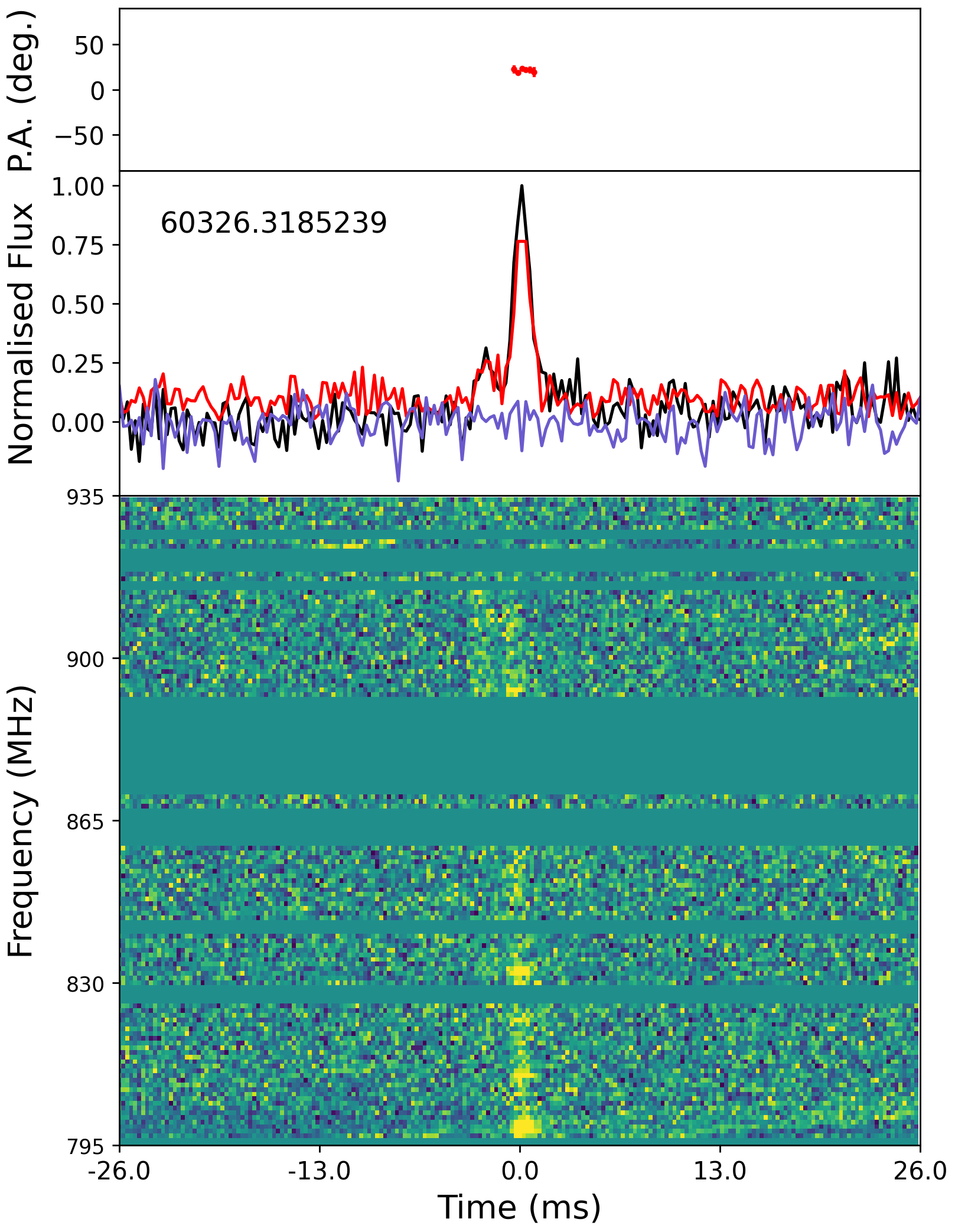}
\\
\end{tabular}
\caption{\textbf{Polarization profiles of \REV{eight} bursts on different days from FRB 20220529, ordered by MJD.} In each subplot, the upper panel displays the position angle of linear polarization at the center frequency. The middle panel shows the polarization pulse profile, where black, red, and blue curves denote total intensity, linear polarization, and circular polarization, respectively. The lower panel presents the dynamic spectra for the total intensity of all pulses, with a frequency resolution of $0.97\,{\rm MHz/channel}$ and a time resolution of 393 or $786\,{\rm \mu s/bin}$. \REV{Note that the burst at 60300.3530282 was detected with one of the highest S/N ratios during the ``RM flare'', although no RM measurement was obtained.}}
\label{fig:burst_pol}
\end{figure*}

\begin{figure*}
\centering
\includegraphics[width=0.75\textwidth]{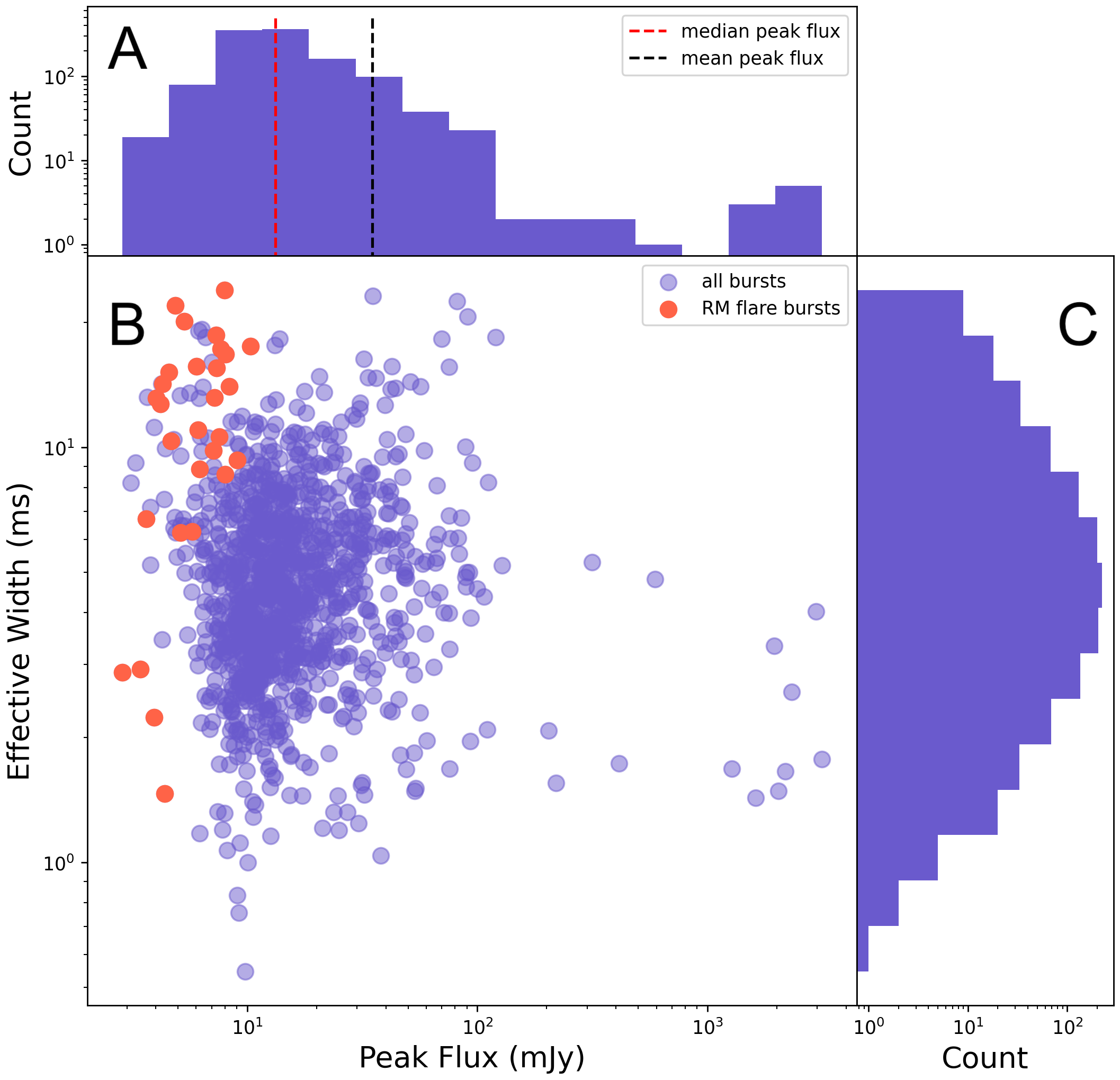}
\caption{\REV{\textbf{Peak flux density and effective width distribution for bursts of FRB 20220529 detected using the FAST telescope.} (A) Histogram showing the peak flux distribution, with lines representing the median peak flux ($\sim$ 13\,mJy) and mean peak flux ($\sim$ 35\,mJy). (B) Two-dimensional distribution of peak flux versus pulse width. (C) Histogram displaying the effective width distribution.}}
\label{fig:flux_wid_FAST}
\end{figure*}

\begin{figure*} 
\centering
\includegraphics[width=0.7\textwidth]{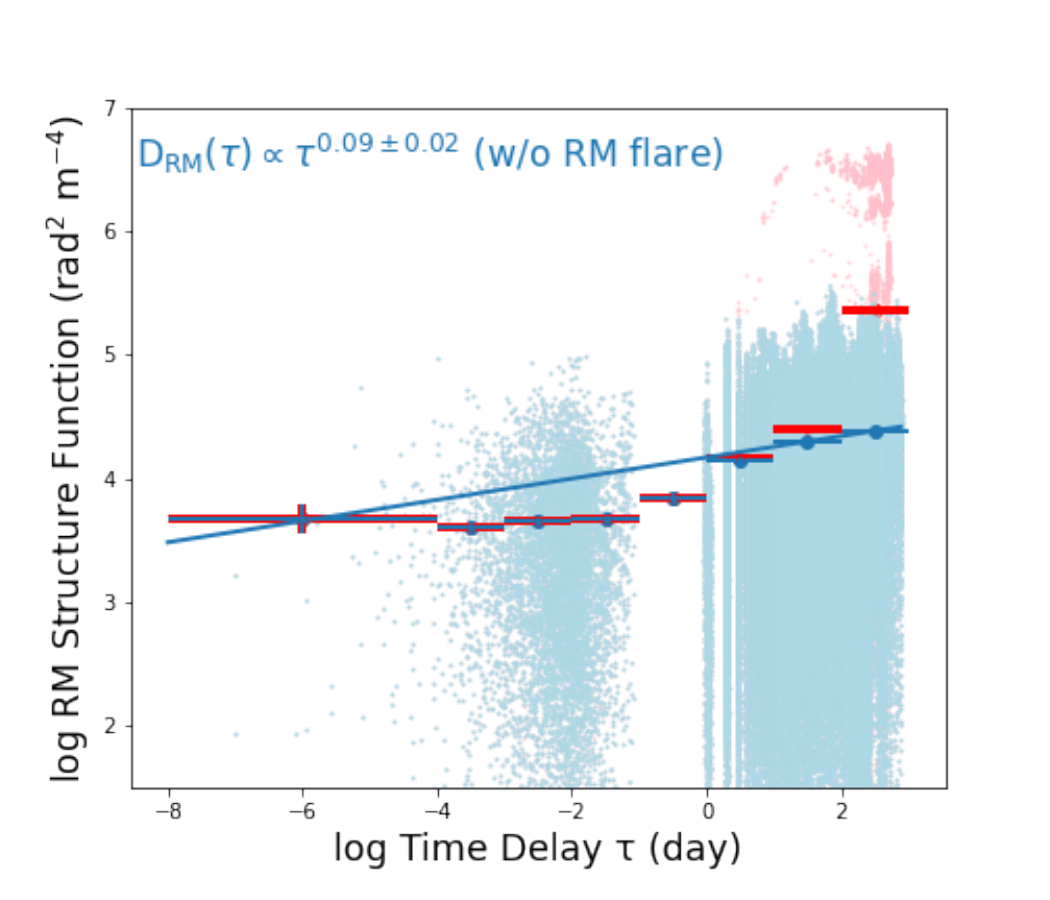}
\caption{\textbf{The structure function of RM}. \REV{The blue and red points indicate the SF excluding and including ``RM flare'', respectively.} The best-fit result outside ``RM flare'' is $D_{\rm RM}(\tau) \propto \tau^{0.09\pm0.02}$. It is consistent with the large scattering within a day and insignificant variability in long time delays. The index is consistent with those of FRB~20121102A and FRB~20180916B, indicating a supersonic turbulence in star-forming regions or shocked stellar winds in massive stars.}
\label{fig:rm_sf}
\end{figure*}

\begin{figure*}
\centering
\includegraphics[width=0.7\textwidth]{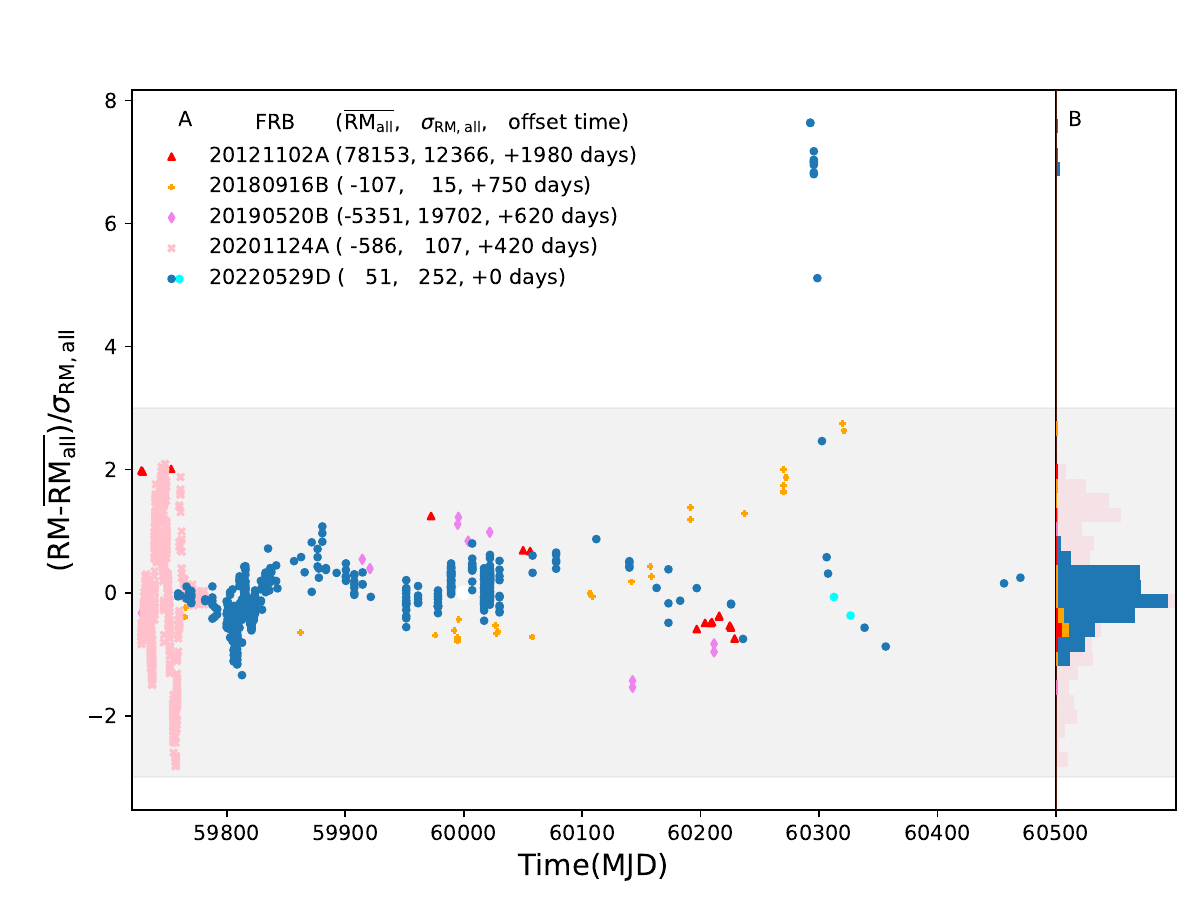}
\caption{\textbf{Significance of RMs for different repeating FRBs.} The significance $({\rm RM}-{\overline{\rm RM_{all}}})/\sigma_{\rm RM,all}$ of the RMs detected in FRB 20220529 (blue and cyan dots), FRB 20121102 (red triangles)\cite{Hilmarsson2021}, FRB 20180916B (orange pluses)\cite{Mckinven2023} , FRB 20190520B (violet diamonds)\cite{annathomas2023} and FRB 20201124A (pink crosses)\cite{xuheng2022} is presented. 
The $\sigma_{\rm RM,all}$ is the standard deviation of the RM for each FRB. The gray region covers three standard deviations.
}
\label{fig:rmflare_snr}
\end{figure*}

\begin{figure*}
\centering
\includegraphics[width = 0.49\linewidth, trim = 0 0 0 0, clip]{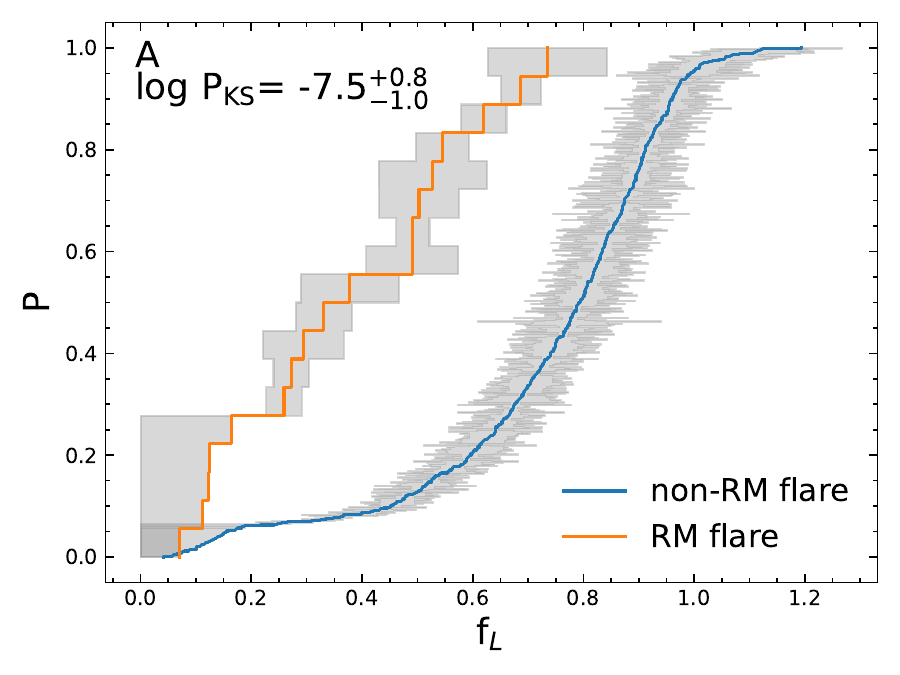}
\includegraphics[width = 0.49\linewidth, trim = 0 0 0 0, clip]{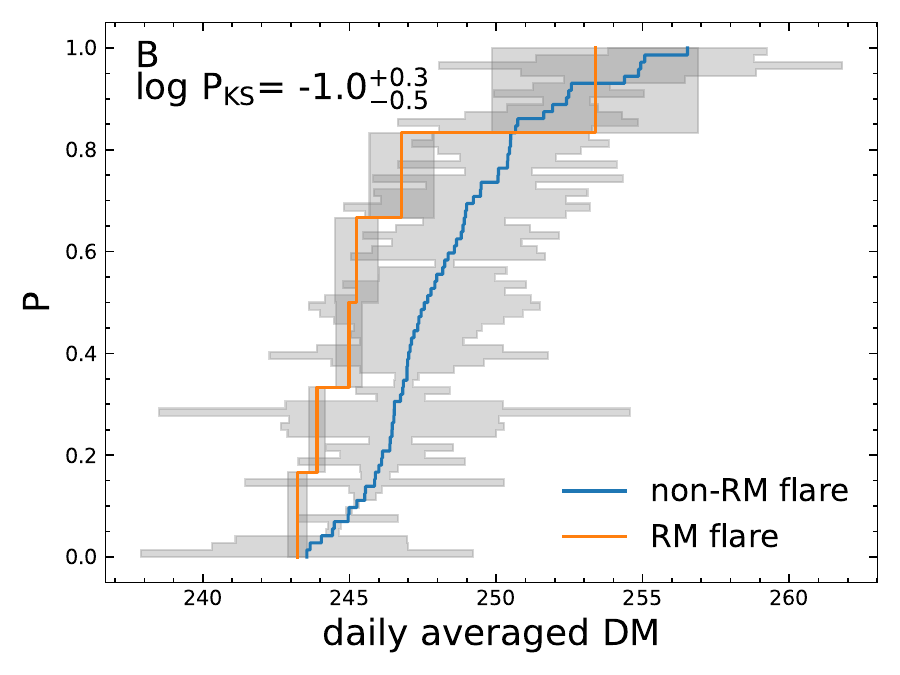}
\includegraphics[width = 0.49\linewidth, trim = 0 0 0 0, clip]{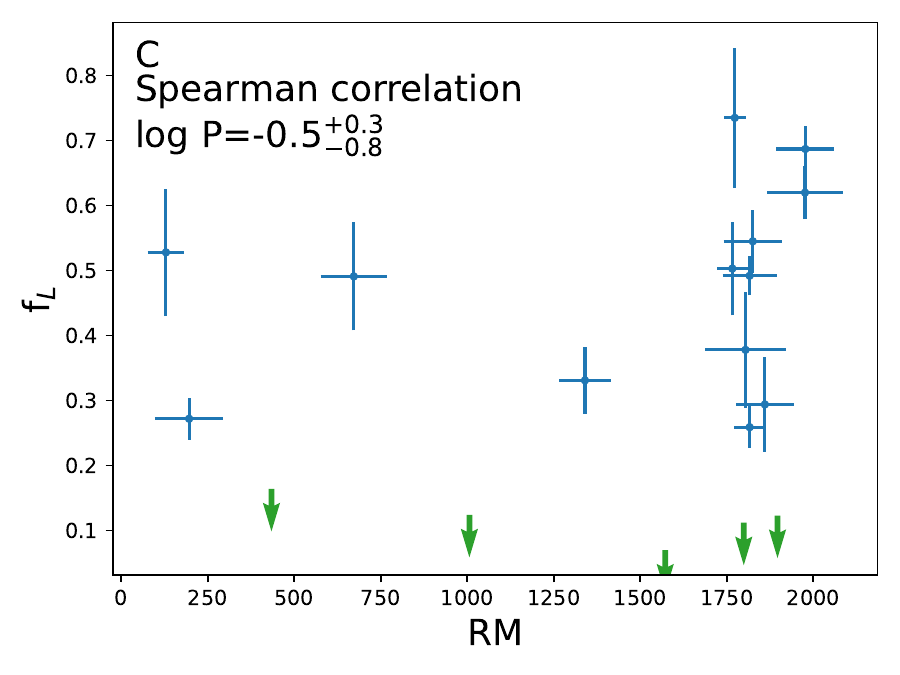}
\includegraphics[width = 0.49\linewidth, trim = 0 0 0 0, clip]{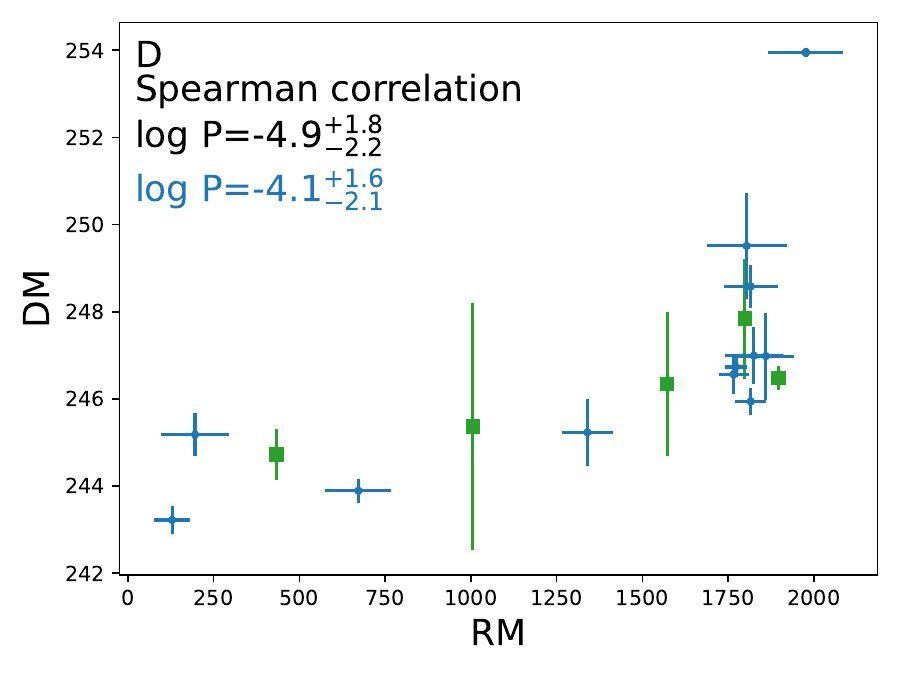}
\caption{
\textbf{Linear polarization degree and DM variations} 
{\bf (A)} The cumulative distribution of the linear polarization degree $f_\mathrm{L}$ 
during (orange) and outside (blue) 
the ``RM flare''. The gray region indicates the uncertainty of each data point. The null probability of the KS test $\mathrm{log}~P_\mathrm{KS}$ is labeled. 
{\bf (B)} The cumulative distribution of daily-averaged DM during the ``RM flare'' (orange) and outside of it (blue). 
{\bf (C)} The relation between RM and the linear polarization degree $f_\mathrm{L}$ within the ``RM flare'' episode. For bursts without detectable RM, $f_\mathrm{L}$ upper limits are presented as green arrows, and RMs are assumed to be the interpolated values of their temporal neighbors. The null probability of the Spearman correlation is labeled, \REVb{with the uncertainties estimated by the bootstrap method}. {\bf (D)} The relation between RM and $\rm DM_{stru}$ within the ``RM flare'' episode. Similar to panel C, the green squares represent unpolarized bursts, whose RMs are estimated by interpolation. \REVb{The $\rm log\ P$ in blue is estimated with the burst with both DM and RM measurements only (blue dots). }
}
\label{fig:flks_dmks}
\end{figure*} 

\begin{figure*}
\centering
\begin{tabular}{cccc}
\includegraphics[width = 1.\linewidth]{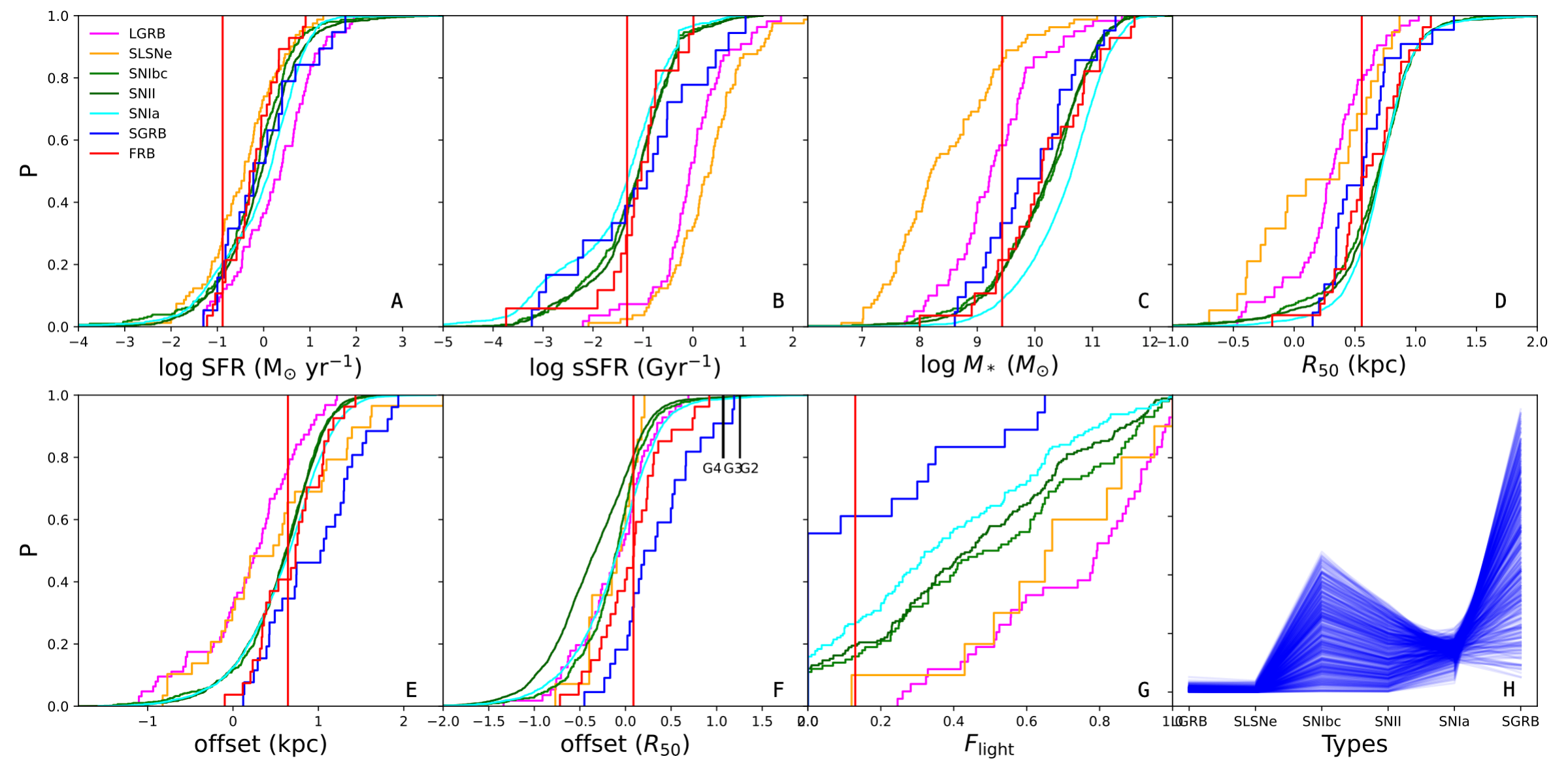} 
\\
\end{tabular}
\caption{
\textbf{Comparison of FRB 20220529 host galaxy properties with those of different types of stellar transients.}
}
\label{fig:host_comparison}
\end{figure*}

\begin{figure*}
\centering
\includegraphics[width = 0.9\linewidth, trim = 50 30 50 50, clip]{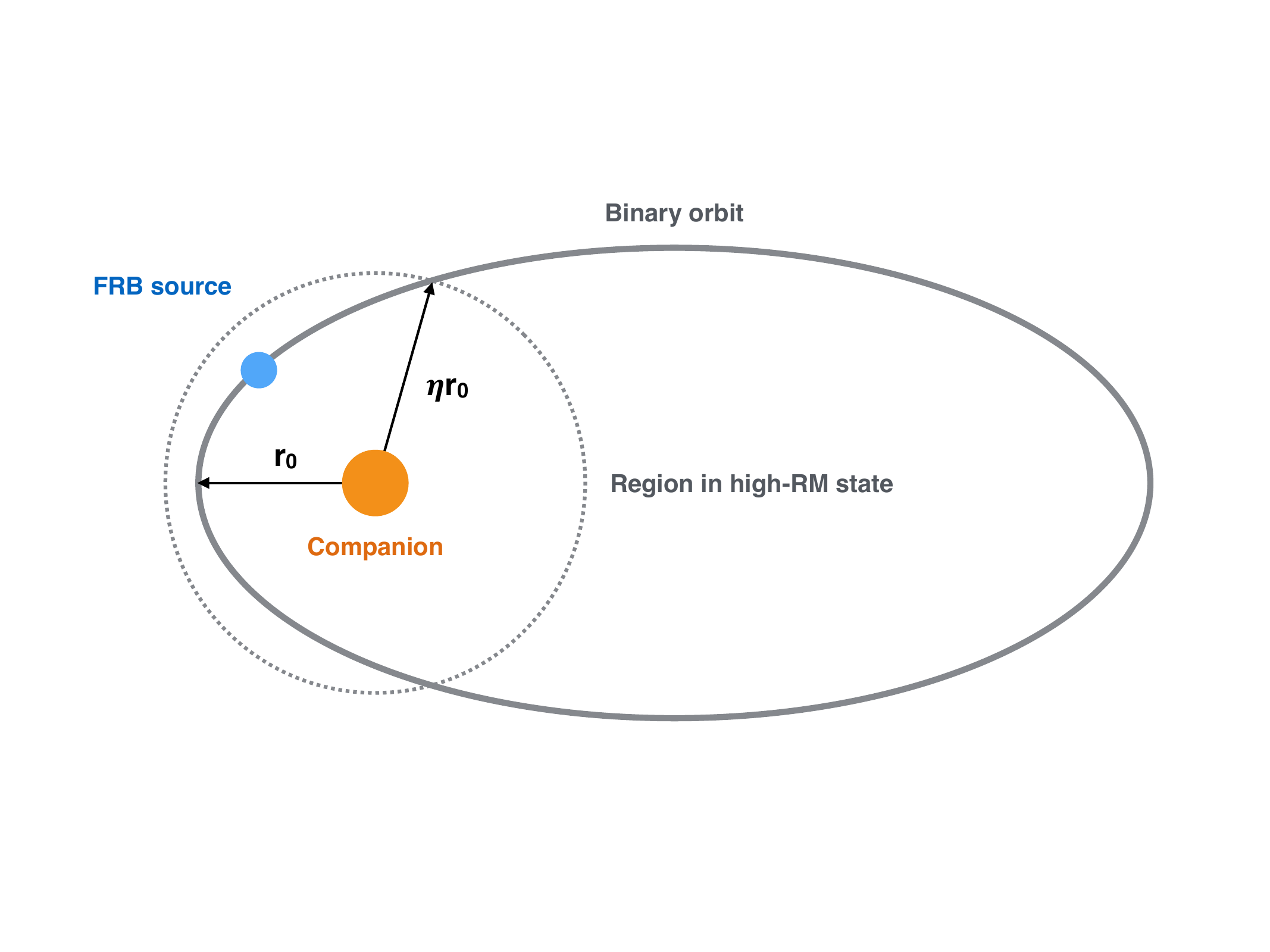}
\caption{\textbf{Schematic configuration of the ``RM flare'' generated by the orbital motion of a binary system.} Near the periastron, the RM could be much larger than that of other orbital phases due to the wind or disk of the companion.
}\label{fig:model_binary}
\end{figure*} 

\begin{figure*}
\centering
\includegraphics[width = 0.9\linewidth, trim = 0 0 0 0, clip]{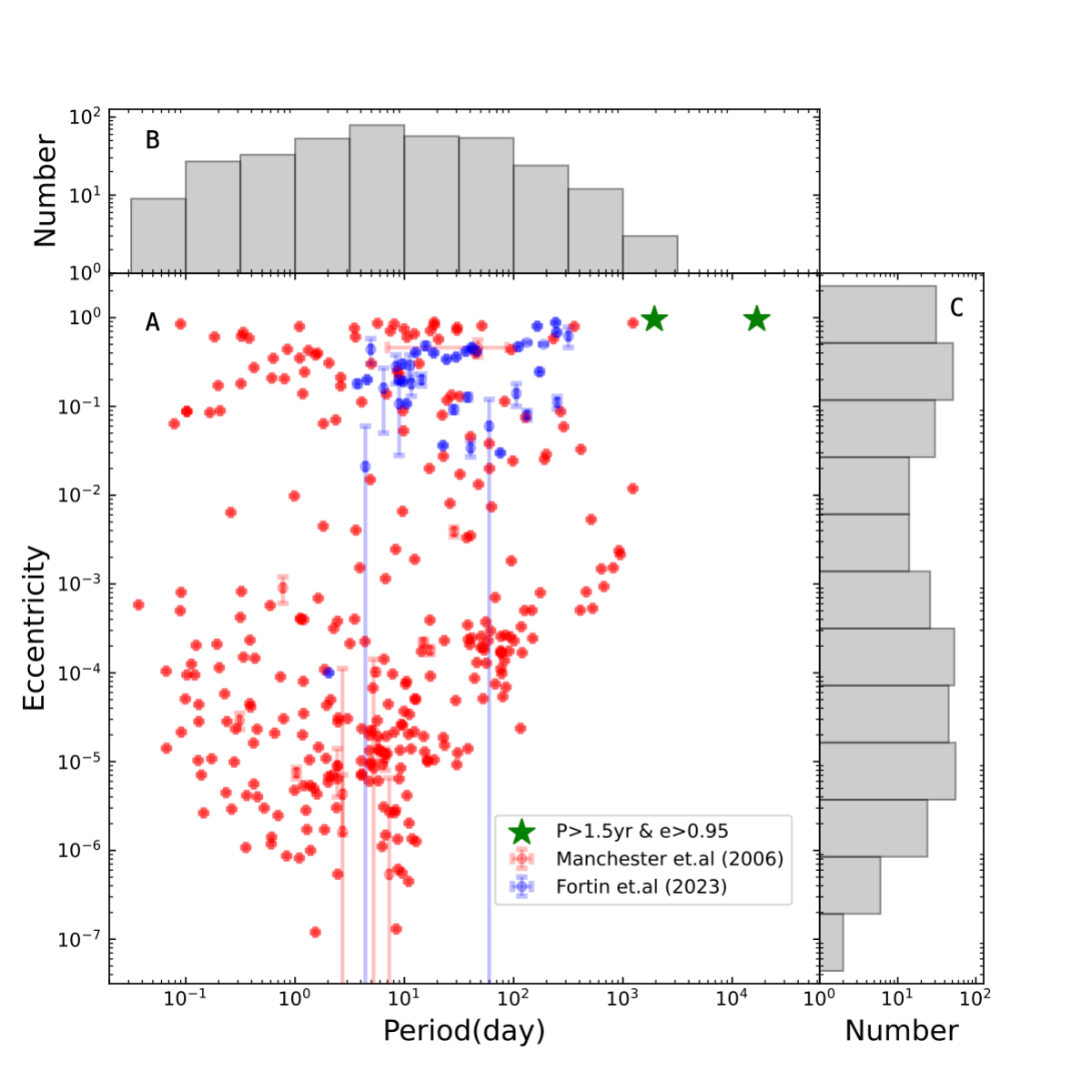}
\caption{\textbf{Two-dimensional distribution of the eccentricity and period of binary systems containing neutron stars.} The red and blue points correspond to the samples from references \cite{Manchester2006} and \cite{Fortin2023}, respectively.
The top and right panels display the distributions of the period and eccentricity of the sample, respectively. Two sources with a period $\gtrsim1.5$ yr and an eccentricity $\gtrsim0.95$ are shown as green star points.}
\label{fig:ecc}
\end{figure*}

\begin{figure*}
\centering
\includegraphics[width = 0.9\linewidth, trim = 50 30 50 50, clip]{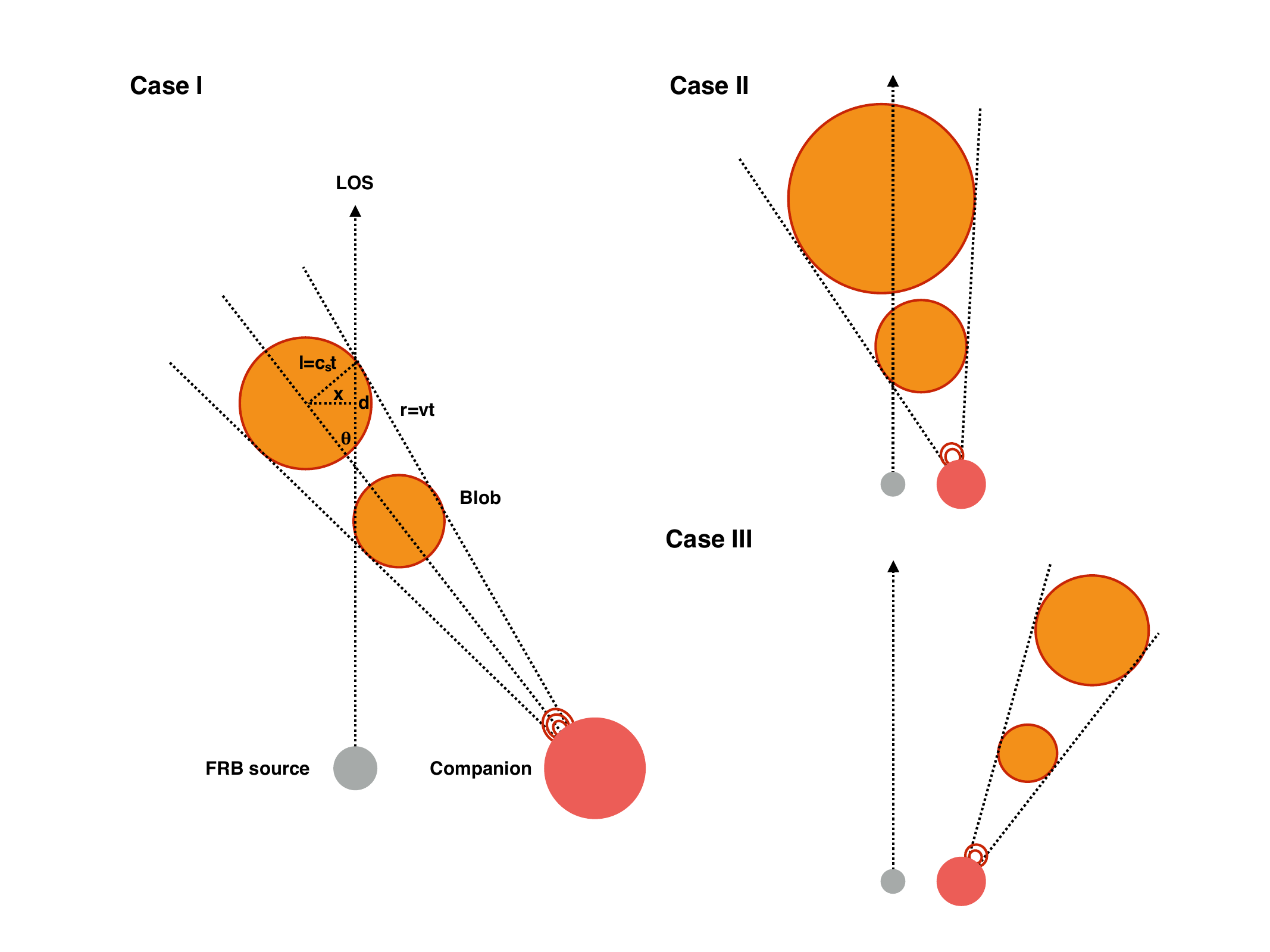}
\caption{\textbf{Schematic configuration of a coronal mass ejection (CME) from a companion star in a binary system}.
The gray circle denotes the FRB source, the red circle denotes the companion star, and the orange circles represent an expanding CME. Three cases are presented.
Case I: a CME moves across the line of sight (LOS) in a limited time. Case II: a CME is always on the LOS. Case III: a CME does not move across the LOS.}\label{fig:model}
\end{figure*}

\begin{figure*}
\centering
\includegraphics[width = 0.49\linewidth, trim = 0 0 0 0, clip]{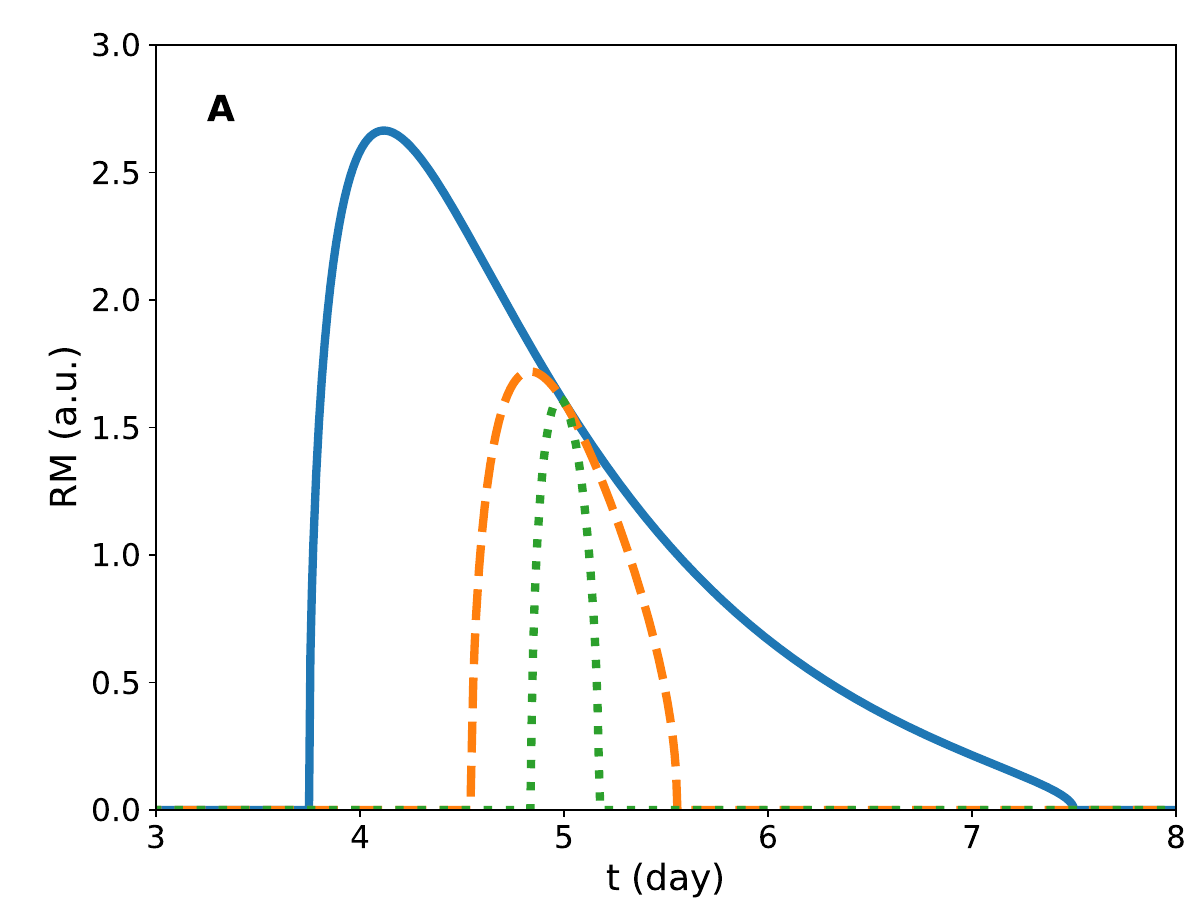}
\includegraphics[width = 0.49\linewidth, trim = 0 0 0 0, clip]{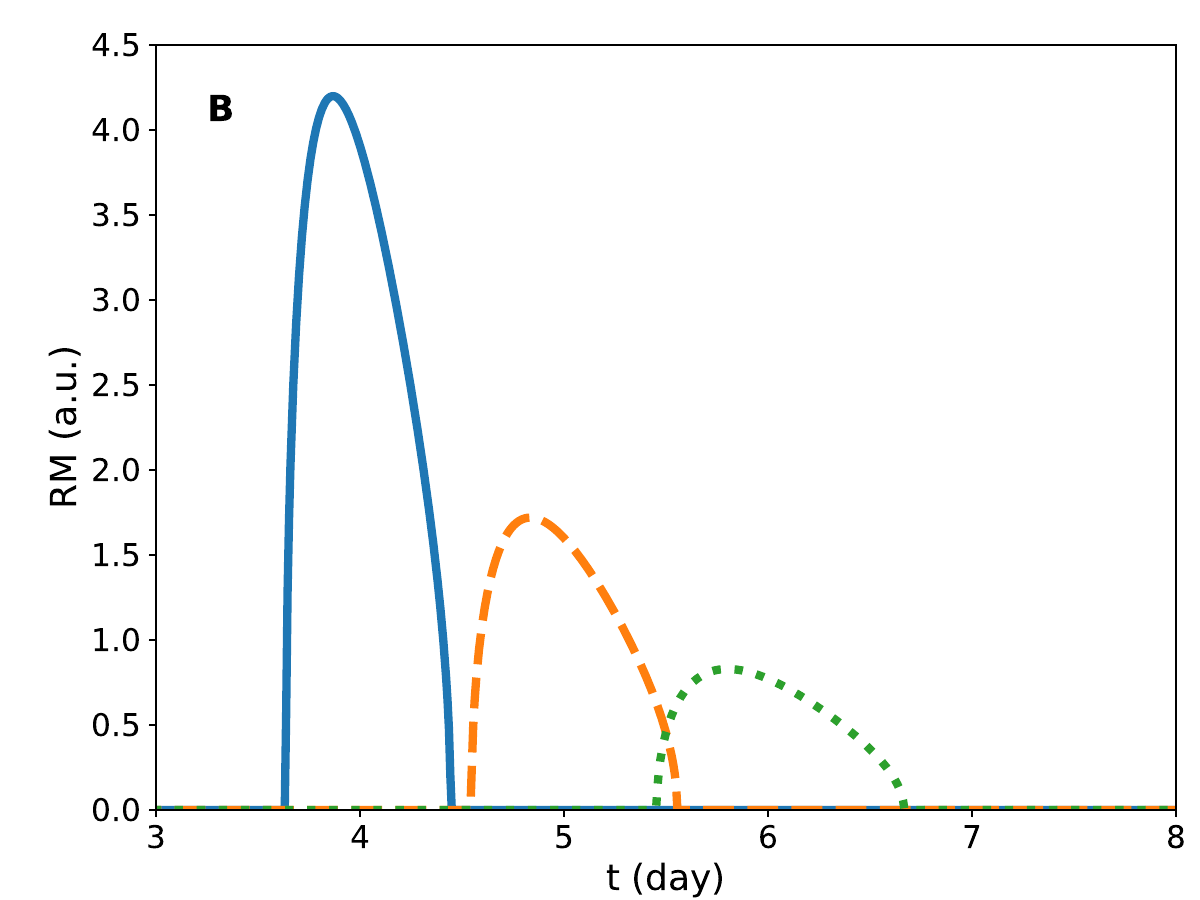}
\caption{\textbf{Predicted RM evolution for CME Case I model}. 
{\bf (A)} The solid, dashed, and dotted lines correspond to the RM evolution with $t_c=5~{\rm days}$ and $v\sin\theta/c_s=3,10,30$, respectively.
{\bf (B)} The solid, dashed, and dotted lines correspond to the RM evolution with $v\sin\theta/c_s=10$ and $t_c=4, 5, 6~ \mathrm{days}$, respectively. The phase-zero time corresponds to the explosion time of the CME.
}\label{result}
\end{figure*}

\begin{table*}[htb]
\centering
\caption{\textbf{Observation log of FRB 20220529 in FAST and Parkes during the ``RM flare'' phase.}}
\setlength{\tabcolsep}{0.7mm}{
\begin{tabular}{cccccccc}
\hline
UTC & observational duration & $N_{\rm FRB}$ & $R_{\rm FRB}$ &  \\
 & seconds & & hour$^{-1}$ \\
\hline
\multicolumn{4}{c}{FAST} \\\hline
2024-01-29  & 1200 & 1 & 3 \\
2024-01-10  & 1800 & 0 & 0 \\
2024-01-04  & 1800 & 0 & 0 \\
2023-12-29 	& 3600 & 1 & 1 \\
2023-12-28 	& 1200 & 4 & 12 \\
2023-12-24 	& 2400 & 4 & 6 \\
2023-12-20 	& 2400 & 2 & 3 \\
2023-12-17 	& 3600 & 15 & 15 \\
2023-12-14 	& 1200 & 4 & 12 \\
2023-11-22 	& 1200 & 0 & 0 \\
2023-11-07 	& 1200 & 0 & 0 \\
2023-10-18 	& 1200 & 3 & 9 \\
\hline
\multicolumn{4}{c}{Parkes} \\\hline
2024-01-29 & 12660 & 0 & 0 \\
2024-01-17 & 8700 & 1 & 0.4 \\
2024-01-11 & 4020 & 0 & 0 \\
2024-01-06 & 6840 & 0 & 0 \\
2024-01-05 & 6420 & 0 & 0 \\
2024-01-03 & 5940 & 1 & 0.6 \\
2023-12-22 & 12000 & 5 & 1.5 \\
2023-12-21 & 9000 & 1 & 0.4 \\
\hline
\end{tabular}
}
\label{table:obs}
\end{table*}

\begin{table*}[htb]
\centering
\caption{\textbf{Host Candidates}}
\setlength{\tabcolsep}{0.9mm}{
\begin{tabular}{cccccccccc}
\hline
ID & $\rm RA$ & $\rm Dec$ & $m_{\rm r}$ & $R_{50}$ & $R_\mathrm{off}$ & $R_\mathrm{off}/R_\mathrm{50}$ & $P_{\rm cc}$ & $\rm P(O|x)$\\
& $\rm deg$ & $\rm deg$ & $\rm mag$ & '' & " & \\\hline
G1 & 19.10450 & 20.63269 & 21.24 $\pm$ 0.03 & 1.17 & 1.4 & 1.2 & 0.013 & 0.999 \\
G2 & 19.10570 & 20.63302 & 21.64 $\pm$ 0.02 & 0.30 & 5.5 & 18.0 & 0.070 & 7e-7 \\
G3 & 19.10277 & 20.63320 & 20.30 $\pm$ 0.01 & 0.48 & 5.7 & 11.9 & 0.028 & 4e-4 \\
G4 & 19.10558 & 20.63399 & 20.05 $\pm$ 0.01 & 0.63 & 7.4 & 11.7 & 0.038 & 4e-4 \\
\hline
\end{tabular}
}
\label{tb_host}
\end{table*}   

\begin{table*}
\centering
\setlength{\tabcolsep}{2pt}
\begin{threeparttable}
\caption{\textbf{Model fitting results.}}
\begin{tabular}{cccccccccc}
\hline
model & $t_0$ & log A & $\Gamma/b^\dagger$ & $t_{\rm c}$ & $-2{\rm log}L$ & dof & AIC & BIC \\
  &  days & & & & days & & \\\hline
power law & -56.1$^{+15.6}_{-1.4}$ & 15.1$^{+0.3}_{-4.2}$ & 6.67$^{+0.03}_{-1.79}$ & - & 44.5 & 11 & 50.5 & 52.4 \\
CME case I & -20 & 8.76$^{+0.02}_{-0.01}$ & 10.8$^{+0.9}_{-3.4}$ & 25.6$^{+1.1}_{-0.3}$ & 5.1 & 11 & 11.1 & 13.0 \\
           & -27 & 9.21$^{+0.01}_{-0.01}$ & 16.8$^{+4.0}_{-2.0}$ & 32.4$^{+0.3}_{-0.6}$ & 3.3 & 11 & 9.3 & 11.2 \\
           & -50 & 10.2$^{+0.01}_{-0.01}$ & 55.5$^{+5.4}_{-18.3}$ & 53.9$^{+0.5}_{-0.6}$ & 5.3 & 11 & 11.3 & 13.2 \\
           & -100 & 11.31$^{+0.02}_{-0.01}$ & 146.6$^{+21.6}_{-70.9}$ & 102.3$^{+0.7}_{-2.7}$ & 5.8 & 11 & 11.8 & 13.7 \\
CME case II & -20.6$^{+0.6}_{-0.6}$ & $8.72^{+0.04}_{-0.05}$ & 4.0 & - & 106.8 & 12 & 110.8 & 112.1 \\
\hline
\end{tabular}
$^\dagger$ $b=v^2\sin^2\theta/c_s^2$ for CME Case I model
\label{table: model_nobkg}
\end{threeparttable}
\end{table*}

\clearpage
\begin{table*}
\centering
\renewcommand\arraystretch{0.8}
\begin{threeparttable}
\caption{\REV{\textbf{Properties of FRB 20220529 bursts without detectable RM during the ``RM flare'' phase.} Each burst was detected with an S/N $\geq$ 10. }}
\begin{tabular}{lccccccc}
\hline
$\rm MJD^a$ & S/N$^b$ & ${\rm DM}_{\rm det}^b$ & $f_{\rm low}^c$ &  $f_{\rm high}^c$  &  ${\rm DM}_{\rm stru}^d$  & $f_{\rm L, lim}$$^e$  \\
  &       &  ($\rm pc~cm^{-3}$) &  (MHz)   &  (MHz)         &   ($\rm pc~cm^{-3}$)                         & ($\%$) \\
\hline
60292.5869571 & 12 & 239.2 & 1000 & 1090 & 246.5 $\pm$ 0.3 & 12 \\
60295.5422626 & 10 & 248.3 & 1000 & 1200 & 247.8 $\pm$ 1.4 & 11 \\
60295.5647276 & 12 & 250.2 & 1000 & 1150 & 246.3 $\pm$ 1.6 & 7 \\
60298.5555579 & 10 & 250.4 & 1000 & 1100 & 245.4 $\pm$ 2.8 & 12 \\
60306.3925821 & 10 & 249.3 & 1120 & 1380 & 244.7 $\pm$ 0.6 & 16 \\
60299.4266043$^f$ & 16 & 250.5 & 750 & 850 & 246.0 $\pm$ 0.2 & 7 \\
60300.3304764$^f$ & 10 & 248.0 & 720 & 840 & 244.6 $\pm$ 0.3 & 9 \\
60300.3530282$^f$ & 18 & 248.0 & 930 & 1200 & 244.3 $\pm$ 0.4 & 6 \\
\hline
\end{tabular}
\begin{tablenotes}
\item [a] MJDs are in barycentric dynamical time (TDB) and are referenced to infinite frequency.
\item [b] The detection DM at the maximum S/N. 
\item [c] $f_{\rm low}$ and $f_{\rm high}$ are the minimum and maximum frequency extent of the bursts, determined visually.
\item [d] The structure-maximizing DM obtained by \emph{\sc DM\_phase} with 1$\sigma$ uncertainties.
\item [e]  The 3$\sigma$ upper limit of linear polarization fraction.
\item [f] Bursts observed by Parkes. 
\end{tablenotes}
\label{table:withoutRM_rf}
\end{threeparttable}
\end{table*}
\clearpage 
\clearpage
\begin{longtable}{lcccccccc}
\caption{\REV{\textbf{Properties of FRB 20220529 bursts with RM measurements outside the ``RM flare'' phase.} Each burst was detected using the FAST telescope with an S/N $\geq$ 10. All errors represent the 1$\sigma$ uncertainties.\\
MJDs are in barycentric dynamical time (TDB) and are referenced to infinite frequency.\\
${\rm DM}_{\rm det}$ is the detected DM at the maximum S/N. \\
$f_{\rm low}$ and $f_{\rm high}$ are the minimum and maximum frequency extent of the bursts, determined visually.\\
${\rm DM}_{\rm stru}$ is the structure-maximizing DM obtained by \emph{\sc DM\_phase} with 1$\sigma$ uncertainties.\\
$f_{\rm L}$ is the percentage of unbiased linear polarization along with 1$\sigma$ uncertainty.}}\\
\hline
\footnotesize $\rm MJD$ & S/N & ${\rm DM}_{\rm det}$ & $f_{\rm low}$ &  $f_{\rm high}$  &  ${\rm DM}_{\rm stru}$  & $\rm RM$ & $f_{\rm L}$$^e$   \\
\footnotesize          &     & ($\rm pc~cm^{-3}$)  &   (MHz)   &  (MHz)         &   ($\rm pc~cm^{-3}$)                         & ($\rm rad~m^{-2}$) & ($\%$) \\
\hline
59758.9282742 & 27 & 255.4 & 1100 & 1500 & 253.5 $\pm$ 1.9 & 35 $\pm$ 14 & 83 $\pm$ 5 \\
59758.9556102 & 12 & 247.5 & 1000 & 1200 & 246.3 $\pm$ 0.3 & 48 $\pm$ 26 & 80 $\pm$ 7 \\
59760.9501019 & 19 & 252.1 & 1000 & 1300 & 251.9 $\pm$ 1.6 & 39 $\pm$ 16 & 96 $\pm$ 4 \\
59766.0138732 & 136 & 246.8 & 1000 & 1250 & 247.0 $\pm$ 0.2 & 26 $\pm$ 3 & 98 $\pm$ 0 \\
59766.0244801 & 10 & 247.2 & 1000 & 1250 & 246.3 $\pm$ 0.5 & 76 $\pm$ 45 & 75 $\pm$ 9 \\
\end{longtable}
\clearpage 
\clearpage
\begin{longtable}{lcccccc}
\caption{\REV{\textbf{Properties of FRB 20220529 bursts without detectable RM outside the ``RM flare'' phase.} Each burst was detected using the FAST telescope with an S/N $\geq$ 10. \\
MJDs are in barycentric dynamical time (TDB) and are referenced to infinite frequency.\\
${\rm DM}_{\rm det}$ is the detected DM at the maximum S/N. \\
$f_{\rm low}$ and $f_{\rm high}$ are the minimum and maximum frequency extent of the bursts, determined visually.\\
${\rm DM}_{\rm stru}$ is the structure-maximizing DM obtained by \emph{\sc DM\_phase} with 1$\sigma$ uncertainties.\\
 $f_{\rm L, lim}$ is the percentage of the 3$\sigma$ upper limit of linear polarization fraction.}}\\
\hline
$\rm MJD$ & S/N & ${\rm DM}_{\rm det}$ & $f_{\rm low}$ &  $f_{\rm high}$  &  ${\rm DM}_{\rm stru}$  & $f_{\rm L, lim}$  \\
          &     & ($\rm pc~cm^{-3}$)  &   (MHz)   &  (MHz)         &   ($\rm pc~cm^{-3}$)                         & ($\%$) \\
\hline
59805.8653515 & 11 & 253.1 & 1000 & 1200 & 251.8 $\pm$ 1.0 & 9 \\
59805.8794861 & 20 & 252.1 & 1000 & 1180 & 243.1 $\pm$ 0.9 & 6 \\
59808.7812704 & 10 & 247.7 & 1000 & 1500 & 247.4 $\pm$ 0.1 & 24 \\
59808.7819025 & 10 & 245.0 & 1000 & 1100 & 246.6 $\pm$ 4.2 & 13 \\
59808.8224942 & 12 & 254.1 & 1000 & 1300 & 261.8 $\pm$ 4.5 & 15 \\
\end{longtable}
\clearpage 

\end{document}